\documentclass[%
  reprint,
  superscriptaddress,  % 著者 affiliations（所属）を脚注や並列表記ではなく、「著者名の上付き番号」で対応付ける形式に変更
  amsmath,amssymb,
  aps,
  pra,
  colorlinks
]{revtex4-2}

\usepackage{dcolumn}% Align table columns on decimal point
\usepackage{bm}% bold math
\usepackage{hyperref} % リンクを機能させる
\hypersetup{
  setpagesize=false,
  bookmarks=true,
  bookmarksdepth=tocdepth,
  colorlinks = true,       % true=文字に色, false=枠色
  linkcolor = blue,
  citecolor = blue,
  urlcolor = blue,
  pdftitle={R. Komatsudaira "Improvement of entanglement generation rate in frequency-multiplexed quantum repeaters using cavity-enhanced SPDC source"},  % unicodeにすることで文字化け防ぐ
  pdfauthor={Ryoma Komatsudaira}
}

\usepackage{dsfont} % mathdsで黒板太字を使えるように、特に1の太字(uplatexの場合)

\usepackage{mathtools}  % := のために導入
\makeatletter
\def\vmargin@#1#2#3{%% #3 with upper-margin #1, lower-margin #2
\setbox0=\hbox{#3}%
\rule[#1]{0pt}{\ht0}%
\lower\dp0\hbox{\rule[-#2]{0pt}{\dp0}}%
\box0%
}
\def\vmargin#1#2#3{%% #3 with upper-margin #1, lower-margin #2
\mathchoice
{\vmargin@{#1}{#2}{$\displaystyle #3$}}
{\vmargin@{#1}{#2}{$\textstyle #3$}}
{\vmargin@{#1}{#2}{$\scriptstyle #3$}}
{\vmargin@{#1}{#2}{$\scriptscriptstyle #3$}}
}
\makeatother
\newcommand{\indexrm}[1]{{\vmargin{0.25ex}{0ex}{\mathrm{#1}}}}

\usepackage{amssymb}    % \thereforeを使うために導入
\allowdisplaybreaks[1]  % align環境などにおける改行でページをまたぐかどうか0~4 \\*で改ページしない改行

\usepackage{comment}    % 複数行をコメントアウトする
\usepackage{tabularx}   % 列幅いっぱいに広げる表
\usepackage{multirow}   % 表中で複数行にまたがる文字を書くとき
\usepackage{listings}   % ソースコードを直接Latexに挿入できる
\usepackage{graphicx}   % 文字サイズをピンポイントで変える\scalebox{h-scale}[v-scale]{text}とか
\usepackage{bxtexlogo}
\usepackage{upgreek}    % ギリシャ文字の立体を使うために導入（\upalpha, \upAlphaなど）
\usepackage{amsmath}
\usepackage{mathrsfs}   % 花文字用
\usepackage{float}
\usepackage{array}
\usepackage{siunitx}    % \SIや\siを使うために導入
\usepackage{physics}    % \vbとか使うために導入
\AtBeginDocument{\RenewCommandCopy\qty\SI}	% qtyコマンドをsiで使いたい
\ExplSyntaxOn
\msg_redirect_name:nnn { siunitx } { physics-pkg } { none }
\ExplSyntaxOff
\usepackage{enumitem}   % enumerate環境の拡張のために導入（descriptionの見出しの長さ決定にも？）
\usepackage{ascmac}     % 枠囲みのために導入
\usepackage{color}      % 文字色を変えるために導入
\usepackage{braket}     % ブラケット記法を使用するために導入
\usepackage[most]{tcolorbox}
\usepackage{blkarray}   % 行列の外側に注釈をつける
\usepackage[normalem]{ulem}       % 二重下線など  → 引用文献欄の斜体が下線になってしまう？  → normalemオプションをつけることで解決
\usepackage{tikz}       % グラフ
\usepackage{pgfplots}
\usetikzlibrary{arrows.meta}
\usetikzlibrary{intersections,calc,arrows.meta,shapes.arrows}
\usetikzlibrary{positioning}  % 数式の背景に色付けするときの下部の補足用
\usepackage{multifootnote}  %複数の脚注をつける用
\usepackage{mhchem}     % 化学式用\ce{}
\usepackage{ragged2e}   % \justifyingで両端揃えにする
\usepackage{multirow}		% 表におけるセルの結合
\usepackage{booktabs}   % 表中に水平な空白行を追加するaddlinespace用
\usepackage{afterpage}  % afterpageコマンドとして追加

\makeatletter
\let\MYcaption\@makecaption
\makeatother
  \usepackage{subcaption}
\makeatletter
\let\@makecaption\MYcaption
\makeatother
\DeclareMathOperator{\arcsinh}{arcsinh}

\DeclareMathOperator{\sinc}{sinc}

\DeclareMathOperator{\Arg}{Arg}

\definecolor{BlueTOL}{HTML}{222255}
\definecolor{BrownTOL}{HTML}{666633}
\definecolor{GreenTOL}{HTML}{225522}
\definecolor{Purple}{HTML}{911146}
\definecolor{Orange}{HTML}{CF4A30}
\definecolor{royalblue}{HTML}{4169e1}
\definecolor{dodgerblue}{HTML}{1e90ff}
\definecolor{orangered}{HTML}{ff4500}
\definecolor{teal}{HTML}{008080}
\definecolor{mediumseagreen}{HTML}{3cb371}
\definecolor{darkslategray}{HTML}{2f4f4f}
\definecolor{myred}{HTML}{ee0066}
\definecolor{mygreen}{HTML}{339999}
\definecolor{textgray}{HTML}{23363A}
\definecolor{mediumturquoise}{HTML}{48d1cc}
\definecolor{myblue}{HTML}{54B2C4}
\definecolor{mypurple}{HTML}{504d71}
\definecolor{PumpBlue}{HTML}{546CE2}
\definecolor{IdlerRed}{HTML}{C34545}
\definecolor{SignalOrange}{HTML}{F88220}
\definecolor{DesmosRed}{HTML}{B44743}
\definecolor{DesmosBlue}{HTML}{2371C0}
\definecolor{DesmosGreen}{HTML}{448352}
\definecolor{SoftGreentea}{HTML}{6DC40E}
\definecolor{EntangleYellow}{HTML}{FFC319}
\definecolor{SoftBlue}{HTML}{7CB7E8}
\definecolor{PMGreen}{HTML}{579D0B}
\definecolor{PumpBlueBlack}{HTML}{3F52B3}
\definecolor{AiryRed}{HTML}{EE0808}
\definecolor{LemmonYellow}{HTML}{F5FF2D}
\definecolor{lossIdlerPurple}{HTML}{825ECA}
\definecolor{lossDetectorPink}{HTML}{F26FF5}
\definecolor{lossSignalGreen}{HTML}{9ADE82}
\definecolor{SubGray}{HTML}{3A3A3A}
\definecolor{TitleBlue}{HTML}{215F9A}
\definecolor{PosterTextGray}{HTML}{404040}
\definecolor{BackGray}{HTML}{F2F2F2}
\definecolor{DisplaceBlue}{HTML}{4169E1}
\definecolor{CovarianceOrange}{HTML}{FF4500}
\definecolor{ComplementGray}{HTML}{AEAEAE}
\definecolor{ReplyPurple}{HTML}{A800FF}

\newcommand{\myeraseall}[1]{}

\newcommand{\absolute}[1]{\left|#1\right|}

\newcommand{\hamiltonian}{H}  % 後で置き換えられるように

\newcommand{\finesse}{\mathcal{F}}  % 後で置き換えられるように
\newcommand{\fwhm}{\varGamma}
\newcommand{\fsr}{\mathrm{FSR}}

\newcommand{\indSignal}{\indexrm{S}}
\newcommand{\indIdler}{\indexrm{I}}
\newcommand{\indPump}{\indexrm{P}}

\newcommand{\Airy}[3][]{\mathcal{A}_#2#1 (#3)}

\newcommand{\fidelity}{F}
\newcommand{\prob}{\mathcal{P}}

\begin{document}

\title{Improvement of entanglement generation rate in frequency-multiplexed\\ quantum repeaters using cavity-enhanced SPDC source}

\author{Ryoma Komatsudaira}
  \email{komatsudaira-ryoma-tp@ynu.jp}
  \affiliation{
    Department of Physics, Yokohama National University, 79-5 Tokiwadai, Hodogaya-ku, Yokohama 240-8501, Japan
  }%
  \affiliation{
    LQUOM Inc.,
    79-5 Tokiwadai, Hodogaya-ku, Yokohama 240-8501, Japan
  }%

\author{Tomoyuki Horikiri}%
  \affiliation{
    Department of Physics, Yokohama National University, 
    79-5 Tokiwadai, Hodogaya-ku, Yokohama 240-8501, Japan
  }%
  \affiliation{
    LQUOM Inc., 
    79-5 Tokiwadai, Hodogaya-ku, Yokohama 240-8501, Japan
  }%
  \affiliation{
    Institute for Multidisciplinary Sciences, 
    Yokohama National University, 79-5 Tokiwadai, Hodogaya-ku, Yokohama 240-8501, Japan
  }%

  \date{March 29, 2026}

\begin{abstract}
  High-rate entanglement generation is essential for the realization of practical quantum repeaters. 
  To this end, frequency multiplexing of the photons employed is an effective approach. 
  In particular, schemes using cavity-enhanced spontaneous parametric down-conversion (cSPDC) as a photon-pair source have been proposed. 
  In this study, toward a theoretical performance evaluation of frequency-multiplexed quantum repeaters based on cSPDC, we derive an approximate expression for the quantum state of the frequency-multiplexed photons, where each frequency mode is treated as an independent two-mode squeezed vacuum (TMSV) state. 
  Using this expression, we calculate the heralding probability and fidelity of entanglement generation for several cases in a single-photon interference scheme using frequency multiplexing with cSPDC.
  Our results demonstrate that by multiplexing approximately 100 modes, the heralding probability improves to approximately \qty{98}{\%} for an elementary link distance $L_\indexrm{EL}=\qty{25}{km}$, even in scenarios where the fidelity exceeds 0.9 across all modes. 
  Furthermore, for $L_\indexrm{EL}=\qty{100}{km}$, we show that the heralding probability, which was approximately \qty{0.7}{\%} in the single-mode case, increases to about \qty{38}{\%} under the condition that the fidelity remains 0.9 or higher. 
  These analytical results demonstrate the effectiveness of employing cSPDC as a photon-pair source for quantum repeaters.
\end{abstract}

\maketitle

%\tableofcontents

\section{\label{sec:intro}Introduction}
  To realize the quantum internet, technologies for sharing entanglement over long distances are being developed worldwide. Overcoming photon loss in optical fibers necessitates the introduction of quantum repeaters \cite{BriegelZoller1998, MuralidharanJiang2016, WehnerHanson2018, Kimble2008}. 
  A typical quantum repeater features quantum memories at multiple intermediate nodes and progressively connects entanglement generated between adjacent nodes through entanglement swapping. 
  This approach allows the suppression of distance-dependent losses to a polynomial rather than an exponential scaling.

  Recently, frequency multiplexing has attracted significant attention as an approach to achieve high-throughput quantum repeaters \cite{PuigibertTittel2017, RielanderRiedmatten2014}. 
  Frequency multiplexing is an effective method that simultaneously utilizes multiple independent spectral modes, thereby scaling the entanglement generation rate with the number of modes. 
  The key to its realization lies in the development of multimode entanglement sources and compatible quantum memory technologies \cite{SeriRiedmatten2019, ItoHorikiri2023}.

  Cavity-enhanced spontaneous parametric down-conversion (cSPDC) is widely used as a source suitable for frequency multiplexing \cite{GotoKobayashi2003,  ScholzBenson2009Statistics, JeronimoURen2010}. 
  In SPDC, a pump beam incident on a nonlinear crystal generates pairs of signal and idler photons that satisfy energy conservation and phase-matching conditions. 
  However, multi-photon pairs can be generated probabilistically, which serves as a primary factor in degrading the fidelity of the generated entanglement. For this reason, multiphoton suppression and optimization of the pair generation probability are critical challenges when utilizing SPDC as an entanglement source.

  Despite its importance, when employing cSPDC as an entanglement source, the influence of its multimode nature on the photon-number distribution and the properties of the generated entanglement is not yet fully understood. 
  Specifically, while the photon-pair distribution in single-mode SPDC is well known to follow a thermal distribution, existing analyses of multimode cSPDC have often been restricted to the weak-pumping regime \cite{ScholzBenson2009Analytical, LuoSilberhorn2015, YamazakiYamamoto2022}. 
  Consequently, a quantitative understanding of inter-mode correlations and a comprehensive theoretical formulation of the photon-pair distribution in multimode SPDC have remained limited.

  In this work, we address this gap by deriving an approximate expression for the quantum state of multimode cSPDC that explicitly includes contributions from multiple photon pairs. 
  We show that the multimode SPDC state can be represented as a tensor product of two-mode squeezed vacuum (TMSV) states, each corresponding to an individual frequency mode. 
  This result ensures that, when utilizing cSPDC as a source, each mode can be treated as an independent photon-pair state.

  Furthermore, based on this representation, we consider, for each frequency mode, an entanglement generation scheme via single-photon interference, where an entangled state with at most one excitation is shared between two nodes \cite{DLCZ2001}.
  For each individual mode, we calculate the heralding probability and fidelity, with the former serving as a metric for the entanglement generation rate.
  Using these mode-wise results, we evaluate the overall heralding probability and fidelity for the multimode case and compare them with those of the single-mode case. This comparison demonstrates that the multimode configuration distributes photon generation across multiple modes, thereby suppressing the multi-photon generation probability within each mode. Consequently, we show that this approach not only mitigates the degradation of entanglement fidelity but also enhances both the heralding probability and the overall generation rate, even in regimes where high fidelity is maintained.

  These results provide design guidelines for frequency-multiplexed quantum repeaters utilizing cSPDC sources and establish a theoretical foundation for realizing highly efficient, high-fidelity multimode entanglement generation systems.

\section{\label{sec:theory}Theory of photon-pair source}
  \subsection{\label{subsec:SPDC} SPDC}
    The effective Hamiltonian for SPDC is given by \cite{ChristSilberhorn2011}:
    \begin{align}
      \hat{\hamiltonian}_\indexrm{eff} \coloneq \alpha\! 
          \int_{0}^{\infty} \!\!\! \dd{\omega_{\indSignal}}
          \int_{0}^{\infty} \!\!\! \dd{\omega_{\indIdler}} \,
          f(\omega_\indSignal, \omega_\indIdler)
          \hat{a}\vphantom{a}^\dagger_\indSignal(\omega_\indSignal) \hat{a}\vphantom{a}^\dagger_\indIdler(\omega_\indIdler) 
          + \mathrm{H.c.}, 
    \end{align}
    where $\alpha$ is a constant, $\omega_\indSignal$ and $\omega_\indIdler$ are the angular frequencies of the signal and idler photons, and $\hat{a}\vphantom{a}^\dagger_\indSignal(\omega_\indSignal)$ and $\hat{a}\vphantom{a}^\dagger_\indIdler(\omega_\indIdler)$ are their respective creation operators. $\mathrm{H.c.}$ denotes the Hermitian conjugate.

    The term $f(\omega_\indSignal, \omega_\indIdler)$ is referred to as the joint spectral amplitude (JSA). The JSA is defined as the product of the pump envelope function (PEF), $s(\omega_\indSignal+ \omega_\indIdler)$, which represents the pump spectrum, and the phase-matching function (PMF), $h(\omega_\indSignal,\omega_\indIdler)\coloneq \sinc(\frac{\Delta k L_\indexrm{c}}{2})e^{i\frac{\Delta k L_\indexrm{c}}{2}}$, such that
    \begin{align}
      f(\omega_\indSignal,\omega_\indIdler) = s(\omega_\indSignal+ \omega_\indIdler)\cdot h(\omega_\indSignal,\omega_\indIdler).
    \end{align}

    Here, $\Delta k \coloneq k_\indPump(\omega_\indPump) - k_\indSignal(\omega_\indSignal) - k_\indIdler(\omega_\indIdler)$ is the phase mismatch, where $k_\epsilon\,(\epsilon \in \{\mathrm{P, S, I}\})$ are the wavenumbers of the pump, signal, and idler fields, respectively. The parameter $L_\indexrm{c}$ denotes the crystal length. Furthermore, energy conservation is assumed to be satisfied, such that $\omega_\indPump=\omega_\indSignal+\omega_\indIdler$.

\begin{figure}[H]
      \centering
      \includegraphics[width=1.0\columnwidth]{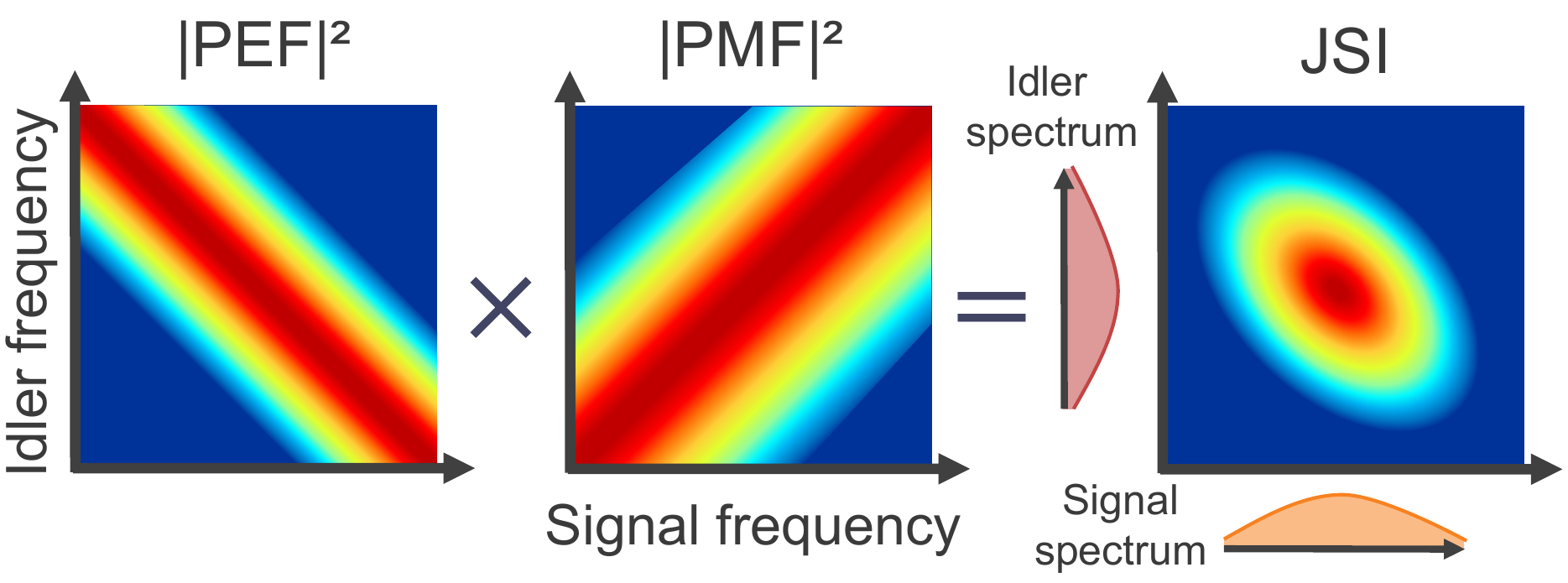}
      \caption{
        Schematic of the JSI. The profile shown here is for illustrative purposes only, as the pump spectral shape $s(\omega_\indSignal, \omega_\indIdler)$ depends on the specific pump source, and the phase-matching function $h(\omega_\indSignal, \omega_\indIdler)$ varies according to the crystal structure and the type of SPDC process.
      }
      \label{fig:JSI_SPDC}
    \end{figure}

    The JSA characterizes the frequency distribution of the signal and idler photons. Its squared magnitude,
    \begin{align}
      S (\omega_\indSignal,\omega_\indIdler) \coloneq \absolute{f (\omega_\indSignal,\omega_\indIdler)}^2,
    \end{align}
    is referred to as the joint spectral intensity (JSI), which represents the spectral intensity or the probability distribution of the generated photon pairs (Fig.~\ref{fig:JSI_SPDC}). 

    When the JSA is separable into a product of functions of the signal and idler frequencies, such that
    \begin{align}
      -\frac{i}{\hbar}\alpha f(\omega_\indSignal, \omega_\indIdler) = r \psi_\indSignal^\ast(\omega_\indSignal)\phi_\indIdler^\ast(\omega_\indIdler),
    \end{align}
    one can define creation and annihilation operators associated with the corresponding spectral modes as \cite{RohdeSilberhorn2007}:
    \begin{align}
      \hat{A} \coloneq \int_0^\infty\!\! \dd{\omega_\indSignal} \psi(\omega_\indSignal) \hat{a}_\indSignal(\omega_\indSignal), \quad
      \hat{B} \coloneq \int_0^\infty\!\! \dd{\omega_\indIdler} \phi(\omega_\indIdler) \hat{a}_\indIdler(\omega_\indIdler),
    \end{align}
    where $r \in \mathbb{R}^+$ is a constant, and the functions $\psi(\omega_\indSignal)$ and $\phi(\omega_\indIdler)$ are normalized spectral mode functions.
    
    Using these definitions, the Hamiltonian can be written as
    \begin{align}
      \hat{\hamiltonian}_\mathrm{eff} = i\hbar r (\hat{A}^\dagger\hat{B}^\dagger-\mathrm{H.c.}).
    \end{align}
    Consequently, the time-evolution operator is given by \cite{WallsMilburn2008, KilmovChumakov2009}:
    \begin{align}
      \hat{U} &= \exp \left[-\frac{i}{\hbar} \hat{\hamiltonian}_\indexrm{eff}\right]\notag\\
      &= \exp \left[r(\hat{A}^\dagger\!\hat{B}^\dagger-\mathrm{H.c.})\right] \notag\\
      &= e^{(\tanh r) \hat{A}^\dagger\hat{B}^\dagger} 
      e^{-\left(\ln(\cosh r)\right)(\hat{A}^\dagger\!\hat{A}+\hat{B}^\dagger\!\hat{B}+1)}
      e^{-(\tanh r) \hat{A}\hat{B}}.
    \end{align}
    
    Therefore, assuming the initial state is the vacuum $\Ket{0}_{\indSignal \indIdler}$, the state $\Ket{\Phi}$ generated by SPDC is expressed as
    \begin{align}
      \Ket{\Phi} &= \hat{U}\Ket{0}_{\indSignal \indIdler}\notag\\
      &= \frac{1}{\cosh r} \sum_{n=0}^\infty \frac{\tanh^n r}{n!}  \hat{A}^\dagger{\vphantom{\hat{A}}}^n \hat{B}^\dagger{\vphantom{\hat{B}}}^n \Ket{0}_{\indSignal \indIdler}\notag\\
      &=  \sum_{n=0}^\infty \frac{\tanh^n r}{\cosh r}  \Ket{n}_\indSignal\Ket{n}_\indIdler.
      \label{eq:thermal_multiPair}
    \end{align}

    The operator
    \begin{align}
      \hat{S}^{\mathrm{SI}}(-r)\coloneq \exp \left[r(\hat{A}^\dagger\!\hat{B}^\dagger-\mathrm{H.c.})\right]
    \end{align}
    is referred to as the two-mode squeezing operator, and the state in Eq.~\eqref{eq:thermal_multiPair} is called the two-mode squeezed vacuum (TMSV).

    Calculating the probability of generating an $n$-photon pair state, we obtain
    \begin{align}
      P(n) &= \absolute{\rule{0pt}{2.0ex}\Bra{n}_\indSignal\!\!\Bra{n}_\indIdler\Ket{\Phi}\,}^2 = \left(\frac{\tanh^n r}{\cosh r}\right)^2 \notag\\
      &= \frac{(\sinh^2 r)^n}{(\sinh^2 r +1)^{n+1}} = \frac{\mu^n}{(\mu+1)^{n+1}},
    \end{align}
    which indicates that the photon-pair distribution follows a thermal distribution with a mean photon number $\mu \coloneq \sinh^2 r$.

  \subsection{\label{subsec:cSPDC} cSPDC}
    \subsubsection{\label{subsubsec:JSAJSI} JSA and JSI}
      Cavity-enhanced SPDC (cSPDC) is a process in which a nonlinear crystal is placed within an optical cavity to induce parametric down-conversion. 
      In addition to enhancing the brightness of the emitted photons \cite{ZYOuYJLu1999}, this configuration allows for spectral narrowing of the linewidth to match the absorption bandwidth of quantum memories.
      
      While we assume a bow-tie cavity configuration in this work, the following discussion can be analogously applied to Fabry--Pérot cavities.

\begin{figure}[H]
        \centering
        \includegraphics[width=0.90\columnwidth]{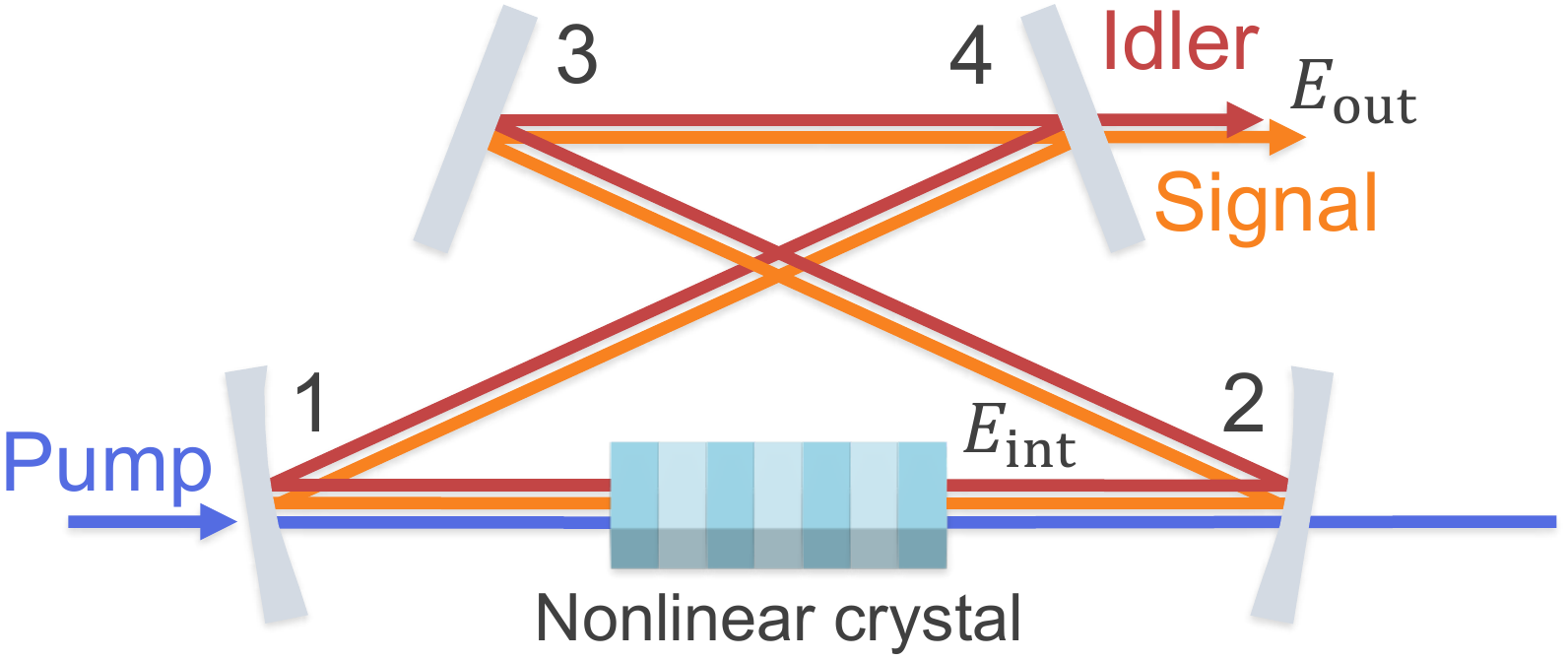}
        \caption{
          Cavity-enhanced SPDC with a bow-tie cavity. A nonlinear crystal is placed within the cavity and pumped externally. The cavity is designed to be resonant only at the signal and idler frequencies generated via SPDC. The output photons are extracted from a mirror that is independent of the pump injection and transmission paths.
        }
        \label{fig:Bow-tie_cavity}
      \end{figure}

      Considering the configuration shown in Fig.~\ref{fig:Bow-tie_cavity}, where photons are emitted through mirror 4, the relationship between the internal electric field immediately after generation within the crystal, $E_\indexrm{int}(\omega_\epsilon)$, and the output electric field transmitted through mirror 4, $E_\indexrm{out}(\omega_\epsilon)$, is given by:
      \begin{align}
        E_\indexrm{int}(\omega_\epsilon) = A_\epsilon(\omega_\epsilon) E_\indexrm{out}(\omega_\epsilon).
      \end{align}
      Here, $A_\epsilon(\omega_\epsilon)$ is the resonance function of the cavity, defined as:
      \begin{align}
        A_\epsilon(\omega_\epsilon)\coloneq \frac{\sqrt{R_{2,\epsilon} R_{3,\epsilon} (1-R_{4,\epsilon})}\,e^{-i\delta_\mathrm{loop}(\omega_\epsilon) \frac{L_{\mathrm{init},\epsilon}}{L_{\mathrm{opt},\epsilon}}}}{1-\sqrt{R_{1,\epsilon}R_{2,\epsilon}R_{3,\epsilon}R_{4,\epsilon}G_\epsilon} \,e^{-i\delta_\mathrm{loop}(\omega_\epsilon)}},
      \end{align}
      where the subscript $\epsilon\in\{\indSignal,\indIdler\}$ refers to the signal and idler, respectively. The terms $R_{j,\epsilon}\ (j=1,2,3,4)$ represent the reflectivity of each mirror, $G_\epsilon$ denotes the round-trip loss, and $\delta_\indexrm{loop}(\omega_\epsilon)\coloneq \omega_\epsilon L_{\indexrm{opt},\epsilon}/c$ is the round-trip phase shift. Furthermore, $L_{\indexrm{opt},\epsilon}$ is the total optical path length for one round trip, $L_{\mathrm{init},\epsilon}$ is the optical path length from the point of generation within the crystal to mirror 4, and $c$ is the speed of light. Here, the reflectivities, losses, and optical path lengths are treated as constants, assuming that their frequency dependence is negligible near the center frequencies of the signal and idler fields.

      Due to these relationships, the effective Hamiltonian for doubly resonant cSPDC, where both the signal and idler fields are resonant, is given by:
      \begin{align}
        \begin{split}
          \hat{\hamiltonian}_\indexrm{cav,eff} &\coloneq \alpha\! 
          \int_{0}^{\infty} \!\!\! \dd{\omega_{\indSignal}}
          \int_{0}^{\infty} \!\!\! \dd{\omega_{\indIdler}} \\
          &\qquad \times
          A_\indSignal(\omega_\indSignal) A_\indIdler(\omega_\indIdler)f(\omega_\indSignal, \omega_\indIdler)
          \hat{a}\vphantom{a}^\dagger_\indSignal(\omega_\indSignal) \hat{a}\vphantom{a}^\dagger_\indIdler(\omega_\indIdler) 
          + \mathrm{H.c.} 
        \end{split}
      \end{align}

      Based on this formulation, the JSA is modified as $f_\indexrm{cav} (\omega_\indSignal, \omega_\indIdler)$. 
      Here, we define $A_\epsilon^0(\omega_\epsilon)$ by normalizing $A_\epsilon(\omega_\epsilon)$ such that its maximum magnitude is unity. 
      Using this normalized function and its squared magnitude $\Airy[^0]{\epsilon}{\omega_\epsilon}\coloneq \absolute{A_\epsilon^0(\omega_\epsilon)}^2$, the JSA and JSI for the doubly resonant cSPDC case are accordingly modified as follows:
      \begin{align}
        f_\indexrm{cav} (\omega_\indSignal,\omega_\indIdler)\coloneq  A_\indSignal^0(\omega_\indSignal) A_\indIdler^0(\omega_\indIdler) f (\omega_\indSignal,\omega_\indIdler),\\
        S_\indexrm{cav} (\omega_\indSignal,\omega_\indIdler)\coloneq  \Airy[^0]{\indSignal}{\omega_\indSignal} \Airy[^0]{\indIdler}{\omega_\indIdler} S (\omega_\indSignal,\omega_\indIdler).
      \end{align}

      In general, the function
      \begin{align}
        \Airy{\epsilon}{\omega_\epsilon} &\coloneq \absolute{A_\epsilon(\omega_\epsilon)}^2\notag\\
        &= T_\mathrm{enh,\epsilon}\, \Airy[^0]{\epsilon}{\omega_\epsilon} 
      \end{align}
      is referred to as the Airy function \cite{JeronimoURen2010, LuoSilberhorn2015}, where $T_\mathrm{enh,\epsilon}$ represents the peak enhancement factor, corresponding to the maximum value of the Airy function:
      \begin{align}
        T_\mathrm{enh,\epsilon} \coloneq \frac{R_{2,\epsilon}R_{3,\epsilon}(1-R_{4,\epsilon})G_{\epsilon}}{(1-\sqrt{R_{1,\epsilon}R_{2,\epsilon}R_{3,\epsilon}R_{4,\epsilon}G_\epsilon})^2}.
      \end{align}
      Note that in some literature, $\Airy[^0]{\epsilon}{\omega_\epsilon}$ itself is referred to as the Airy function. Furthermore, it should also be cautioned that this resonance profile is entirely distinct from the Airy function $\mathrm{Ai}(\cdot)$ from the field of special functions, as they represent unrelated mathematical concepts.

      Furthermore, near the center frequencies of the signal and idler fields, $\Airy[^0]{\epsilon}{\omega_\epsilon}$ can be expressed in terms of the free spectral range (FSR),
      \begin{align}
        \fsr_\epsilon = \frac{c}{L_{\indexrm{opt},\epsilon}}  
      \end{align} 
      (in units of Hz), and the finesse,
      \begin{align}
        \finesse_\epsilon \simeq \frac{\pi \left(R_{1,\epsilon}R_{2,\epsilon}R_{3,\epsilon}R_{4,\epsilon}G_\epsilon\right)^{1/4}}{1-\left(R_{1,\epsilon}R_{2,\epsilon}R_{3,\epsilon}R_{4,\epsilon}G_\epsilon\right)^{1/2}},   
      \end{align}
      as follows:
      \begin{align}
        \Airy[^0]{\epsilon}{\omega_\epsilon} \simeq \frac{1}{1+\left(\frac{2\finesse_\epsilon}{\pi}\right)^2 \sin^2 \left(\frac{\pi}{\fsr_\epsilon}\nu_\epsilon\right)},
      \end{align}
      where $\nu_\epsilon = \omega_\epsilon/2\pi$.
      
      Depending on the literature, the JSA and JSI for cSPDC are sometimes defined as $A_\indSignal(\omega_\indSignal) A_\indIdler(\omega_\indIdler) f (\omega_\indSignal,\omega_\indIdler)$ and $\Airy{\indSignal}{\omega_\indSignal} \Airy{\indIdler}{\omega_\indIdler} S (\omega_\indSignal,\omega_\indIdler)$, respectively. In this work, however, we define the JSA and JSI for cSPDC such that the multiplicative resonance factors, $A_\epsilon^0$ and $\mathcal{A}_\epsilon^0$, are normalized to a peak value of unity to ensure consistency in our treatment.

      Under this convention, the effective Hamiltonian for the cSPDC case is expressed as:
      \begin{align}
        \begin{split}
          \hat{\hamiltonian}_\indexrm{cav,eff} \!&\coloneq \! \alpha^\prime \!\! 
          \int_{0}^{\infty} \!\!\! \dd{\omega_{\indSignal}} \!
          \int_{0}^{\infty} \!\!\! \dd{\omega_{\indIdler}}
          f_\indexrm{cav}(\omega_\indSignal, \omega_\indIdler)
          \hat{a}\vphantom{a}^\dagger_\indSignal(\omega_\indSignal) \hat{a}\vphantom{a}^\dagger_\indIdler(\omega_\indIdler) 
          \! + \! \mathrm{H.c.} 
        \end{split}
      \end{align}
      Here, $\alpha^\prime$ is a complex constant satisfying $\absolute{\alpha^\prime}^2 = \absolute{\alpha}^2 T_{\mathrm{enh},\indSignal} T_{\mathrm{enh},\indIdler}$.

\begin{figure}[H]        
        \centering
        \begin{tikzpicture}[xscale=0.5, yscale=2.25, samples=1000]
          \pgfmathsetmacro{\tikL}{10.0}
          \draw[SignalOrange, semithick] plot[domain=-8:8] (\x,{1/(1+100*(sin(\x r))^2)});
          \draw[->,>=stealth,semithick] (-8.0,0)--(8.0,0) node[right]{\scalebox{1.0}[1.0]{$x$}};
          \draw[->,>=stealth,semithick] (0,-0.2)--(0,1.25);
          \draw (0,0) node[below left]{\scalebox{1.0}[1.0]{$\mathrm{O}$}}
                (0,1) node[above left] {\scalebox{1.0}[1.0]{$1$}};
          \draw[solid,very thin] (5,0.025)--(5,-0.025)
            node[below]{\scalebox{1.0}[1.0]{$5$}};
          \draw[solid,very thin] (-5,0.025)--(-5,-0.025)
            node[below]{\scalebox{1.0}[1.0]{$-5$}};
          \node(P) at (3,1.15) [SignalOrange]{\scalebox{1.3}[1.3]{$\frac{1}{1+100\sin^2 x}$}};
        \end{tikzpicture}
        \vspace{-1ex}
        \caption{
          Typical profile of the Airy function.
        }
        \label{fig:AiryFunc}
      \end{figure}
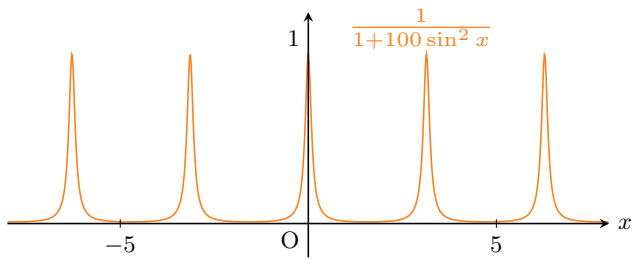

\begin{figure}[H]
        \vspace{-2ex}
        \centering
        \includegraphics[width=0.95\columnwidth]{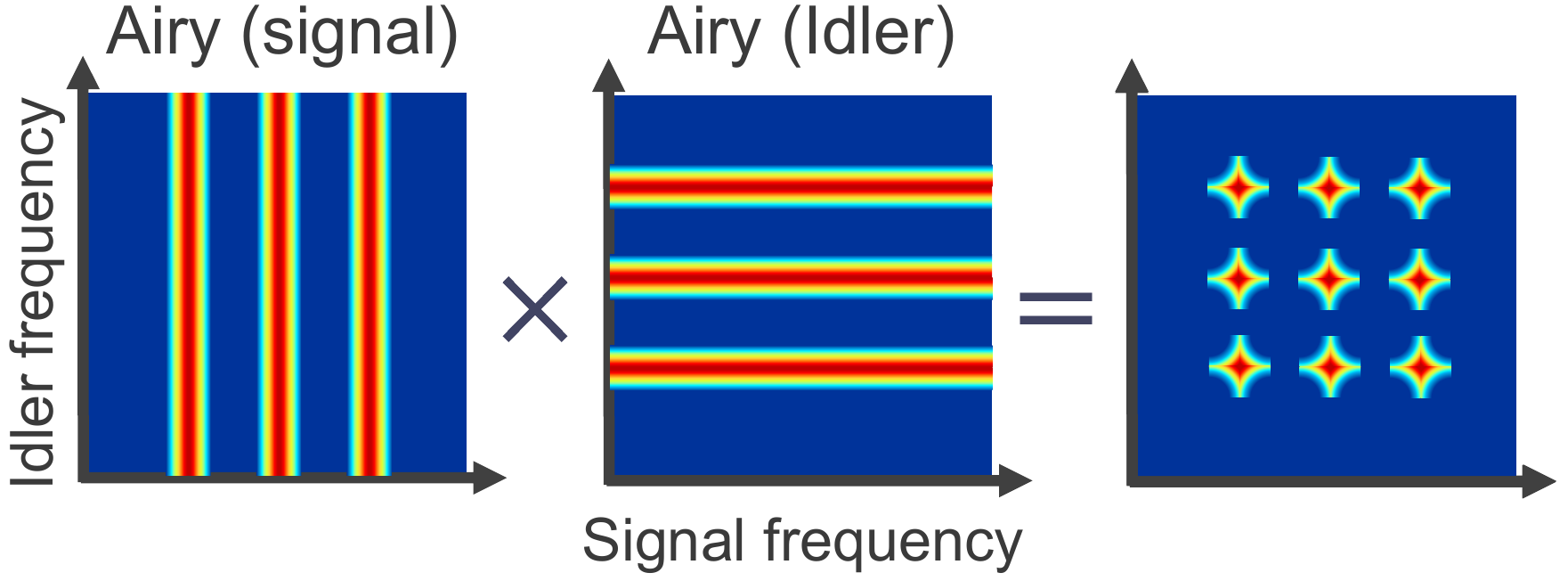}
        \vspace{-1ex}
        \caption{
          Schematic of the frequency spectrum in a doubly resonant cavity. The resulting spectral distribution is obtained by taking the product of the Airy functions for the signal and idler fields. Significant values are observed only at frequencies where the resonance conditions for both fields are simultaneously satisfied.
        }
        \label{fig:JSI_Airy}
      \end{figure}

      The Airy function exhibits a periodic profile with sharp peaks at the resonance frequencies, as illustrated in Fig.~\ref{fig:AiryFunc}. Consequently, the joint resonance structure---representing the product of the Airy functions for the signal and idler fields---forms a two-dimensional grid of peaks, as shown in Fig.~\ref{fig:JSI_Airy}.

    \subsubsection{\label{subsubsec:cluster}Cluster effect}
      As described above, in cavity-enhanced SPDC under doubly resonant conditions, both the intensity and generation probability of the signal-idler photon pairs are constrained by the profile of the Airy functions. 
      In this configuration, the signal and idler photons propagate through the nonlinear crystal within the cavity. Due to the frequency dependence of the refractive index (dispersion) within the crystal, the optical path lengths for the two fields differ even when they share the same physical cavity length. Consequently, the signal and idler fields exhibit distinct FSR values.

\begin{figure}[ht]
        \centering
        \includegraphics[width=1.0\columnwidth]{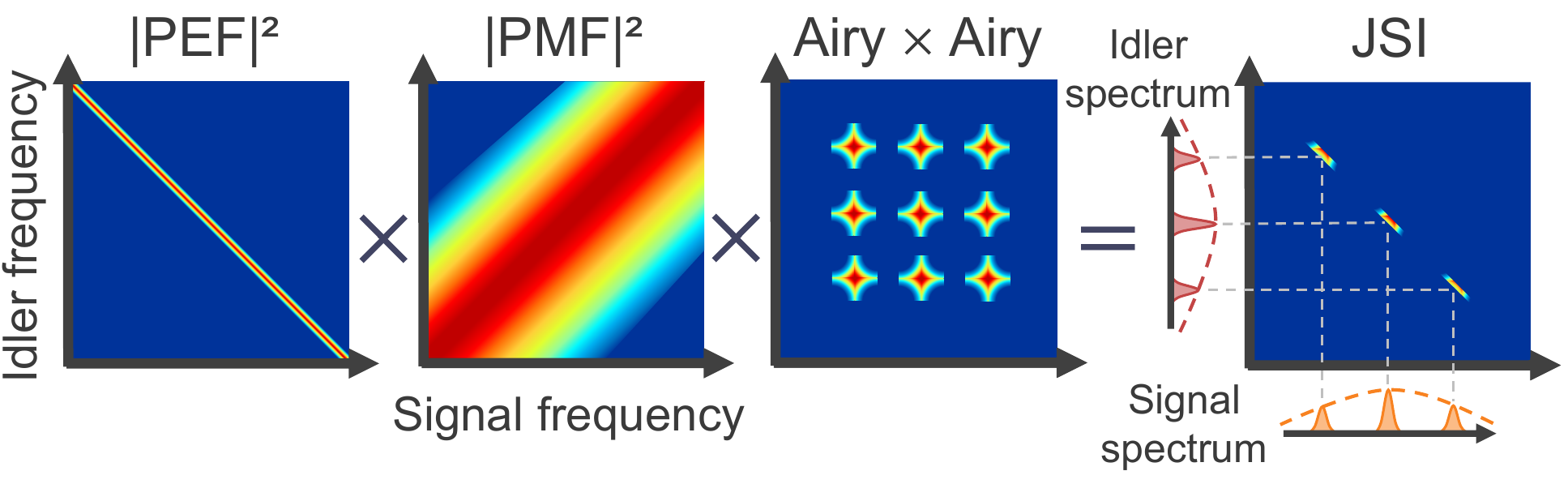}
        % \vspace{-6ex}
        \caption{
          Schematic of the JSI for cavity-enhanced SPDC (cSPDC) pumped by a monochromatic CW laser. The pump envelope function acts as a Dirac delta function representing energy conservation. Consequently, the resulting joint spectral distribution is restricted to the resonance peaks that overlap with this delta function.
        }
        \label{fig:JSI_cSPDC}
        % \vspace{-20ex}
      \end{figure}

      Furthermore, when the system is pumped by a monochromatic continuous-wave (CW) laser at frequency $\nu_\indPump^0$, the signal and idler photons are generated in pairs that must satisfy the energy conservation law:
      \begin{align}
        \nu_\indSignal + \nu_\indIdler = \nu_\indPump^0.
      \end{align}
      As a result, the spectral distribution of the photon-pair intensity and generation probability is further restricted by this additional constraint, as illustrated in Fig.~\ref{fig:JSI_cSPDC}.      

\begin{figure}[ht]
        \centering
        \includegraphics[width=1\columnwidth]{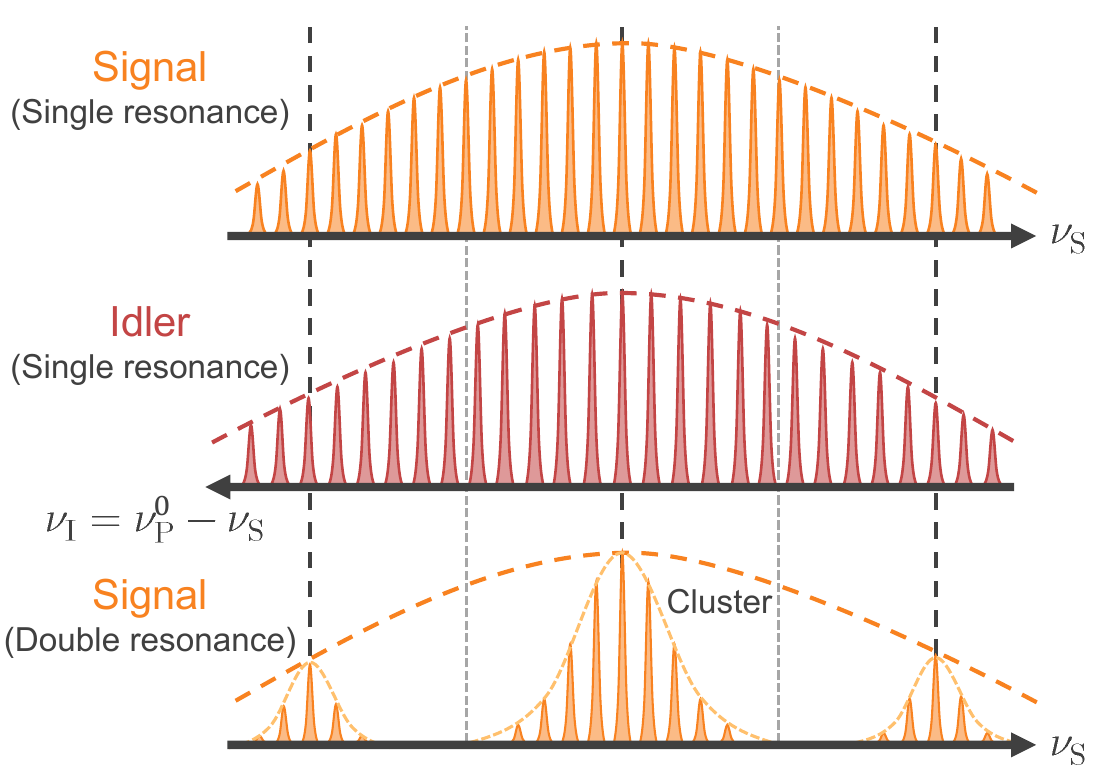}
        % \vspace{-6ex}
        \caption{
          Schematic illustration of the cluster effect. In non-degenerate SPDC within a doubly resonant cavity pumped by a monochromatic CW laser, the difference between the signal and idler FSRs results in a spectral distribution characterized by a clustered structure for both fields.
        }
        \label{fig:cluster_effect}
      \end{figure}

      As shown in the bottom panel of Fig.~\ref{fig:cluster_effect}, the frequency spectrum of the generated photons exhibits a distribution where the resonance peaks are grouped into distinct clusters. This phenomenon is known as the cluster effect \cite{LuoSilberhorn2015}. In this work, we refer to the cluster located closest to the point satisfying the phase-matching condition, $h(\omega_\indSignal, \omega_\indIdler) = 1$, as the main cluster.

\section{\label{sec:decompToCavity}decomposition to cavity-modes}
  \subsection{\label{subsec:decompOfJSI}Decomposition of JSI and JSA}
    The objective of this study is to quantitatively evaluate the impact of frequency multiplexing on the entanglement generation rate and the fidelity. Therefore, it is necessary to derive a formulation that accounts for the states generated in each frequency mode of the cSPDC---a photon-pair source---without neglecting the occurrence of multiple photon pairs. Even if a perfectly rigorous expression is unattainable, our goal is to find the closest possible representation of the quantum state within a framework that describes it as a composite state of frequency modes.

    As a first assumption, we limit our scope to the region where the phase-matching condition is satisfied.
    This is justified because our proposed scheme utilizes an atomic frequency comb (AFC)---prepared within the inhomogeneously broadened transition of a $\ce{Pr^{3+}}$-doped $\ce{Y2SiO5}$ (Pr:YSO) crystal---as a quantum memory, which limits the operational bandwidth of the photon source to approximately $\qty{10}{GHz}$ \cite{EquallCone1995}. Since the phase-matching bandwidth typically ranges from several hundred $\unit{GHz}$ to several $\unit{THz}$, the phase-matching condition $\Delta k = 0$ can be assumed to hold throughout the entire photon bandwidth of interest, yielding
    \begin{align}
      h(\omega_\indSignal, \omega_\indIdler)\simeq 1.
    \end{align}

    Furthermore, we assume that the SPDC process occurs in a periodically poled crystal. We utilize quasi-phase matching (QPM), where the phase-matching condition is defined as $\Delta k = 0$ with
    \begin{align}
      \Delta k = k_\indPump(\omega_\indPump) - k_\indSignal(\omega_\indSignal) - k_\indIdler(\omega_\indIdler) - \frac{2\pi}{\Lambda},
    \end{align}
    where $\Lambda$ denotes the poling period.

    In this case, the JSI is given by
    \begin{align}
      S_\indexrm{cav}(\omega_\indSignal, \omega_\indIdler) \simeq \absolute{s(\omega_\indSignal+\omega_\indIdler)}^2\, \Airy[^0]{\indSignal}{\omega_\indSignal}\, \Airy[^0]{\indIdler}{\omega_\indIdler}.
    \end{align}

    In the regime where the peaks of the Airy function are well separated---specifically, when the full width at half maximum (FWHM) $\fwhm_\epsilon = \fsr_\epsilon / \finesse_\epsilon$ is much smaller than the free spectral range ($\fwhm_\epsilon \ll \fsr_\epsilon$; e.g., $\finesse_\epsilon \gtrsim 10$ \cite{Andreas2019phd})---the Airy function can be approximated as a sum of Lorentzians centered at $m_\epsilon \fsr_\epsilon$ \cite{ScholzBenson2009Analytical, Andreas2019phd, SanchezSoto2016, Prakash2021phd}:
    \begin{align}
      \Airy[^0]{\epsilon}{\omega_\epsilon} &\simeq \sum_{m_\epsilon=-\infty}^{\infty} \frac{1}{1+\left(\frac{2\finesse_\epsilon}{\pi}\right)^2 \left(\frac{\pi}{\fsr_\epsilon}\nu_\epsilon -m_\epsilon \pi\right)^2}\notag\\
      &= \sum_{m_\epsilon=-\infty}^{\infty} \frac{1}{1+\frac{4}{\fwhm_\epsilon^2} \left(\nu_\epsilon -m_\epsilon \fsr_\epsilon\right)^2}.
      \label{eq:approx_Airy}
    \end{align}

    By applying this expression,
    \begin{widetext}
      \begin{align}
        S_\indexrm{cav} (\omega_\indSignal,\omega_\indIdler) &\simeq \absolute{s(\omega_\indSignal+\omega_\indIdler)}^2
        \left(
          \sum_{m_\indSignal=-\infty}^{\infty} \frac{1}{1+\frac{4}{\fwhm_\indSignal^2} \left(\rule{0pt}{2ex}\nu_\indSignal -m_\indSignal \fsr_\indSignal\right)^2}
        \right)
        \left(
          \sum_{m_\indIdler=-\infty}^{\infty} \frac{1}{1+\frac{4}{\fwhm_\indIdler^2} \left(\rule{0pt}{2ex}\nu_\indIdler -m_\indIdler \fsr_\indIdler\right)^2}
        \right)\notag\\
        &= \absolute{s(\omega_\indSignal+\omega_\indIdler)}^2
        \sum_{m_\indSignal=-\infty}^{\infty} \sum_{m_\indIdler=-\infty}^{\infty}
        \left(
          \frac{1}{1+\frac{4}{\fwhm_\indSignal^2} \left(\rule{0pt}{2ex}\nu_\indSignal -m_\indSignal \fsr_\indSignal\right)^2}
          \frac{1}{1+\frac{4}{\fwhm_\indIdler^2} \left(\rule{0pt}{2ex}\nu_\indIdler -m_\indIdler \fsr_\indIdler\right)^2}
        \right)\notag\\
        &= \absolute{s(\omega_\indSignal+\omega_\indIdler)}^2
        \sum_{m_\indSignal=-\infty}^{\infty} \sum_{m_\indIdler=-\infty}^{\infty}
        \Xi_{m_\indSignal, m_\indIdler}(\nu_\indSignal, \nu_\indIdler)
      \end{align}    
    \end{widetext}
    where we have introduced
    \begin{align}
      \begin{split}
        \Xi_{m_\indSignal, m_\indIdler}(\nu_\indSignal, \nu_\indIdler) 
        \coloneq\,
        & \frac{1}{1+\frac{4}{\fwhm_\indSignal^2} \left(\rule{0pt}{2ex}\nu_\indSignal -m_\indSignal \fsr_\indSignal\right)^2}\\
        & \times\frac{1}{1+\frac{4}{\fwhm_\indIdler^2} \left(\rule{0pt}{2ex}\nu_\indIdler -m_\indIdler \fsr_\indIdler\right)^2}.
      \end{split}
    \end{align}

    As a second assumption, we consider pumping with a narrow-linewidth laser, which can be approximated as monochromatic. In this case, the pump envelope function is given by
    \begin{align}
      s(\omega_\indSignal+\omega_\indIdler) \simeq \delta(\omega_\indSignal+\omega_\indIdler-\omega_\indPump^0) = \delta(\nu_\indSignal+\nu_\indIdler-\nu_\indPump^0),
    \end{align}
    where $\delta(\cdot)$ denotes Dirac's delta function.

    \begin{widetext}
      Consequently, we obtain
      \begin{align}
        S_\indexrm{cav} (\omega_\indSignal,\omega_\indIdler)
        &\simeq \delta(\nu_\indSignal+\nu_\indIdler-\nu_\indPump^0)
        \sum_{k=-\infty}^{\infty} \sum_{j=-\infty}^{\infty}
        \Xi_{m_\indSignal, m_\indIdler}(\nu_\indSignal, \nu_\indIdler)
        \label{eq:cSPDC_JSI_2}
      \end{align}    
    \end{widetext}
    where $\nu_\indPump^0 = \omega_\indPump^0/2\pi$ is the central frequency of the pump field.

    Owing to the delta function $\delta(\nu_\indSignal+\nu_\indIdler-\nu_\indPump^0)$ in Eq.~\eqref{eq:cSPDC_JSI_2}, only the terms $\Xi_{m_\indSignal, m_\indIdler}(\nu_\indSignal, \nu_\indIdler)$ that overlap with the line defined by $\nu_\indSignal + \nu_\indIdler = \nu_\indPump^0$ take non-zero values. Since each $\Xi_{m_\indSignal, m_\indIdler}(\nu_\indSignal, \nu_\indIdler)$ is localized in the vicinity of $(\nu_\indSignal, \nu_\indIdler)=(m_\indSignal\fsr_\indSignal, m_\indIdler\fsr_\indIdler)$, the indices $(m_\indSignal, m_\indIdler)$ for which $\Xi_{m_\indSignal, m_\indIdler}$ is non-zero in Eq.~\eqref{eq:cSPDC_JSI_2} must satisfy the condition
    \begin{align}
      m_\indSignal\fsr_\indSignal +  m_\indIdler\fsr_\indIdler \simeq \nu_\indPump^0.
      \label{eq:cw_bound_1}
    \end{align}

    Under CW pumping, the cluster effect results in a photon spectrum characterized by a series of distinct clusters. In the configuration considered here, both the required photon bandwidth and the individual cluster width are on the order of several GHz. Consequently, we assume that only the main cluster is utilized. As a third assumption, we restrict our analysis to the spectral region within the main cluster, a condition that can be experimentally realized by filtering out all side clusters.

    We define $(K_\indSignal, K_\indIdler)$ as the indices corresponding to the peak with the maximum intensity in the main cluster. By substituting $m_\indSignal = K_\indSignal + k$ and $m_\indIdler = K_\indIdler - j$ into Eq.~\eqref{eq:cw_bound_1}, we obtain
    \begin{align}
      (K_\indSignal + k)\fsr_\indSignal +  (K_\indIdler - j)\fsr_\indIdler \simeq \nu_\indPump^0.
    \end{align}
    Given that $K_\indSignal\fsr_\indSignal + K_\indIdler\fsr_\indIdler \simeq \nu_\indPump^0$, this relation simplifies to
    \begin{align}
      j \simeq \frac{\fsr_\indSignal}{\fsr_\indIdler}k\qquad \label{eq:cluster_index}
    \end{align}
    Assuming that there are $N_\indSignal$ peaks in the signal spectrum between the centers of adjacent clusters, the following relationship holds \cite{LuoSilberhorn2015}:
    \begin{align}
      \fsr_\indSignal N_\indSignal = \fsr_\indIdler (N_\indSignal \pm 1).
    \end{align}
    This yields
    \begin{align}
      \frac{\fsr_\indSignal}{\fsr_\indIdler} = (1\pm \frac{1}{N_\indSignal}).
    \end{align}

    Thus, in the vicinity of the main cluster ($k \ll N_\indSignal$), Eq.~\eqref{eq:cluster_index} reduces to
    \begin{align}
      j \simeq (1\pm \frac{1}{N_\indSignal})k = k \pm \frac{k}{N_\indSignal} \simeq k. 
      \label{eq:cluster_index_proof}
    \end{align}

\begin{figure*}[thbp]
      \begin{tabular}{c}
        \begin{subfigure}[t]{0.475\hsize}
          \centering
          \caption{}
          \vspace{-4.75ex}
          \includegraphics[width=\columnwidth]{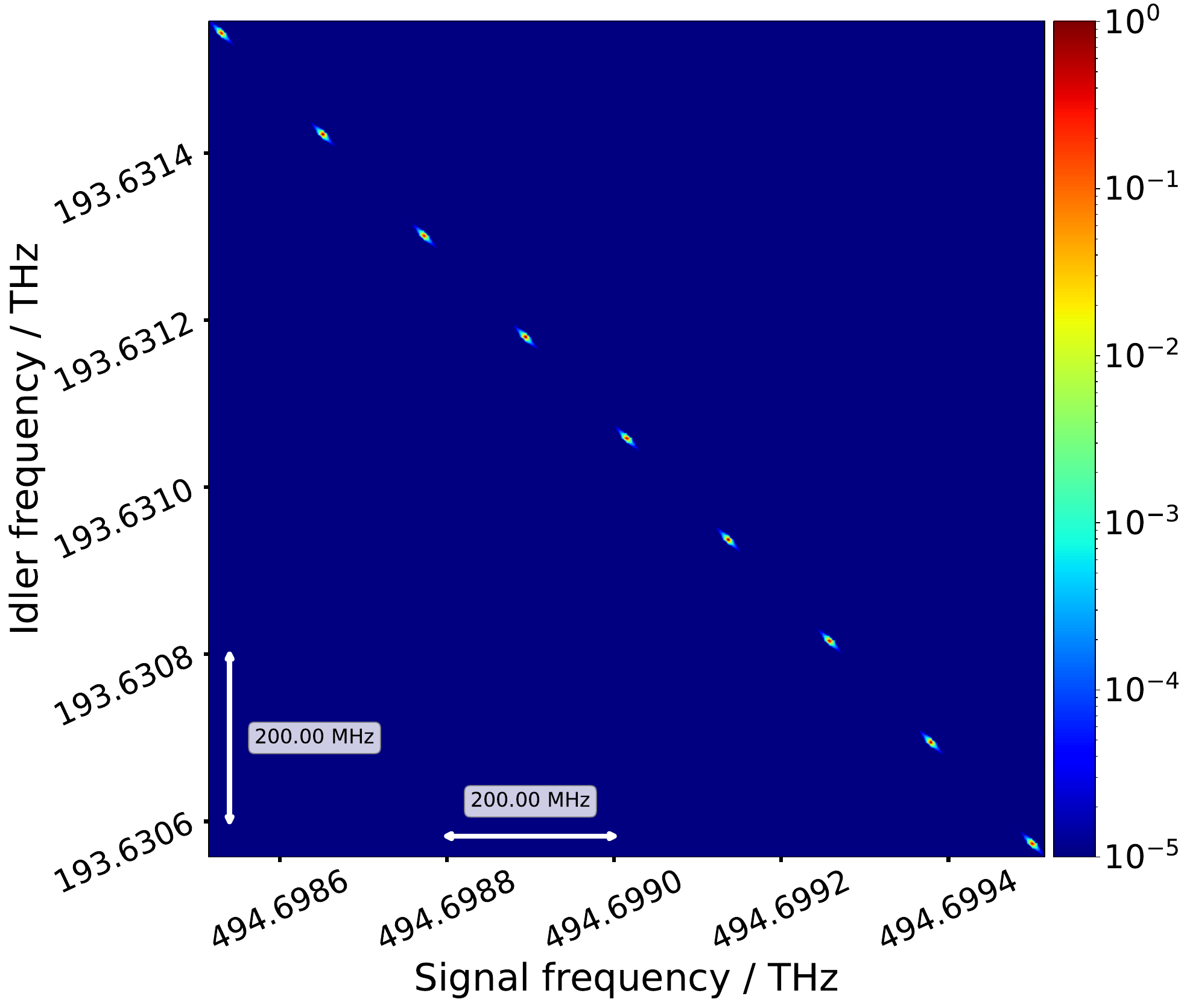}
          \label{subfig:plot_JSI_Airy}
        \end{subfigure}
        \begin{subfigure}[t]{0.475\hsize}
          \centering
          \caption{}
          \vspace{-4.75ex}
          \includegraphics[width=\columnwidth]{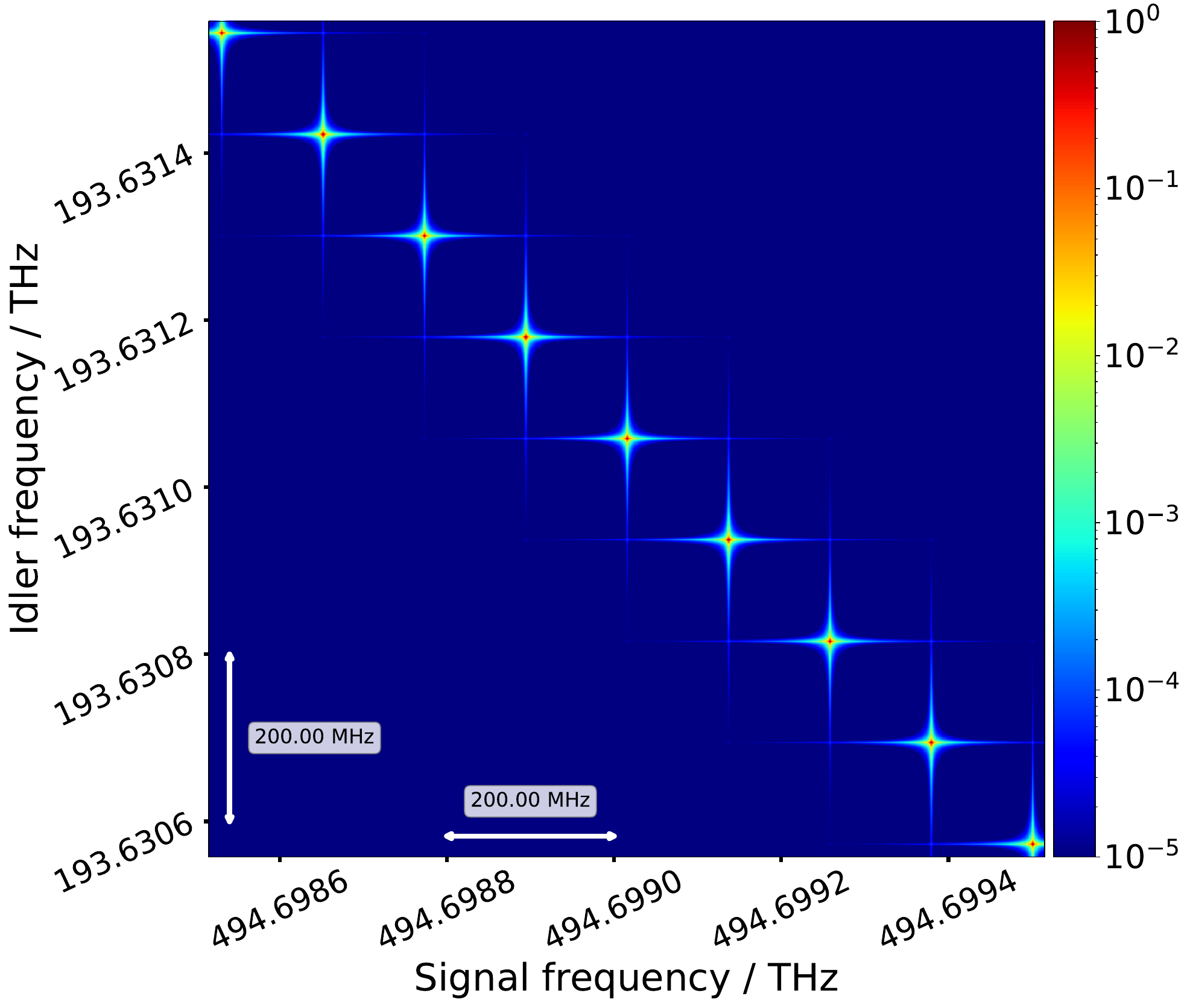}
          \label{subfig:plot_JSI_decompose}
        \end{subfigure}\\[-4.0ex]
        \begin{subfigure}[t]{0.475\hsize}
          \centering
          \caption{}
          \vspace{-4.75ex}
          \includegraphics[width=\columnwidth]{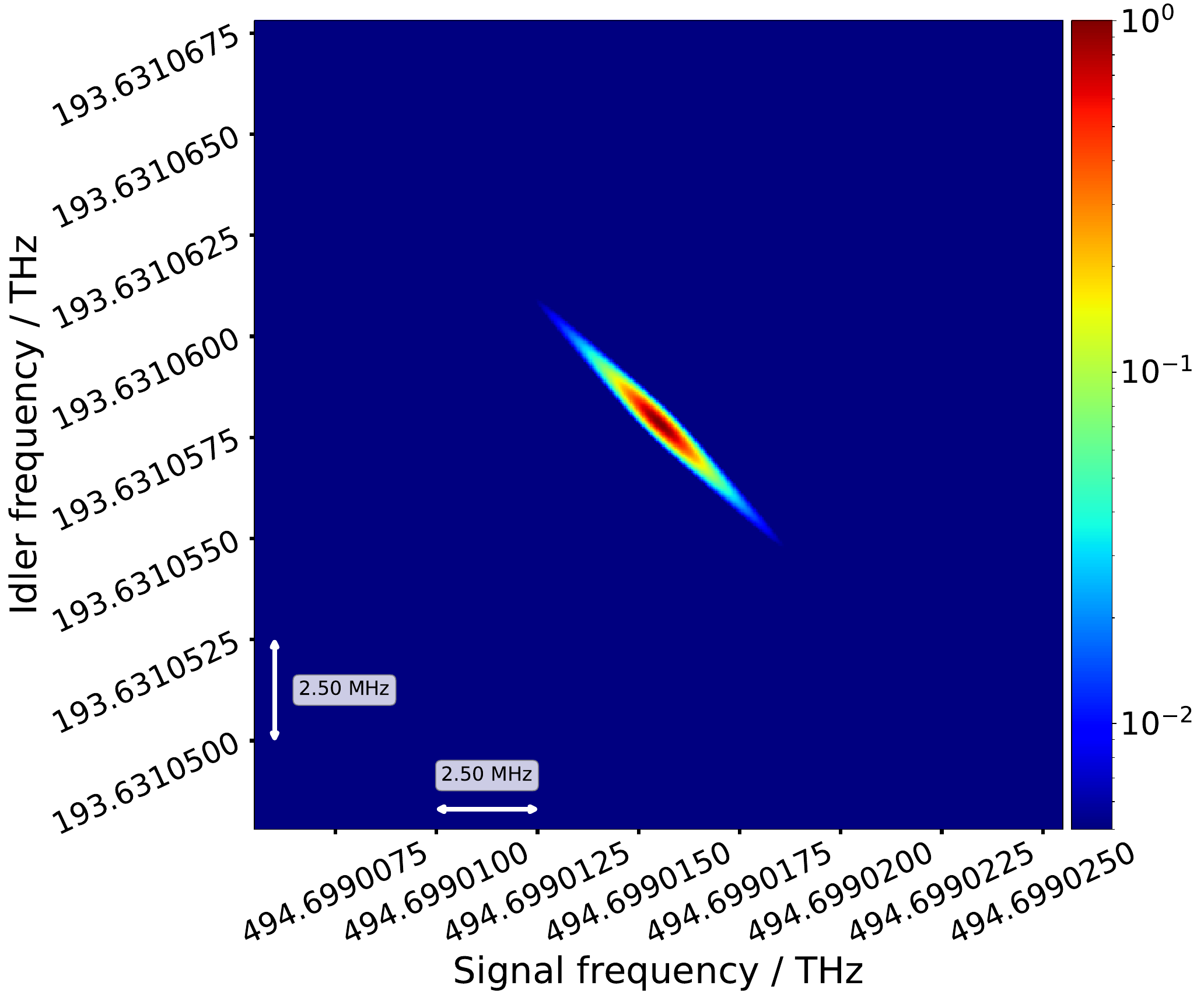} % heightをcolomnwidthの9/16倍
          \label{subfig:plot_JSI_mode0_Airy}
        \end{subfigure}
        %
        % \hspace{0.05\hsize}
        %
        \begin{subfigure}[t]{0.475\hsize}
          \centering
          \caption{}
          \vspace{-4.75ex}
          \includegraphics[width=\columnwidth]{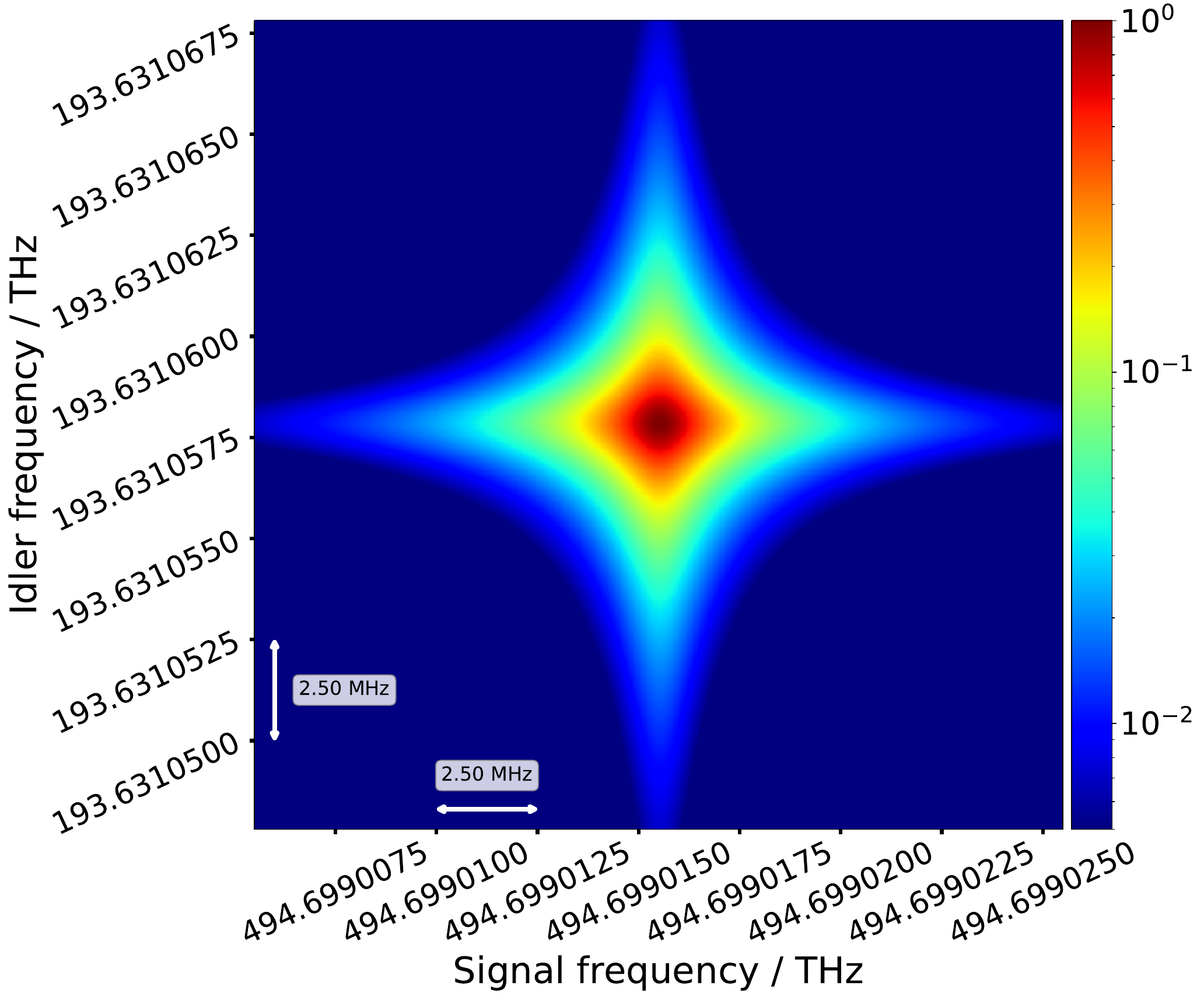}
          \label{subfig:plot_JSI_mode0_decompose}
        \end{subfigure}\\[-4.0ex]
        \begin{subfigure}[t]{0.475\hsize}
          \centering
          \caption{}
          \vspace{-4.75ex}
          \includegraphics[width=\columnwidth]{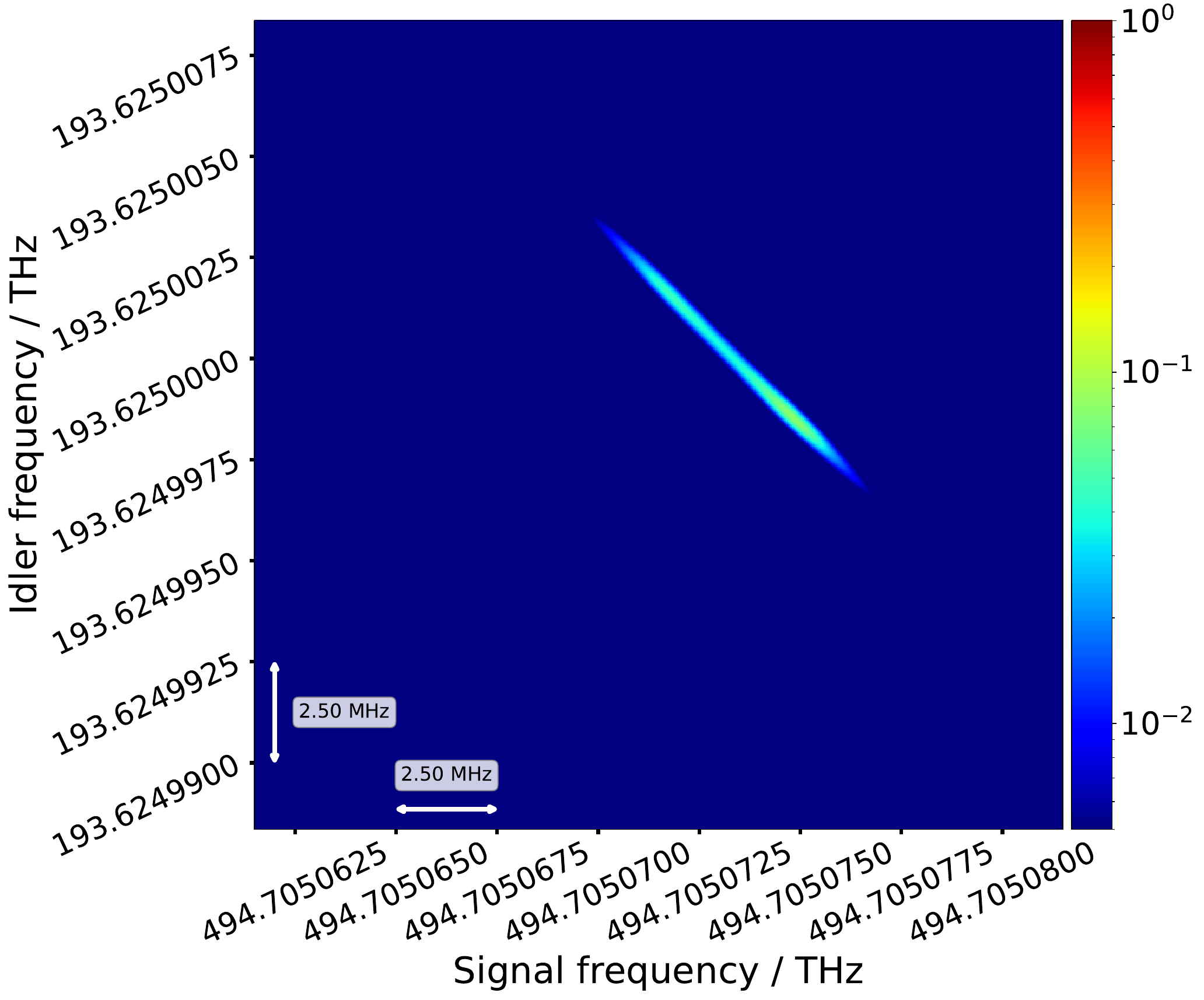} % heightをcolomnwidthの9/16倍
          \label{subfig:plot_JSI_mode50_Airy}
        \end{subfigure}
        %
        % \hspace{0.05\hsize}
        %
        \begin{subfigure}[t]{0.475\hsize}
          \centering
          \caption{}
          \vspace{-4.75ex}
          \includegraphics[width=\columnwidth]{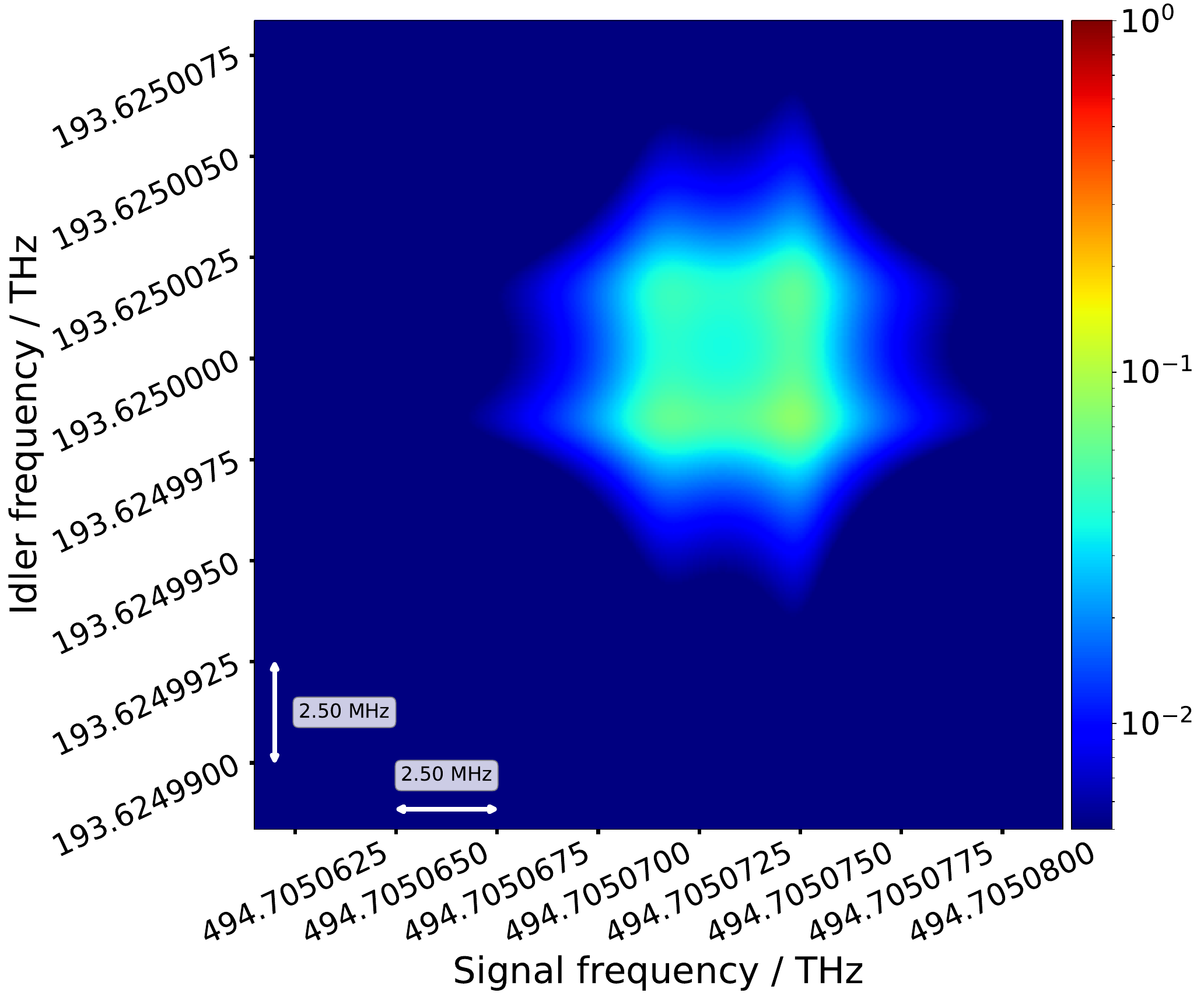}
          \label{subfig:plot_JSI_mode50_decompose}
        \end{subfigure}
      \end{tabular}
      \vspace{-3.5ex}
      \caption{
        Plots of the joint spectral intensity (JSI). Panels (\subref{subfig:plot_JSI_Airy}), (\subref{subfig:plot_JSI_mode0_Airy}), and (\subref{subfig:plot_JSI_mode50_Airy}) show $S_\indexrm{cav}(\omega_\indSignal, \omega_\indIdler)=\absolute{s(\omega_\indSignal, \omega_\indIdler)}^2\Airy{\indSignal}{\omega_\indSignal}\Airy{\indIdler}{\omega_\indIdler}$. For CW excitation, the pump envelope is given by $s(\omega_\indSignal, \omega_\indIdler)=\delta(\omega_\indSignal+\omega_\indIdler-\omega_\indPump^0)$; however, for the purpose of plotting, we use a Gaussian form $s(\omega_\indSignal, \omega_\indIdler)=\exp[-\left(\nu_\indSignal+\nu_\indIdler-\nu_\indPump^0\right)^2/(2\sigma_\indPump^2)]$, with $\sigma_\indPump = \qty{2.5}{MHz}$ in (\subref{subfig:plot_JSI_Airy}) and $\sigma_\indPump = \qty{0.25}{MHz}$ in (\subref{subfig:plot_JSI_mode0_Airy}) and (\subref{subfig:plot_JSI_mode50_Airy}). Panels (\subref{subfig:plot_JSI_decompose}), (\subref{subfig:plot_JSI_mode0_decompose}), and (\subref{subfig:plot_JSI_mode50_decompose}) show $S_\indexrm{cav}^\indexrm{(approx)}(\omega_\indSignal, \omega_\indIdler)=\sum_{k=-50}^{50}\absolute{\psi_{k}(\omega_\indSignal)}^2\absolute{\phi_{k}(\omega_\indIdler)}^2$. Panels (\subref{subfig:plot_JSI_Airy}) and (\subref{subfig:plot_JSI_decompose}) plot the eight modes near the center ($k=-4,-3,\dots,3$), (\subref{subfig:plot_JSI_mode0_Airy}) and (\subref{subfig:plot_JSI_mode0_decompose}) show the central mode ($k=0$), and (\subref{subfig:plot_JSI_mode50_Airy}) and (\subref{subfig:plot_JSI_mode50_decompose}) show the 50th mode from the center ($k=50$).
      }
      \label{fig:plot_JSI}
    \end{figure*}

    Consequently, the JSI is expressed as
    \begin{widetext}
      \begin{align}
        S_\indexrm{cav} (\omega_\indSignal,\omega_\indIdler)
        &\simeq \delta(\nu_\indSignal+\nu_\indIdler-\nu_\indPump^0)
        \sum_{k=-M}^{M} \sum_{j=-M}^{M} 
        \Xi_{K_\indSignal + k, K_\indIdler - j}(\nu_\indSignal, \nu_\indIdler)\notag\\
        &= \delta(\nu_\indSignal+\nu_\indIdler-\nu_\indPump^0)
        \sum_{k=-M}^{M} \sum_{j=-M}^{M} \delta_{kj}\,
        \Xi_{K_\indSignal + k, K_\indIdler - j}(\nu_\indSignal, \nu_\indIdler)\notag\\
        &= \delta(\nu_\indSignal+\nu_\indIdler-\nu_\indPump^0)
        \sum_{k=-M}^{M} 
        \Xi_{K_\indSignal + k, K_\indIdler - k}(\nu_\indSignal, \nu_\indIdler).
      \end{align}    
    % \end{widetext}
      Here, $M$ denotes the number of peaks considered on each side of the central peak within the main cluster.

      The term $\sum_{k=-M}^{M} \Xi_{K_\indSignal + k, K_\indIdler - k}(\nu_\indSignal, \nu_\indIdler)$ refers to those resonance peaks---among the collection shown in Fig.~\ref{fig:JSI_Airy}---that lie on the line of energy conservation. When the delta function representing energy conservation is applied to this term, the photon spectrum becomes even more strictly constrained, as illustrated in Fig.~\ref{fig:JSI_cSPDC}; this additional spectral filtering constitutes the cluster effect. 
      \newpage
      
      Incorporating this effect, the expression can be rewritten as
      \begin{align}
        S_\indexrm{cav} (\omega_\indSignal,\omega_\indIdler)
        &\simeq \delta(\nu_\indSignal+\nu_\indIdler-\nu_\indPump^0)
          \sum_{k=-M}^{M} 
          \frac{1}{1+\frac{4}{\fwhm_\indSignal^2} \left(\rule{0pt}{2ex}\nu_\indSignal -(K_\indSignal + k) \fsr_\indSignal\right)^2}
          \frac{1}{1+\frac{4}{\fwhm_\indIdler^2} \left(\rule{0pt}{2ex}\nu_\indIdler -(K_\indIdler - k) \fsr_\indIdler\right)^2}\notag\\
        \begin{split}
          &= \delta(\nu_\indSignal+\nu_\indIdler-\nu_\indPump^0)\!
          \sum_{k=-M}^{M} \!\!
          \left(
            \frac{1}{1+\frac{4}{\fwhm_\indSignal^2} \left(\rule{0pt}{2ex}\nu_\indSignal -(K_\indSignal + k) \fsr_\indSignal\right)^2}
          \right)^{\!\!\! 1/2}\!\!\!
          \left(
            \frac{1}{1+\frac{4}{\fwhm_\indSignal^2} \left(\rule{0pt}{2ex}\nu_\indPump^0 - \nu_\indIdler -(K_\indSignal + k) \fsr_\indSignal\right)^2}
          \right)^{\!\!\! 1/2}\\
          &\hspace{22ex} \times\!
          \left(
            \frac{1}{1+\frac{4}{\fwhm_\indIdler^2} \left(\rule{0pt}{2ex}\nu_\indIdler -(K_\indIdler - k) \fsr_\indIdler\right)^2}
          \right)^{\!\!\! 1/2}\!\!\!
          \left(
            \frac{1}{1+\frac{4}{\fwhm_\indIdler^2} \left(\rule{0pt}{2ex}\nu_\indPump^0 - \nu_\indSignal -(K_\indIdler - k) \fsr_\indIdler\right)^2}
          \right)^{\!\!\! 1/2}
        \end{split}\notag\\
        \begin{split}
          &= \delta(\nu_\indSignal+\nu_\indIdler-\nu_\indPump^0)\!
          \sum_{k=-M}^{M} \!\!
          \left(
            \frac{1}{1+\frac{4}{\fwhm_\indSignal^2} \left(\rule{0pt}{2ex}\nu_\indSignal -(K_\indSignal + k) \fsr_\indSignal\right)^2}
          \right)^{\!\!\! 1/2}\!\!\!
          \left(
            \frac{1}{1+\frac{4}{\fwhm_\indIdler^2} \left(\rule{0pt}{2ex}\nu_\indPump^0 - \nu_\indSignal -(K_\indIdler - k) \fsr_\indIdler\right)^2}
          \right)^{\!\!\! 1/2}\\
          &\hspace{22ex} \times\!
          \left(
            \frac{1}{1+\frac{4}{\fwhm_\indSignal^2} \left(\rule{0pt}{2ex}\nu_\indPump^0 - \nu_\indIdler -(K_\indSignal + k) \fsr_\indSignal\right)^2}
          \right)^{\!\!\! 1/2}\!\!\!
          \left(
            \frac{1}{1+\frac{4}{\fwhm_\indIdler^2} \left(\rule{0pt}{2ex}\nu_\indIdler -(K_\indIdler - k) \fsr_\indIdler\right)^2}
          \right)^{\!\!\! 1/2}.
        \end{split}    
      \end{align}
    % \end{widetext}

      On the line representing energy conservation, the terms within this summation yield values identical to those of $\Xi_{K_\indSignal + k, K_\indIdler - k}(\nu_\indSignal, \nu_\indIdler)$. However, as will be confirmed later in Fig.~\ref{fig:plot_JSI}, the range over which these terms take non-zero values is more restricted than that of a simple product of Lorentzians.

      Our objective here is to derive a closed-form expression that can describe the quantum states generated by cSPDC, including multi-photon pairs, based on the JSI and JSA. To facilitate the subsequent Schmidt decomposition, we relax the strict constraints imposed by the delta function and adopt the following as an approximate expression for the JSI:
    % \begin{widetext}
      \begin{align}
        \begin{split}
          S_\indexrm{cav}^\indexrm{(approx)} (\omega_\indSignal,\omega_\indIdler)
          &\coloneq 
          \sum_{k=-M}^{M} 
          \left(
            \frac{1}{1+\frac{4}{\fwhm_\indSignal^2} \left(\rule{0pt}{2ex}\nu_\indSignal -(K_\indSignal + k) \fsr_\indSignal\right)^2}
            \frac{1}{1+\frac{4}{\fwhm_\indIdler^2} \left(\rule{0pt}{2ex}\nu_\indPump^0 - \nu_\indSignal -(K_\indIdler - k) \fsr_\indIdler\right)^2}
          \right)^{\! 1/2}\\
          &\hspace{10ex} \times
          \left(
            \frac{1}{1+\frac{4}{\fwhm_\indSignal^2} \left(\rule{0pt}{2ex}\nu_\indPump^0 - \nu_\indIdler -(K_\indSignal + k) \fsr_\indSignal\right)^2}
            \frac{1}{1+\frac{4}{\fwhm_\indIdler^2} \left(\rule{0pt}{2ex}\nu_\indIdler -(K_\indIdler - k) \fsr_\indIdler\right)^2}
          \right)^{\! 1/2}.
        \end{split}
        \label{eq:JSI_cav_approx}      
      \end{align}
    % \end{widetext}

      By letting $\alpha^\prime$ be a complex number satisfying $|\alpha^\prime|^2 = \beta$, we define the following expression:
    % \begin{widetext}
      \begin{align}
        \begin{split}
          f_\indexrm{cav}^\indexrm{(approx)} (\omega_\indSignal,\omega_\indIdler)
          &\coloneq  
          \sum_{k=-M}^{M} 
          \left(
            \frac{1}{1+i\frac{2}{\fwhm_\indSignal} \left(\rule{0pt}{2ex}\nu_\indSignal -(K_\indSignal + k) \fsr_\indSignal\right)}
            \frac{1}{1+i\frac{2}{\fwhm_\indIdler} \left(\rule{0pt}{2ex}\nu_\indPump^0 - \nu_\indSignal -(K_\indIdler - k) \fsr_\indIdler\right)}
          \right)^{\! 1/2}\\
          &\hspace{10ex} \times
          \left(
            \frac{1}{1+i\frac{2}{\fwhm_\indSignal} \left(\rule{0pt}{2ex}\nu_\indPump^0 - \nu_\indIdler -(K_\indSignal + k) \fsr_\indSignal\right)}
            \frac{1}{1+i\frac{2}{\fwhm_\indIdler} \left(\rule{0pt}{2ex}\nu_\indIdler -(K_\indIdler - k) \fsr_\indIdler\right)}
          \right)^{\! 1/2}
        \end{split}
        \label{eq:JSA_cav_approx}
      \end{align}
    % \end{widetext}
      Since this expression satisfies $\absolute{f_\indexrm{cav}^\indexrm{(approx)} (\omega_\indSignal,\omega_\indIdler)}^2 \simeq S_\indexrm{cav}^\indexrm{(approx)} (\omega_\indSignal, \omega_\indIdler)$ (as shown in Appendix \ref{sec:proof_approx}), we adopt it as our approximate expression for the JSA.

      Here, letting
    % \begin{widetext}
      \begin{subequations}
        \begin{align}
            \tilde{\psi}_{k} (\omega_\indSignal) &\coloneq 
            \left(
              \frac{1}{1-i\frac{2}{\fwhm_\indSignal} \left(\rule{0pt}{2ex}\nu_\indSignal -(K_\indSignal + k) \fsr_\indSignal\right)}
              \frac{1}{1-i\frac{2}{\fwhm_\indIdler} \left(\rule{0pt}{2ex}\nu_\indPump^0 - \nu_\indSignal -(K_\indIdler - k) \fsr_\indIdler\right)}
            \right)^{\! 1/2}
        \end{align}\vspace{-2ex}
        \begin{align}
            \tilde{\phi}_{k} (\omega_\indIdler) &\coloneq 
            \left(
              \frac{1}{1-i\frac{2}{\fwhm_\indSignal} \left(\rule{0pt}{2ex}\nu_\indPump^0 - \nu_\indIdler -(K_\indSignal + k) \fsr_\indSignal\right)}
              \frac{1}{1-i\frac{2}{\fwhm_\indIdler} \left(\rule{0pt}{2ex}\nu_\indIdler -(K_\indIdler - k) \fsr_\indIdler\right)}
            \right)^{\! 1/2}
        \end{align}
      \end{subequations}   
    \end{widetext}
    and using the constants
    \begin{subequations}
      \begin{align}
        C_{\indSignal,k} \coloneq 
        \left(
          \int_{0}^{\infty} \dd{\omega_\indSignal} \tilde{\psi}_{k}^\ast (\omega_\indSignal) \tilde{\psi}_{k} (\omega_\indSignal)
        \right)^{\! 1/2}
      \end{align}
      \begin{align}
        C_{\indIdler,k} \coloneq 
        \left(
          \int_{0}^{\infty} \dd{\omega_\indIdler} \tilde{\phi}_{k}^\ast (\omega_\indIdler) \tilde{\phi}_{k} (\omega_\indIdler)
        \right)^{\! 1/2} \ 
      \end{align}
      \label{eq:normalizationConstant}
    \end{subequations}
    to define the normalized functions
    \begin{subequations}
      \begin{align}
        \psi_{k} (\omega_\indSignal) \coloneq \frac{1}{ C_{\indSignal,k}} \tilde{\psi}_{k} (\omega_\indSignal),
      \end{align}
      \begin{align}
        \phi_{k} (\omega_\indIdler) \coloneq \frac{1}{ C_{\indIdler,k}} \tilde{\phi}_{k} (\omega_\indIdler),
      \end{align}
    \end{subequations}
    we obtain
    % \newpage
    % \begin{widetext}
    \begin{align}
      f_\indexrm{cav}^\indexrm{(approx)} (\omega_\indSignal,\omega_\indIdler)
        &= \sum_{k=-M}^{M}  
        C_{\indSignal,k} C_{\indIdler,k} \psi_{k}^\ast (\omega_\indSignal)
         \phi_{k}^\ast (\omega_\indIdler).
        \label{eq:cSPDC_JSA_approximate}
    \end{align} 
    % \end{widetext}
    
    Now, defining
    \begin{align}
      r_k \coloneq - \frac{i}{\hbar} \alpha^\prime\, C_{\indSignal,k} C_{\indIdler,k}
      \label{eq:squeezingParameter}
    \end{align}
    and selecting $\alpha^\prime$ to ensure $r_k \in \mathbb{R}^+$, we obtain
    \begin{align}
      - \frac{i}{\hbar} \alpha^\prime f_\indexrm{cav}^\indexrm{(approx)} (\omega_\indSignal,\omega_\indIdler)
      &= \sum_{k=-M}^{M} r_k  
        \psi_{k}^\ast (\omega_\indSignal)
        \phi_{k}^\ast (\omega_\indIdler)
      \label{eq:cSPDC_JSA_Schmidt}
    \end{align}
    Since $\psi_{k} (\omega_\indSignal)$ and $\phi_{k} (\omega_\indIdler)$ satisfy the orthonormality conditions
    \begin{subequations}
      \begin{align}
        \int_{0}^{\infty} \dd{\omega_\indSignal} \psi_{k}^\ast (\omega_\indSignal) \psi_{k^\prime} (\omega_\indSignal) = \delta_{k,k^\prime},
      \end{align}\vspace{-2ex}
      \begin{align}
        \int_{0}^{\infty} \dd{\omega_\indIdler} \phi_{k}^\ast (\omega_\indIdler) \phi_{k^\prime} (\omega_\indIdler) = \delta_{k,k^\prime},
      \end{align}
    \end{subequations}
    the sets $\{\psi_{k} (\omega_\indSignal)\}$ and $\{\phi_{k} (\omega_\indIdler)\}$ form complete orthonormal bases for the respective subspaces to which the JSA under consideration belongs. Thus, Eq.~\eqref{eq:cSPDC_JSA_Schmidt} constitutes a Schmidt decomposition \cite{ChristSilberhorn2011}.

    \begin{figure*}[th]
      \begin{tabular}{c}
        \begin{subfigure}[t]{0.5075\hsize}
          \centering
          \caption{}
          \vspace{-2ex}
          \includegraphics[height=0.5630\columnwidth]{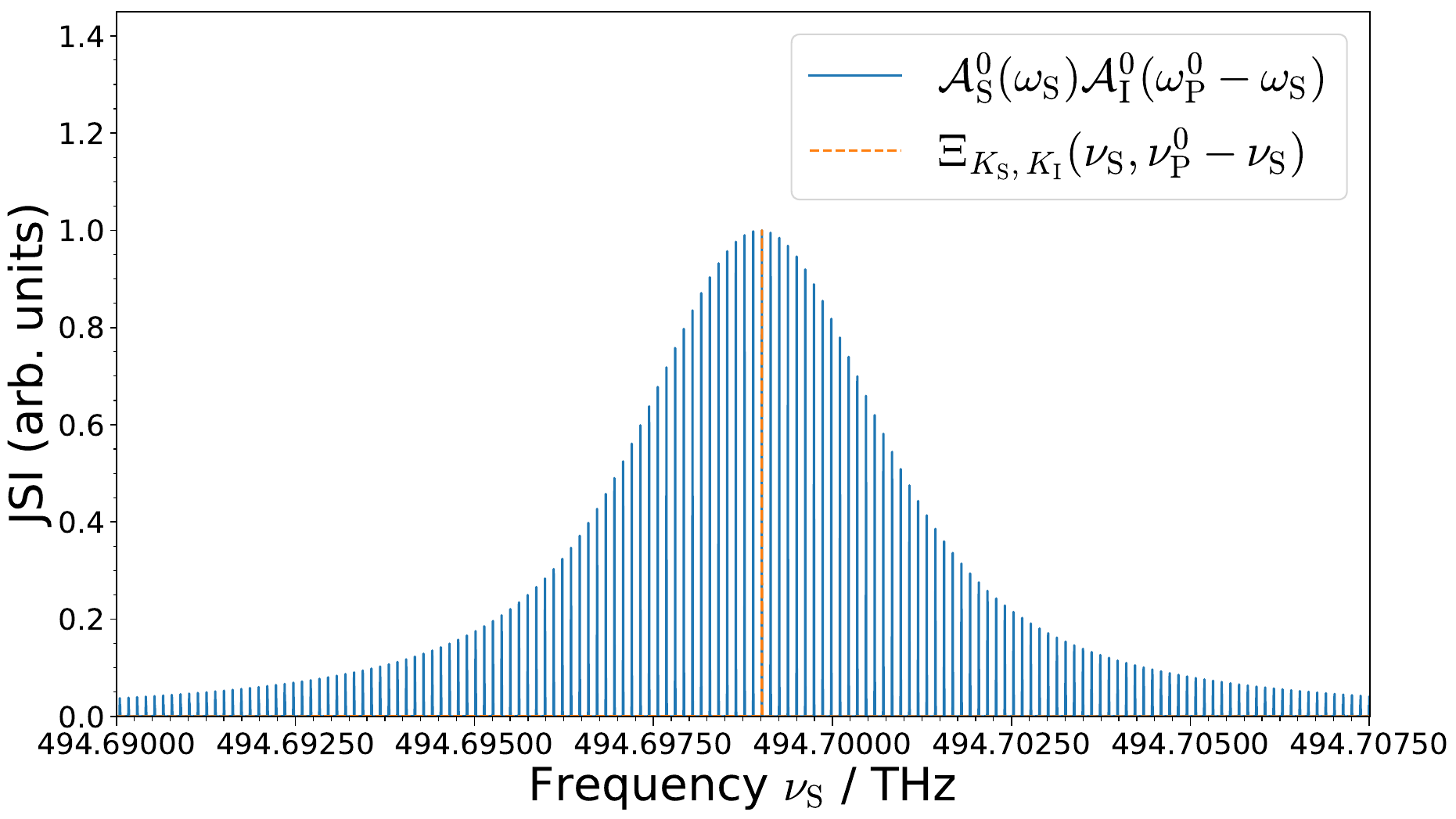}
          \label{subfig:spectrum_signal_mode0_all}
        \end{subfigure}
        \hspace{-0.015\hsize}
        \begin{subfigure}[t]{0.5075\hsize}
          \centering
          \caption{}
          \vspace{-2ex}
          \includegraphics[height=0.5630\columnwidth]{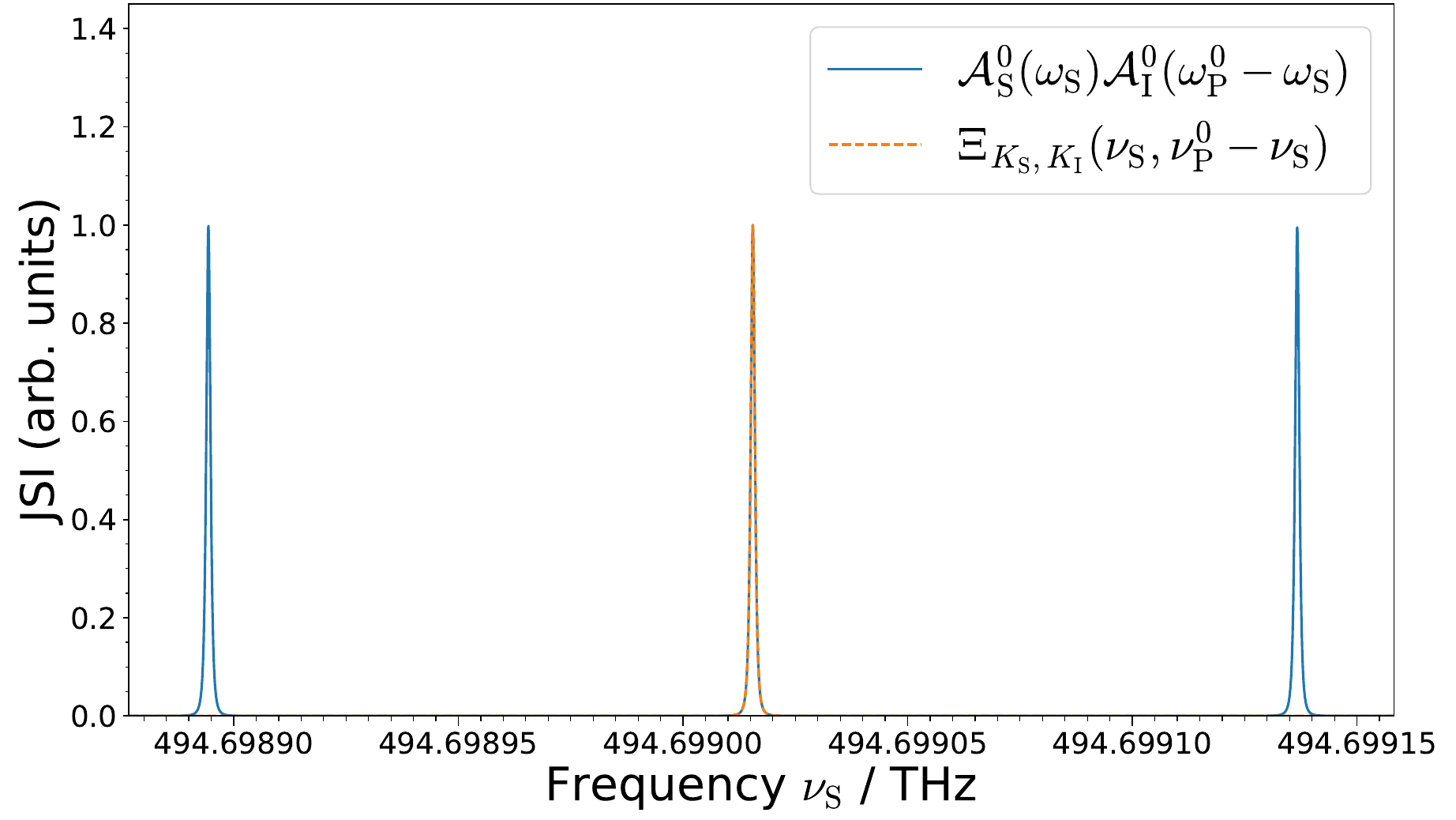}
          \label{subfig:spectrum_signal_mode0_center}
        \end{subfigure}
      \end{tabular}
      \vspace{-4ex}
      \caption{
        Signal spectrum. The plots are shown for $\finesse_\indSignal = \num{61.0}$ and $\finesse_\indIdler = \num{83.0}$. The solid blue line and the dotted orange line represent $\Airy[^0]{\indSignal}{\omega_\indSignal}\Airy[^0]{\indIdler}{\omega_\indPump^0-\omega_\indSignal}$ and $\Xi_{K_\indSignal, K_\indIdler}(\nu_\indSignal,\nu_\indPump^0-\nu_\indSignal)$, respectively. Panel (\subref{subfig:spectrum_signal_mode0_all}) shows the main cluster, while (\subref{subfig:spectrum_signal_mode0_center}) provides a magnified view of the three modes around the central peak.
      }
      \label{fig:spectrum_signal}
    \end{figure*}

    Furthermore, the photon annihilation operators corresponding to the frequency modes $\psi_{k}$ and $\phi_{k}$ are defined as \cite{RohdeSilberhorn2007}
    \begin{align}
      \hat{A}_k\! \coloneq\! \int_0^\infty\!\! \dd{\omega_\indSignal} \psi_{k}(\omega_\indSignal) \hat{a}_\indSignal(\omega_\indSignal), \ 
      \hat{B}_k\! \coloneq\! \int_0^\infty\!\! \dd{\omega_\indIdler} \phi_{k}(\omega_\indIdler) \hat{a}_\indIdler(\omega_\indIdler).
    \end{align}
    These operators satisfy the commutation relations
    \begin{subequations}
      \begin{align}
        \left[\hat{A}_k, \hat{A}_{k^\prime}^\dagger\right] = \delta_{k,k^\prime},\  \left[\hat{A}_k, \hat{A}_{k^\prime}\right] = 0,
      \end{align}\vspace{-2ex}
      \begin{align}
        \left[\hat{B}_k, \hat{B}_{k^\prime}^\dagger\right] = \delta_{k,k^\prime},\ \left[\hat{B}_k, \hat{B}_{k^\prime}\right] = 0.
      \end{align}
      \label{eq:commutation_relations}
    \end{subequations}

  \subsection{\label{subsec:decompOfState}Decomposition of state}
    Based on the approximate expression of the JSA for cSPDC given in Eq.~\eqref{eq:cSPDC_JSA_Schmidt}, the generated quantum state can be approximated as:
    \begin{align}
      \Ket{\Psi} &= \hat{U}_\indexrm{cav}\Ket{0}_{\indSignal \indIdler} = \exp\left[-\frac{i}{\hbar}\hat{\hamiltonian}_\indexrm{cav,eff}\right]\Ket{0}_{\indSignal \indIdler} \notag\\
      &\simeq \exp\left[\sum_{k=-M}^{M}\! r_k(\hat{A}_k^\dagger\hat{B}_k^\dagger-\mathrm{H.c.})\right]\Ket{0}_{\indSignal \indIdler}\notag\\
      &= \bigotimes_{k=-M}^{M}\! \exp\left[r_k(\hat{A}_k^\dagger\hat{B}_k^\dagger-\mathrm{H.c.})\right]\Ket{0}_{\indSignal \indIdler} \quad\left(\text{from Eq.\eqref{eq:commutation_relations}}\right)\notag\\
      &= \bigotimes_{k=-M}^{M}\! \hat{S}_k^{(\mathrm{SI})}(-r_k)\Ket{0}_{\indSignal \indIdler}\notag\\
      &= \bigotimes_{k=-M}^{M}\! \left( \sum_{n_k=0}^\infty \frac{\tanh^{n_k} r_k}{\cosh r_k}  \Ket{n_k}_{\indSignal,k} \Ket{n_k}_{\indIdler,k}\right),
      \label{eq:cSPDC_state}
    \end{align}
    where $k$ denotes each discrete frequency mode defined by the cavity. This yields an approximate representation where the state in each individual frequency mode corresponds to a two-mode squeezed vacuum (TMSV).

    It is important to note that the JSA for the case of CW pumping generally cannot undergo Schmidt decomposition in its original form. This is because, while pulsed pumping results in a JSA with a two-dimensional spectral distribution that can be decomposed into a finite number of Schmidt modes, CW pumping yields a JSA that is one-dimensionally distributed along the line of energy conservation. Consequently, even within each discrete frequency mode defined by the cavity, an expansion into an infinite number of Schmidt modes would be required.
    In this study, to evaluate the entanglement-heralding rate while accounting for multi-photon pair generation, we utilize an approximate JSA representation that facilitates Schmidt decomposition into the discrete frequency modes of the cavity. By relaxing the strict one-dimensional nature of the JSA under CW pumping, we describe each frequency mode of interest as a single TMSV state, thereby providing a formulation of the quantum state that incorporates multi-photon contributions.

  \subsection{\label{subsec:checkValidity} Validation of the approximate JSI and JSA}
    As discussed in the previous subsection, we have employed an approximate representation for the JSA. While this expression is not strictly exact, we confirm that it is reasonably close to the original JSA and JSI by presenting numerical plots for comparison.

    In our simulations, we assume a signal center wavelength of approximately \qty{606}{nm} and an idler center wavelength of approximately \qty{1550}{nm}, with a pump wavelength around \qty{435}{nm}. Specifically, the pump wavelength is set to $\lambda_\indPump^0 = \qty{435.5359}{nm}$. The free spectral range (FSR) and finesse $\finesse_\epsilon = \fsr_\epsilon / \fwhm_\epsilon$ ($\epsilon \in \{\indSignal, \indIdler\}$) are set to the design values used in our group's previous experimental work \cite{TateishiHorikiri2025}: $\fsr_\indSignal = \qty{121.120}{MHz}$, $\fsr_\indIdler = \qty{121.189}{MHz}$, $\finesse_\indSignal = \num{61.0}$, and $\finesse_\indIdler = \num{83.0}$. Note that the measured FSR for the signal in that experiment was $\fsr_\indSignal \simeq \qty{123.0}{MHz}$, which is close to the design value.

    To perform these plots, it is necessary to determine the values of $K_\indSignal$ and $K_\indIdler$. Given that the center frequencies of the signal and idler are $\nu_\indSignal^0$ and $\nu_\indIdler^0$, respectively, we utilize the following relations:
    \begin{align}
      K_\indSignal \fsr_\indSignal \simeq \nu_\indSignal^0, \quad
      K_\indIdler \fsr_\indIdler \simeq \nu_\indIdler^0.
      \label{eq:constK_centFreq}
    \end{align}
    Specifically, we first plot the individual spectra of the signal and idler to identify the precise center frequencies and determine the values of $K_\indSignal$ and $K_\indIdler$. We then verify these values by plotting $\Xi_{K_\indSignal, K_\indIdler} (\nu_\indSignal, \nu_\indIdler)$ and confirming that its peak aligns with the central peak of the overall spectrum.

    Performing this procedure yields $K_\indSignal=\num{4084371}$ and $K_\indIdler=\num{1597761}$. Indeed, by plotting the original spectrum for the signal, $\Airy{\indSignal}{\omega_\indSignal}\, \Airy{\indIdler}{\omega_\indPump^0 - \omega_\indSignal}$, alongside the function $\Xi_{K_\indSignal+k, K_\indIdler-k} (\nu_\indSignal, \nu_\indPump^0 - \nu_\indSignal)$ corresponding to the determined values of $K_\indSignal$ and $K_\indIdler$, we confirm that the central peaks coincide, as shown in Fig.~\ref{fig:spectrum_signal}.
    
    The joint spectral intensity (JSI) plotted with these parameters is presented in Fig.~\ref{fig:plot_JSI}. Here, $S_\indexrm{cav}(\omega_\indSignal, \omega_\indIdler)=\absolute{s(\omega_\indSignal, \omega_\indIdler)}^2\Airy{\indSignal}{\omega_\indSignal}\Airy{\indIdler}{\omega_\indIdler}$ is shown in Figs.~\ref{fig:plot_JSI}(\subref{subfig:plot_JSI_Airy}), (\subref{subfig:plot_JSI_mode0_Airy}), and (\subref{subfig:plot_JSI_mode50_Airy}), while Figs.~\ref{fig:plot_JSI}(\subref{subfig:plot_JSI_decompose}), (\subref{subfig:plot_JSI_mode0_decompose}), and (\subref{subfig:plot_JSI_mode50_decompose}) represent the approximation $S_\indexrm{cav}^\indexrm{(approx)}(\omega_\indSignal, \omega_\indIdler)=\sum_{k=-50}^{50}\absolute{\psi_{k}(\omega_\indSignal)}^2\absolute{\phi_{k}(\omega_\indIdler)}^2$.

    Ideally, the pump envelope under CW pumping takes the form $s(\omega_\indSignal, \omega_\indIdler)=\delta(\nu_\indSignal+\nu_\indIdler-\nu_\indPump^0)$. For numerical plotting purposes, however, we employ the approximation $s(\omega_\indSignal, \omega_\indIdler)=\exp[-\left(\nu_\indSignal+\nu_\indIdler-\nu_\indPump^0\right)^2/(2\sigma_\indPump^2)]$.

    In Figs.~\ref{fig:plot_JSI}(\subref{subfig:plot_JSI_Airy}) and (\subref{subfig:plot_JSI_decompose}), the nine modes in the vicinity of the central mode ($k=-4, -3, \dots, 4$) are shown.
    While the logarithmic scale leads to slight visual differences, both plots can be seen to exhibit similar discrete structures.

    Next, Figs.~\ref{fig:plot_JSI}(\subref{subfig:plot_JSI_mode0_Airy}) and (\subref{subfig:plot_JSI_mode0_decompose}) plot the central mode ($k=0$), and Figs.~\ref{fig:plot_JSI}(\subref{subfig:plot_JSI_mode50_Airy}) and (\subref{subfig:plot_JSI_mode50_decompose}) plot the 50th mode from the center ($k=50$).
    For the calculations in Figs.~\ref{fig:plot_JSI}(\subref{subfig:plot_JSI_mode0_Airy}) and (\subref{subfig:plot_JSI_mode50_Airy}), we set $\sigma_\indPump=\qty{1}{MHz}$. 
    Although the actual JSI is narrower than shown in these plots because the pump envelope function is a delta function rather than a Gaussian, the approximated representations in Figs.~\ref{fig:plot_JSI}(\subref{subfig:plot_JSI_mode0_decompose}) and (\subref{subfig:plot_JSI_mode50_decompose}) still capture the characteristic features along the lines.
    Furthermore, the regions where intensity is distributed in Figs.~\ref{fig:plot_JSI}(\subref{subfig:plot_JSI_mode0_decompose}) and (\subref{subfig:plot_JSI_mode50_decompose}) are contained within those of Figs.~\ref{fig:plot_JSI}(\subref{subfig:plot_JSI_mode0_Airy}) and (\subref{subfig:plot_JSI_mode50_Airy}) along both the signal and idler axes, respectively. 
    Consequently, it can be concluded that the approximation derived and employed in this work provides a valid description.

\section{\label{sec:evalRate} Evaluation of heralding probability and fidelity}
  \subsection{Theory\label{subsec:probFid}}
    In the quantum repeater scheme considered in this work, single-photon entanglement is generated at the links between repeaters (elementary links, ELs) through single-photon interference \cite{DLCZ2001, SimonGisin2007, SangouardGisin2011}.
    By arranging these links and performing entanglement swapping, entanglement is distributed between two distant locations. Therefore, the entanglement generation rate and fidelity in the EL are crucial metrics for performance evaluation, and this study evaluates these indicators.

    Furthermore, the entanglement generation rate $\mathcal{R}$ is expressed as $\mathcal{R}=\prob/\tau$, where $\tau$ denotes the time required for a single entanglement generation trial and $\prob$ represents the heralding probability. It therefore suffices to evaluate the heralding probability to assess the generation rate. Accordingly, this study evaluates the rate improvement by focusing on the enhancement of the heralding probability through multiplexing, without specifying a particular value for $\tau$.

    \subsubsection{Single-mode case\label{subsubsec:singleMode}}
\begin{figure}[H]
        \centering
        \includegraphics[width=0.95\columnwidth]{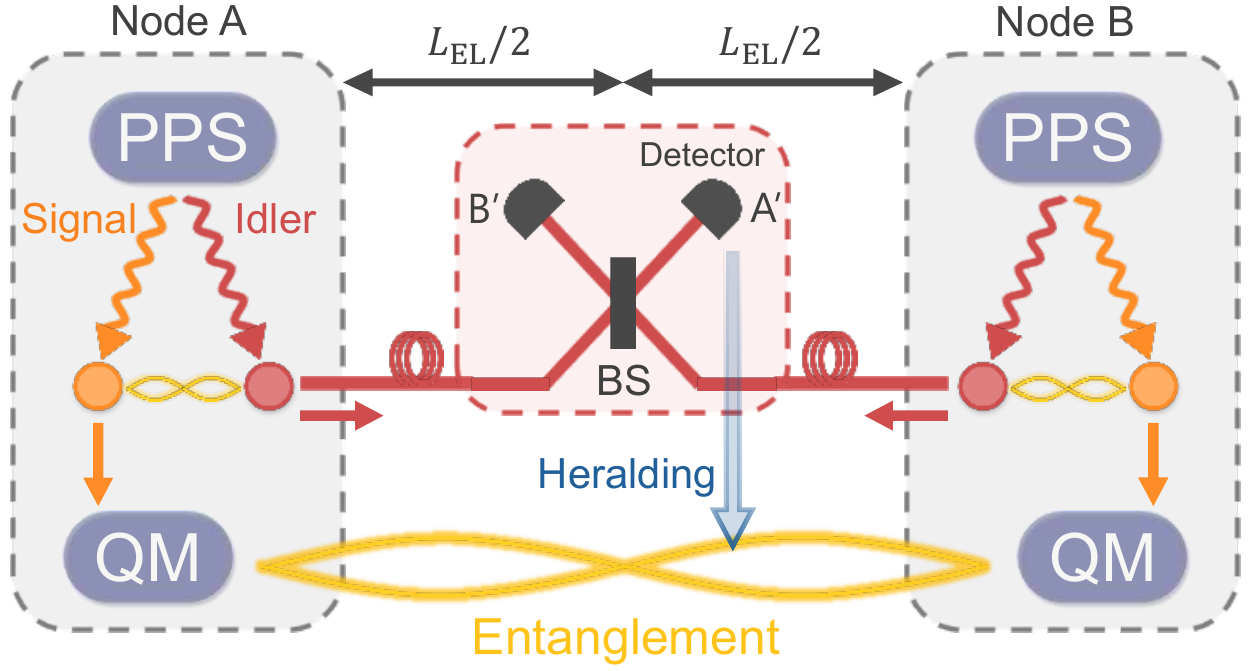}
        \caption{
          Schematic of single-mode entanglement generation via single-photon interference in an elementary link. PPS: photon-pair source; QM: quantum memory; BS: beam splitter. Using SPDC at each PPS, the signal photon is stored in the QM while the idler photon is sent to a central station for interference with the idler photon from the other node. A click at only one of the detectors heralds the generation of entanglement between the signal photons stored in the respective QMs.
        }
        \label{fig:single_photon_interference}
      \end{figure}

      We first consider the single-mode case. As illustrated in Fig.~\ref{fig:single_photon_interference}, we assume a single-photon interference scheme (also referred to as a single-photon detection scheme) in which nodes A and B are each equipped with a photon-pair source (PPS) and a quantum memory (QM). In this setup, the idler photons emitted from the PPS at each node are transmitted to a central station for interference and detection; this process generates photon-number entanglement between the signal photons stored in the respective QMs. Furthermore, the PPS is assumed to produce a two-mode squeezed vacuum (TMSV) state with a small mean photon number, ensuring that the probability of generating multiple photon pairs remains sufficiently low.
      
      In the single-photon interference scheme, entanglement is heralded when a click occurs in only one of the two detectors following the beam splitter. As illustrated in Fig.~\ref{fig:single_photon_interference}, we denote the detectors at the central station as $A'$ and $B'$. Letting $\prob^{(\indexrm{A^{\smash{\!\prime}}})}$ and $\prob^{(\indexrm{B^{\smash{\prime}}})}$ be the probabilities that a click occurs only at detector $A'$ and only at detector $B'$, respectively, the heralding probability in the single-mode case is given by
      \begin{align}
        \prob_\indexrm{single} = \prob^{(\indexrm{A^{\smash{\!\prime}}})} + \prob^{(\indexrm{B^{\smash{\prime}}})}.
      \end{align}
      
      We make several assumptions here. First, we assume that the mean photon numbers of the TMSVs generated at both nodes, the transmission losses of the optical paths from each node, and the detection efficiencies of the two idler detectors at the central station are identical, i.e.,
      $(\mu^{\smash{(\indexrm{A})}}=\mu^{\smash{(\indexrm{B})}}=\mu, 
      \eta_{\indexrm{att}}^{\smash{(\indexrm{A})}}=\eta_{\indexrm{att}}^{\smash{(\indexrm{B})}}=\eta_{\indexrm{att}}, 
      \eta_{\indexrm{det}}^{\smash{\indexrm{(A^{\smash{\!\prime}})}}}=\eta_{\indexrm{det}}^{\smash{\indexrm{(B^{\smash{\prime}})}}}=\eta_{\indexrm{det}})$.
      That is, we assume a symmetric configuration with respect to nodes A and B. Under these conditions, the click probabilities satisfy $\prob^{(\indexrm{\!A^{\smash{\!\prime}}})}=\prob^{(\indexrm{B^{\smash{\prime}}})}\eqcolon\prob_\indexrm{either}$, and the heralding probability in the single-mode case is expressed as
      \begin{align}
        \prob_\indexrm{single} = 2 \prob_\indexrm{either}.
        \label{eq:prob_single}
      \end{align}

      Furthermore, assuming that the central station for idler interference is located at the midpoint of an elementary link of distance $L_\indexrm{EL}$ as shown in Fig.~\ref{fig:single_photon_interference}, the transmission efficiency is expressed as $\eta_\indexrm{att}=10^{-\alpha_\indexrm{att}(L_\indexrm{EL}/2)/ 10}$, where $\alpha_\indexrm{att}$ denotes the attenuation coefficient of the optical path.

      Subsequently, we assume an ideal scenario in which the quantum memories can absorb photons with \qty{100}{\%} efficiency. We also assume that the spectral demultiplexing efficiency---the efficiency of directing frequency-multiplexed idler photons to their respective detectors---is \qty{100}{\%}.
      Additionally, the detectors for the idler photons are taken to be on-off detectors without photon-number-resolving capability.

      Under these assumptions, we utilize the results from Ref.~\cite{RazaviShapiro2006}, in which the heralding probability and fidelity for a DLCZ-type scheme---generating photon-number entanglement via single-photon interference using TMSV sources---are evaluated by accounting for multi-photon pair generation as well as transmission and detection efficiencies. Accordingly, the probability is expressed as
      \begin{align}
        \prob_\indexrm{either} = \frac{\mu^\prime}{(\mu^\prime+1)^2},
        \label{eq:prob_either}
      \end{align}
      where $\mu^\prime\coloneq\eta_{\indexrm{att}}\eta_{\indexrm{det}}\mu$.

      Next, the target state to be generated in this scheme is the photon-number entangled state defined as
      \begin{align}
        \Ket{\psi_{\mathrm{ideal}}^\pm} = 
          \frac{1}{\sqrt{2}}\Ket{01}_{\indSignal_\indexrm{\!A},\indSignal_\indexrm{B}}
          \pm \frac{1}{\sqrt{2}}e^{i\theta}\Ket{10}_{\indSignal_\indexrm{\!A},\indSignal_\indexrm{B}},
      \end{align}
      where $\theta$ is the phase difference acquired by the idler photons from nodes A and B until they reach the detectors, and the subscripts $\indSignal_\indexrm{\!A}$ and $\indSignal_\indexrm{B}$ denote the signal photons at nodes A and B, respectively. 
      We consider the fidelity of the state heralded on the condition that only one of the idler detectors clicks, relative to this ideal state. 
      
      Assuming that the phase difference $\theta$ is perfectly locked, we can again utilize the results from Ref.~\cite{RazaviShapiro2006} to obtain the fidelity as
      \begin{align}
        \fidelity = \frac{(\mu^\prime+1)^2}{(\mu+1)^3}.
        \label{eq:fidelity}
      \end{align}

      \begin{figure*}[ht]
        \centering
        \includegraphics[width=0.8\linewidth]{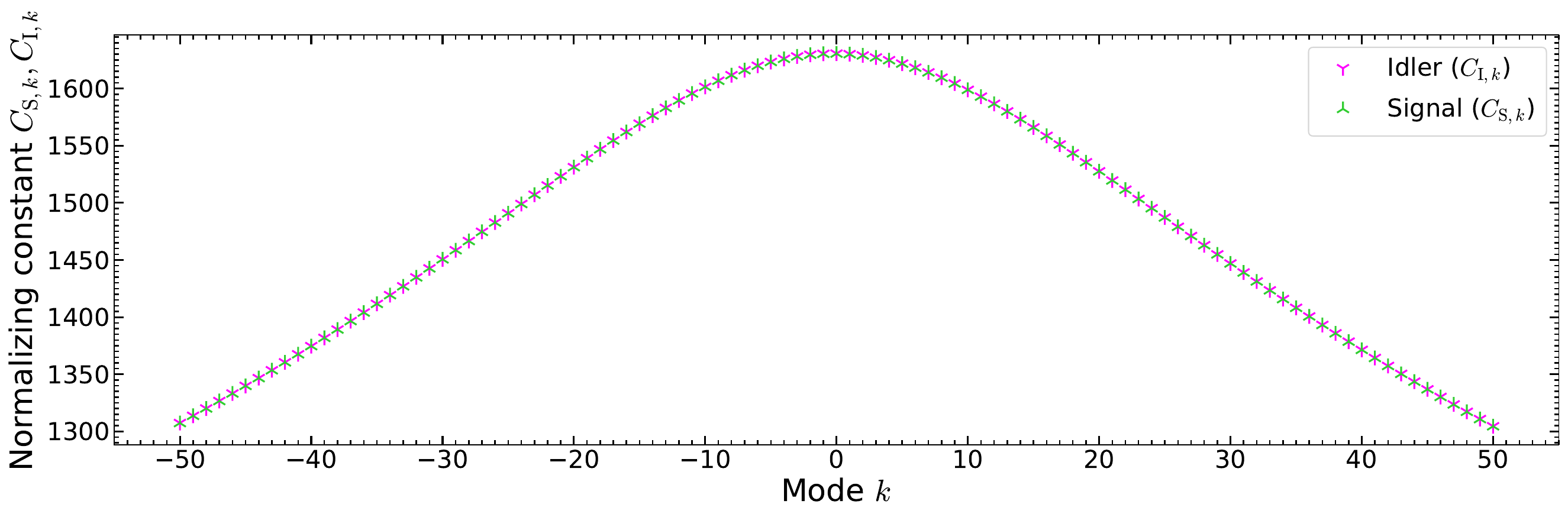}
        \caption{
          Calculated normalization constants obtained using Eq.~\eqref{eq:normalizationConstant}. The calculation was performed with the following parameters: $\fsr_\indSignal = \qty{121.120}{MHz}$, $\fsr_\indIdler = \qty{121.189}{MHz}$, $\finesse_\indSignal = \num{61.0}$, $\finesse_\indIdler = \num{83.0}$, $\lambda_\indPump^0 = \qty{435.5359}{nm}$, $K_\indSignal = \num{4084371}$, and $K_\indIdler = \num{1597761}$.
        }
        \label{fig:plot_normConst}
      \end{figure*}
    
    \subsubsection{Multimode case\label{subsubsec:multiModes}}
      We next turn our attention to the multimode case. In Sec.~\ref{sec:decompToCavity}, we derived an expression for the state generated via cavity-enhanced SPDC (cSPDC) [Eq.~\eqref{eq:cSPDC_state}], demonstrating that each frequency mode discretized by the cavity can be approximated as an independent TMSV state. This implies that a PPS based on cSPDC is capable of generating photon pairs independently across multiple frequency modes.
      Furthermore, we assume a scenario where photon detection and storage in the quantum memories are performed independently for each frequency mode. 
      
      Given these assumptions, our proposed frequency-multiplexing scheme allows for independent entanglement generation in each mode. Therefore, we first determine the heralding probability and fidelity for each individual mode under the assumption of independent entanglement generation using TMSV sources. Based on these single-mode values, we then evaluate the overall heralding probability and fidelity for the entire multiplexed system.

      First, for the state described by Eq.~\eqref{eq:cSPDC_state}, since $r_k \propto C_{\indSignal,k} C_{\indIdler,k}$ from Eq.~\eqref{eq:squeezingParameter}, the squeezing parameter $r_k$ for each TMSV mode can be expressed as
      \begin{align}
        r_k = \frac{C_{\indSignal,k} C_{\indIdler,k}}{C_{\indSignal,0} C_{\indIdler,0}} r_0.
      \end{align}
      Therefore, by specifying the mean photon number $\mu_0$ of the reference mode ($k=0$), the mean photon number $\mu_k$ for each TMSV mode can be determined as follows:
      \begin{align}
        \mu_k = \sinh^2 r_k = \sinh^2 \left(\frac{C_{\indSignal,k} C_{\indIdler,k}}{C_{\indSignal,0} C_{\indIdler,0}} \arcsinh \sqrt{\mu_0}  \right).
        \label{eq:meanPhotonNum}
      \end{align}

      By substituting these expressions for the mean photon number into Eqs.~\eqref{eq:prob_single}, \eqref{eq:prob_either}, and \eqref{eq:fidelity}, we can determine the heralding probability $\prob_{\indexrm{single}, k}$ and fidelity $\fidelity_k$ for entanglement generation via the single-photon interference scheme in each mode $k$.

      Assuming that the entanglement generation trials in each mode are performed independently, an overall heralding event is successful if heralding occurs in at least one of the available modes. Accordingly, the total heralding probability $\prob_\indexrm{multi}$ is expressed as
      \begin{align}
        \prob_\indexrm{multi} =  1 - \prod_{k}(1 - \prob_{\indexrm{single},k}).
      \end{align}

      Furthermore, when considering a range of $M$ modes on either side of the center, we evaluate the overall fidelity using the minimum value among the fidelities of all individual single modes. As will be demonstrated in the following section, this minimum value corresponds to the fidelity of the reference mode, $\fidelity_0$, in our case.

      Additionally, the total mean photon number for the multiplexed system, $\mu_\indexrm{multi}$, is given by
      \begin{align}
        \mu_\indexrm{multi}\coloneq\sum_{k}\mu_k.
      \end{align}

  \subsection{\label{subsec:calcProbFidelity}Calculation}
    In this part, we calculate the heralding probability and fidelity of frequency-multiplexed entanglement generation for elementary link distances of \qty{25}{km}, \qty{50}{km}, and \qty{100}{km}. For each distance, we consider three scenarios with different mean photon numbers for the reference mode and compare the results with the single-mode case that utilizes only the reference mode.

    In our model, the number of available modes is determined by the spectral acceptance bandwidth of the quantum memory. Assuming that the inhomogeneous broadening of the Pr:YSO crystal is approximately \qty{10}{GHz} and the free spectral range (FSR) of the signal photons is approximately \qty{120}{MHz}, the number of accessible modes is around 100. Accordingly, we perform our calculations assuming the multiplexing of a total of 101 modes, consisting of the center mode and $\pm 50$ side modes.
    
    To provide a specific example of the evaluation, we perform calculations using the same parameters as in Sec.~\ref{subsec:checkValidity}: $\fsr_\indSignal = \qty{121.120}{MHz}$, $\fsr_\indIdler = \qty{121.189}{MHz}$, $\finesse_\indSignal = \num{61.0}$, $\finesse_\indIdler = \num{83.0}$, and $\lambda_\indPump^0 = \qty{435.5359}{nm}$.

    The normalization constants calculated via Eq.~\eqref{eq:normalizationConstant} using Mathematica are shown in Fig.~\ref{fig:plot_normConst}. As observed in the figure, the normalization constant is maximal at the center mode and decreases toward the outer modes. This behavior arises because the functions $\tilde{\psi}_k(\omega_\indSignal)$ and $\tilde{\phi}_k(\omega_\indIdler)$ incorporate the cluster effect.

    Subsequently, the mean photon number $\mu_k$ for each mode, derived from Eq.~\eqref{eq:meanPhotonNum} using these normalization constants, is plotted in Fig.~\ref{fig:plot_compare}(\subref{subfig:plot_compare_mean}). Reflecting the trend of the normalization constants, the mean photon number is highest at the center mode and decreases for modes further from the center.
    
    For each case of $L_\indexrm{EL}$, the resulting heralding probability $\prob_{\indexrm{single}, k}$ and fidelity $\fidelity_k$ for entanglement generation in each mode are shown in Figs.~\ref{fig:plot_compare}(\subref{subfig:plot_compare_prob_25})--(\subref{subfig:plot_compare_fidelity_100}), where the detector efficiency is set to $\eta_\indexrm{det}=\num{0.9}$. 
    These results indicate that, for the mean photon numbers considered here, the heralding probability is highest for the center mode, whereas the fidelity is lowest at the center.
   
    % \afterpage{%
      % \clearpage % ←「次ページ頭」でフロートを吐き出す（ここがポイント）
      \begingroup
        % このブロック内だけ効かせる
        \setcounter{dbltopnumber}{7}            % 上部に置ける全幅フロート数
        \setcounter{totalnumber}{9}             % 1ページの総フロート数
        \renewcommand{\dbltopfraction}{1.0}    % 上部を占められる最大割合
        \renewcommand{\dblfloatpagefraction}{0.00} % 全幅フロートだけで構成される「フロート専用ページ」を作ってよいと判定するための最低充足率。
        \renewcommand{\textfraction}{0.00}      % 本文の最小割合を緩める
        \setlength{\dblfloatsep}{4pt plus 0pt minus 0pt}     % 全幅フロート同士   plus, minusはそれぞれ最大いくつずれていいかを表す
        \setlength{\dbltextfloatsep}{12pt plus 2pt minus 2pt} % 全幅フロートと本文

        \begin{figure*}[t]
          \begin{tabular}{c}
            \begin{subfigure}[t]{0.666\hsize}
              \centering
              \caption{}
              \vspace{-1.75ex}
              \includegraphics[height=0.495\columnwidth]{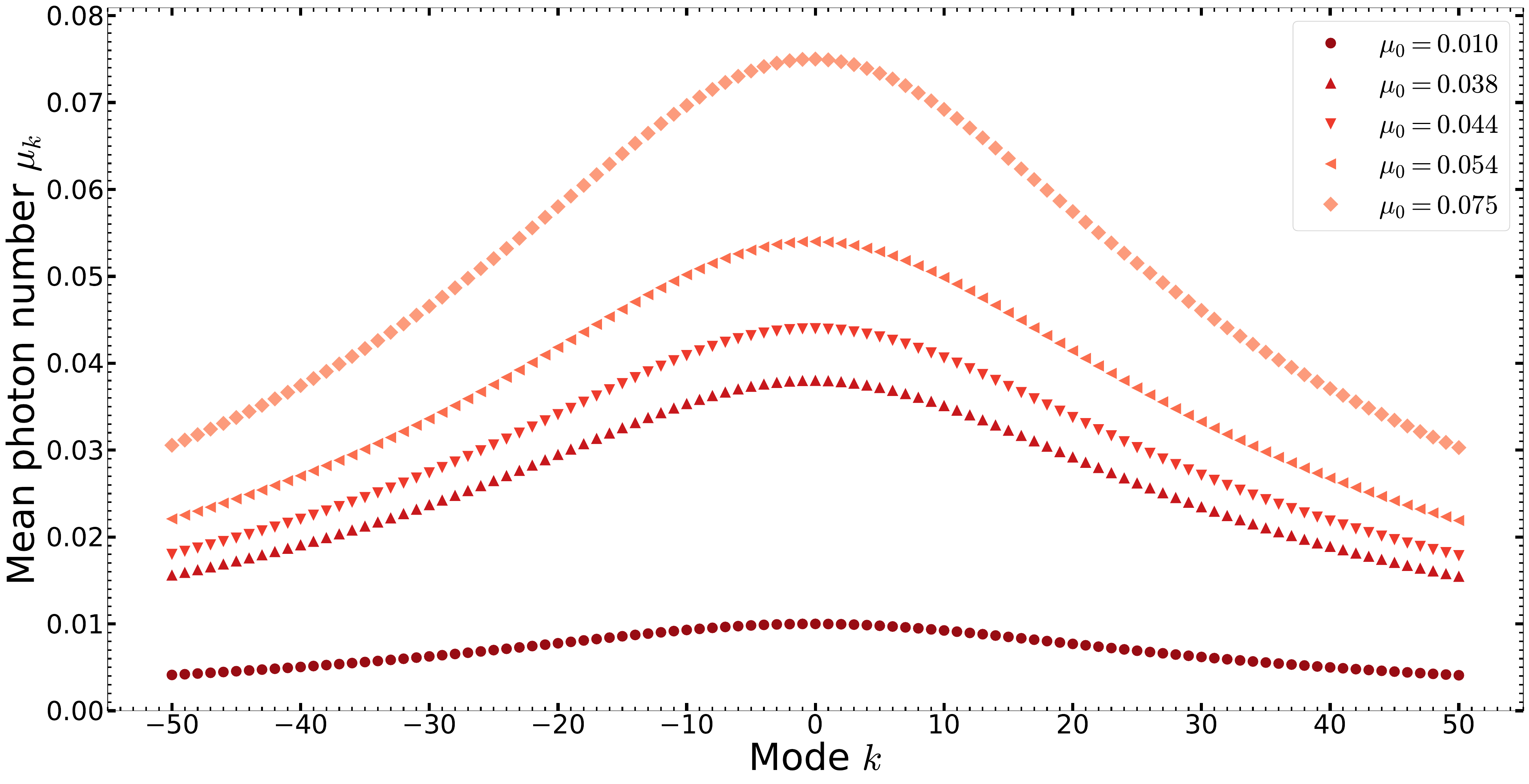}
              \label{subfig:plot_compare_mean}
            \end{subfigure}\\[-2.5ex]
            \begin{subfigure}[t]{0.337\hsize}
              \centering
              \caption{}
              \vspace{-3.75ex}
              \includegraphics[height=0.990\columnwidth]{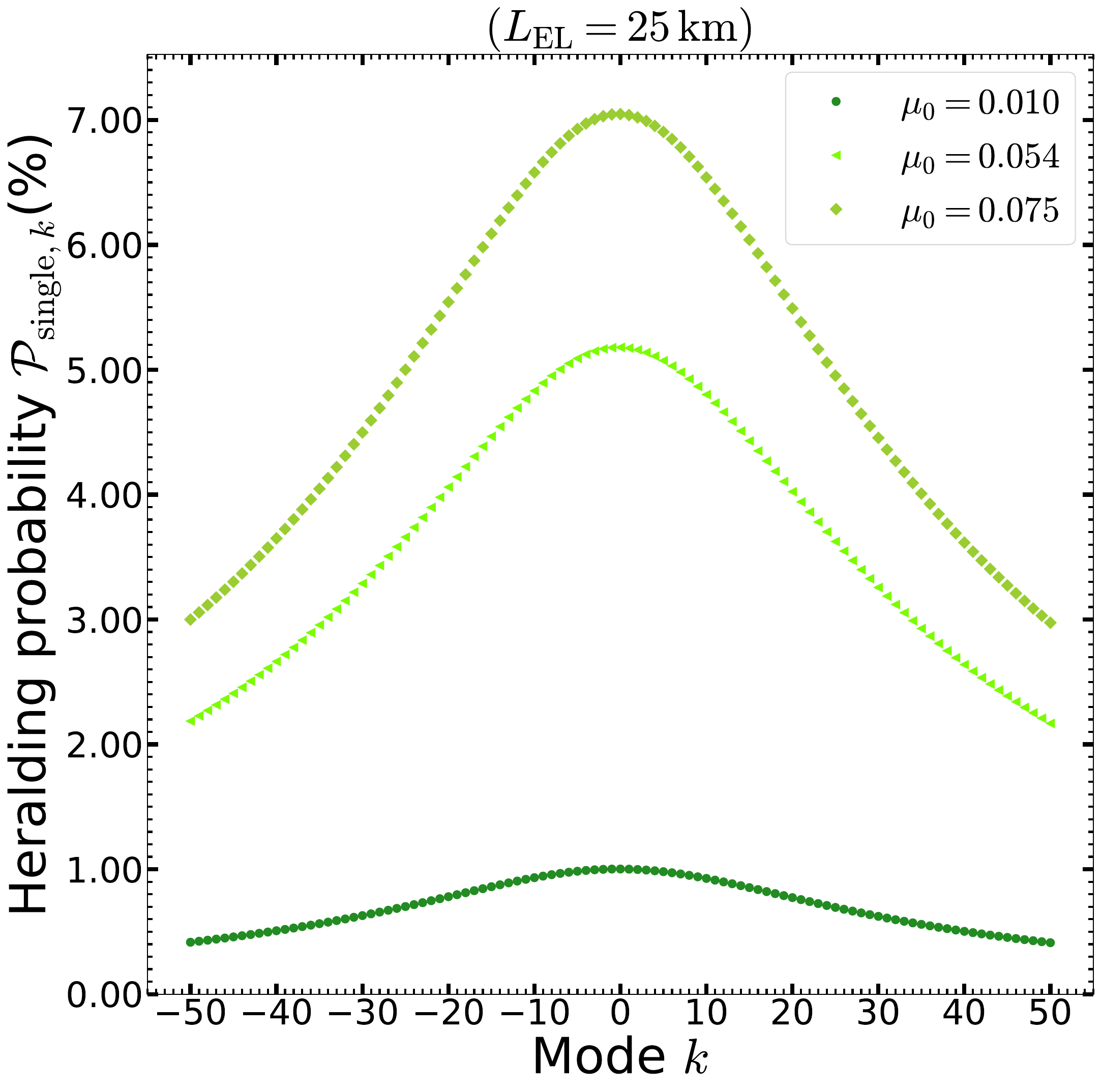}
              \label{subfig:plot_compare_prob_25}
            \end{subfigure}
            %
            % \hspace{0.0075\hsize}
            %
            \begin{subfigure}[t]{0.337\hsize}
              \centering
              \caption{}
              \vspace{-3.75ex}
              \includegraphics[height=0.990\columnwidth]{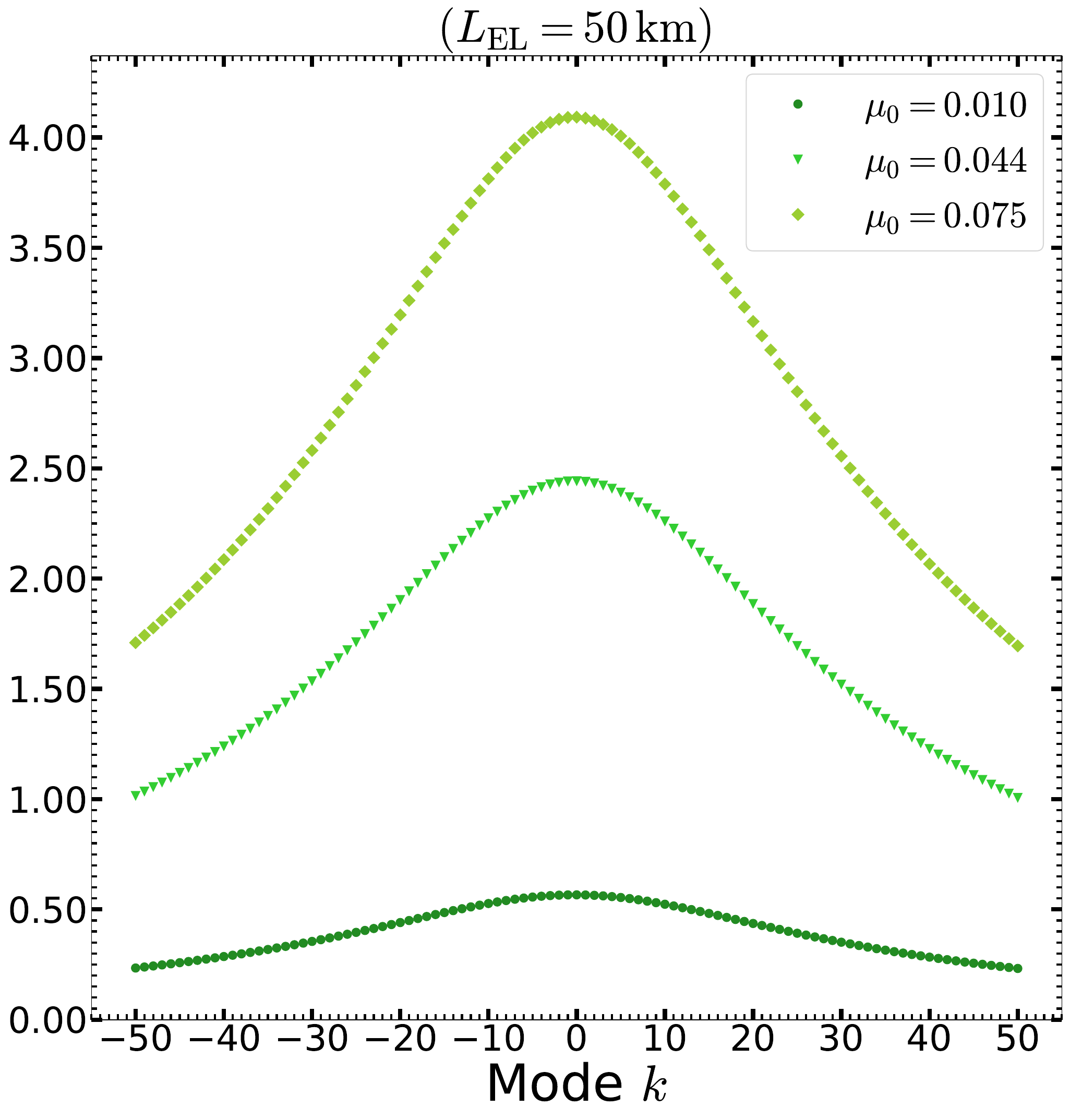} % heightをcolomnwidthの9/16倍
              \label{subfig:plot_compare_prob_50}
            \end{subfigure}
            \hspace{-0.0175\hsize}
            \begin{subfigure}[t]{0.337\hsize}
              \centering
              \caption{}
              \vspace{-3.75ex}
              \includegraphics[height=0.990\columnwidth]{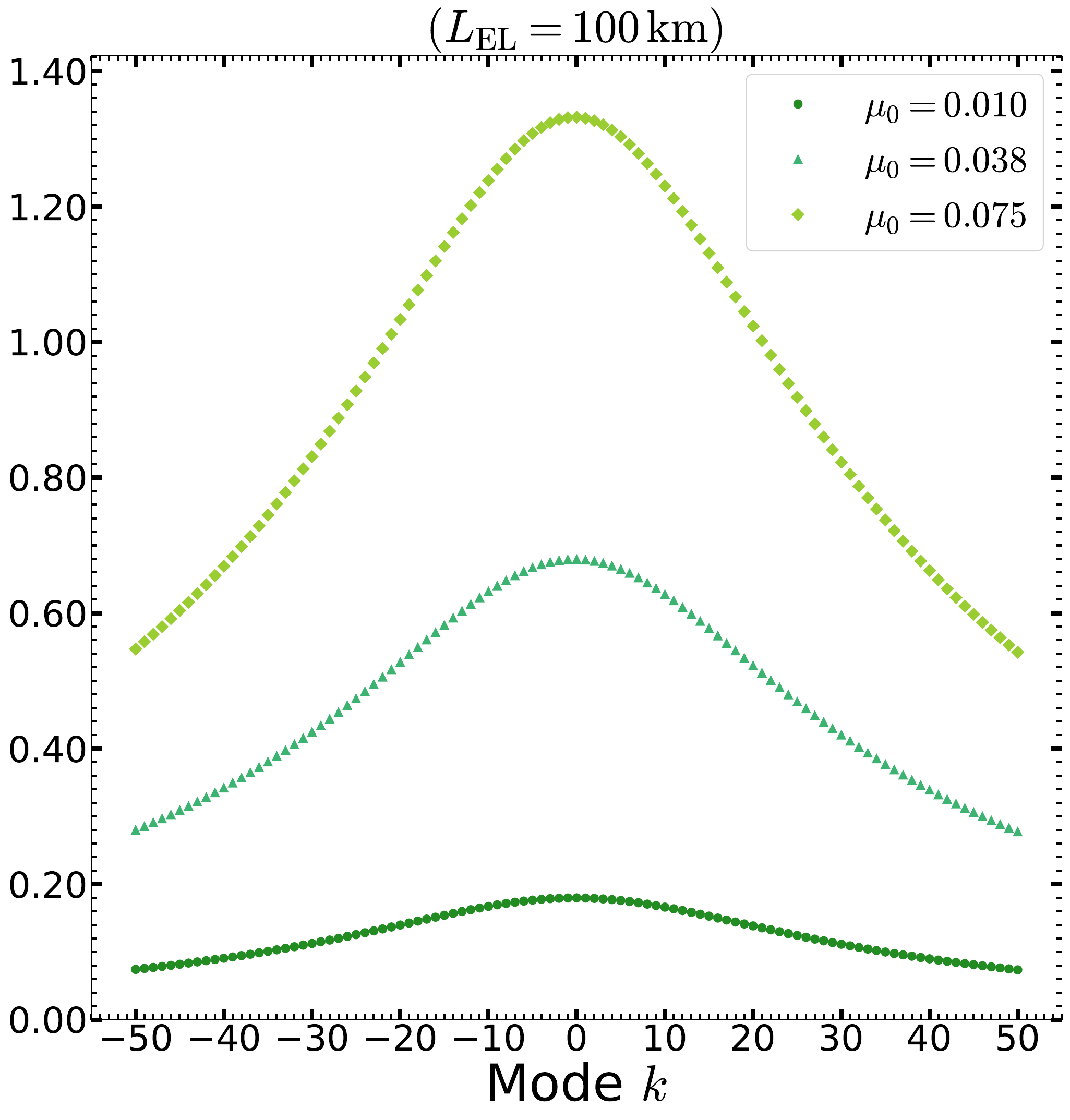}
              \label{subfig:plot_compare_prob_100}
            \end{subfigure}\\[-3.5ex]
            \begin{subfigure}[t]{0.337\hsize}
              \centering
              \caption{}
              \vspace{-3.75ex}
              \includegraphics[height=0.990\columnwidth]{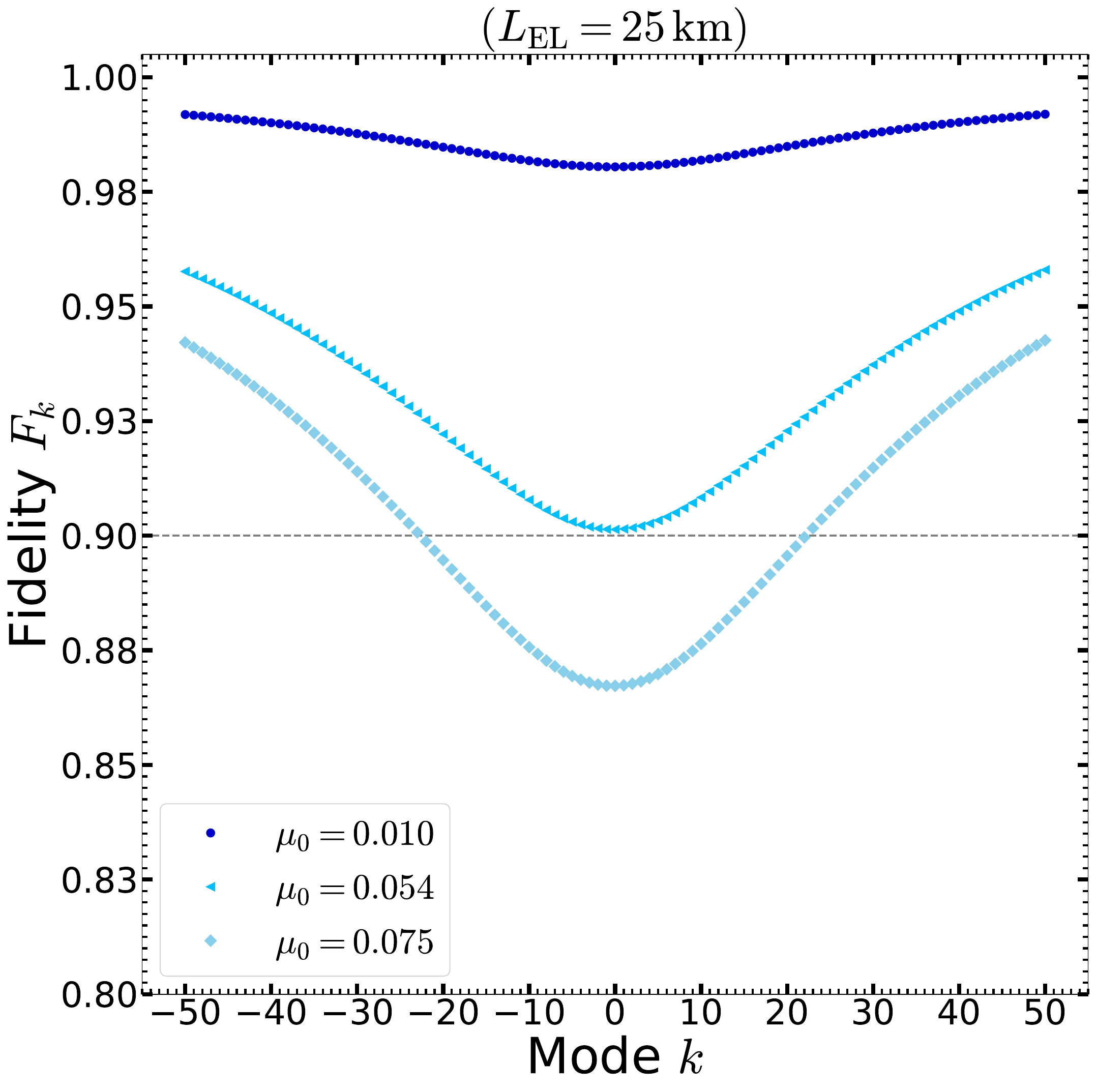}
              \label{subfig:plot_compare_fidelity_25}
            \end{subfigure}
            %
            % \hspace{0.0075\hsize}
            %
            \begin{subfigure}[t]{0.337\hsize}
              \centering
              \caption{}
              \vspace{-3.75ex}
              \includegraphics[height=0.990\columnwidth]{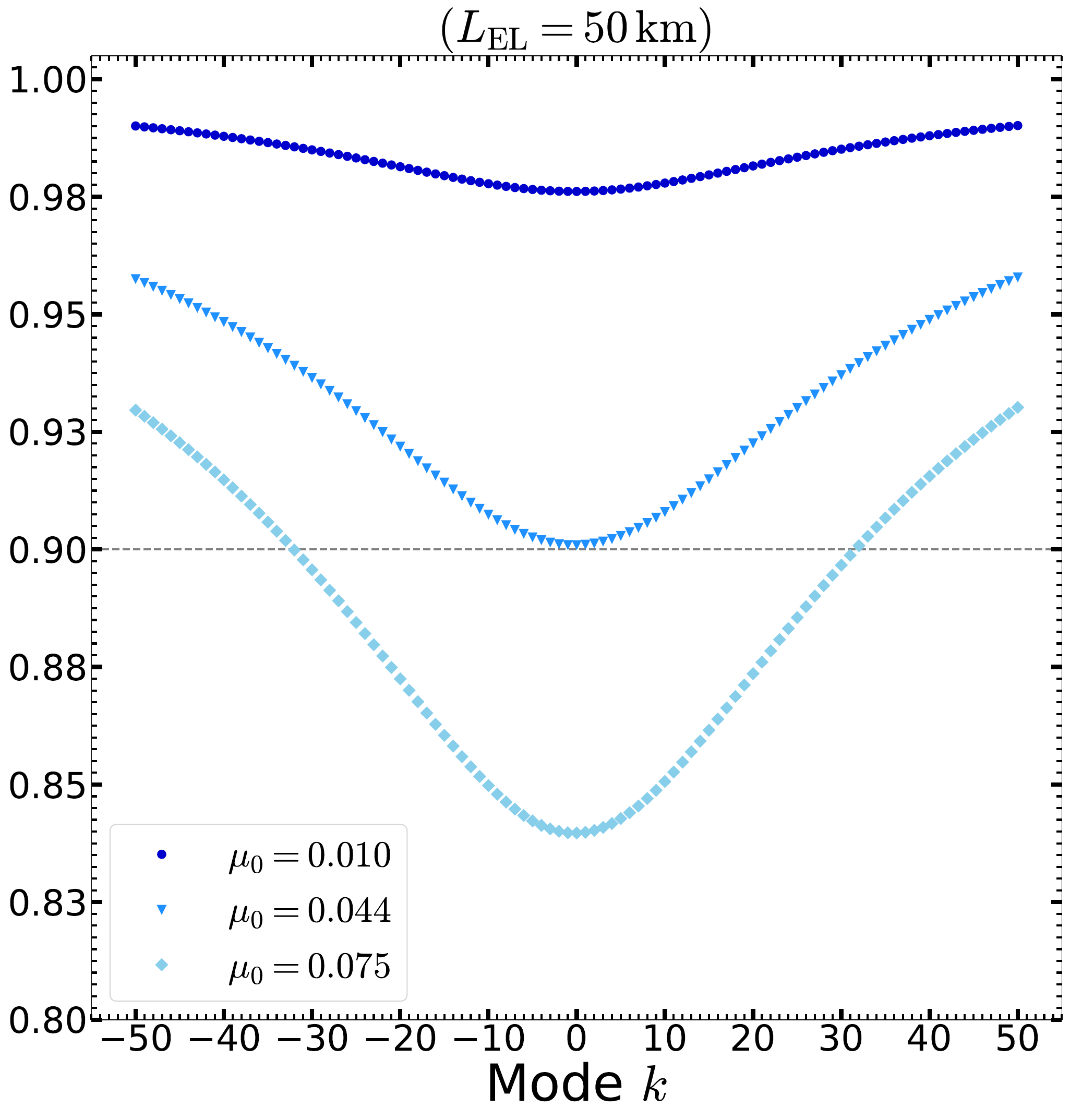} % heightをcolomnwidthの9/16倍
              \label{subfig:plot_compare_fidelity_50}
            \end{subfigure}
            \hspace{-0.0175\hsize}
            \begin{subfigure}[t]{0.337\hsize}
              \centering
              \caption{}
              \vspace{-3.75ex}
              \includegraphics[height=0.990\columnwidth]{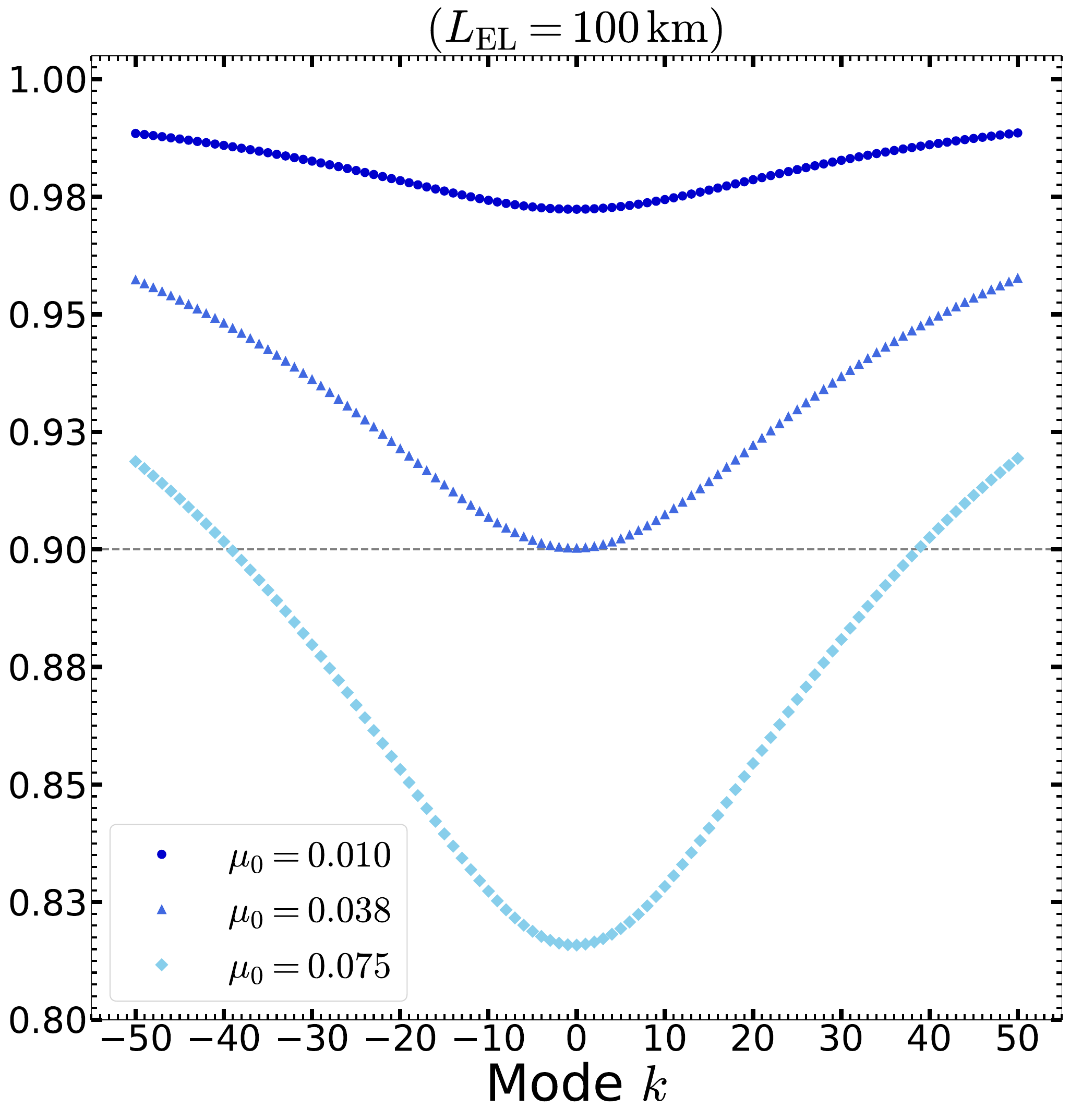}
              \label{subfig:plot_compare_fidelity_100}
            \end{subfigure}
          \end{tabular}
          \vspace{-2.5ex}
          \caption{
            Numerical results calculated using the same parameters as in Sec.~\ref{subsec:checkValidity}. (\subref{subfig:plot_compare_mean}) Mean photon numbers $\mu_k$ for the 101 modes plotted against the reference mode mean photon number $\mu_0$. (\subref{subfig:plot_compare_prob_25}), (\subref{subfig:plot_compare_prob_50}), and (\subref{subfig:plot_compare_prob_100}) Heralding probabilities $\prob_{\indexrm{single}, k}$ for each mode in the single-photon interference scheme for $L_\indexrm{EL} = 25, 50$, and \qty{100}{km}. For each distance, three cases are shown where the reference fidelity $\fidelity_0$ is is well above, marginally above, or below 0.9. (\subref{subfig:plot_compare_fidelity_25}), (\subref{subfig:plot_compare_fidelity_50}), and (\subref{subfig:plot_compare_fidelity_100}) Corresponding fidelities $\fidelity_k$ under the same conditions.
          }
          \label{fig:plot_compare}
        \end{figure*}
      \endgroup
      % \clearpage  % ←フロートだけページの後に本文を再開させる
      % figure, tableのオプションをtからpに変えることで対応できた
    % }

    In this evaluation, we consider a fidelity exceeding $0.9$ to be sufficient for compatibility with applications such as quantum key distribution (QKD) \cite{SangouardGisin2011}, while accounting for the impact of multi-photon pair generation. For each $L_\indexrm{EL}$, we define three scenarios based on the reference mode fidelity: high fidelity ($F_0 > 0.9$), marginal fidelity ($F_0 \geq 0.9$ and $F_0 \simeq 0.9$), and insufficient fidelity ($F_0 < 0.9$).
    We then determine the reference mode mean photon numbers corresponding to these conditions and calculate the resulting fidelity and heralding probability for each case.

    For these scenarios, we calculate the overall mean photon number and heralding probability for the multiplexed system. We also evaluate the degree of improvement relative to the single-mode case---where only the reference mode is utilized---and these results are shown in Table~\ref{tab:data_Fs61Fi83}.

    % \afterpage{%
      % \clearpage % ←「次ページ頭」でフロートを吐き出す（ここがポイント）
      \begingroup
        % このブロック内だけ効かせる
        \setcounter{dbltopnumber}{5}            % 上部に置ける全幅フロート数
        \setcounter{totalnumber}{9}             % 1ページの総フロート数
        \renewcommand{\dbltopfraction}{1.0}    % 上部を占められる最大割合
        \renewcommand{\dblfloatpagefraction}{0.00} % 全幅フロートだけで構成される「フロート専用ページ」を作ってよいと判定するための最低充足率。
        \renewcommand{\textfraction}{0.00}      % 本文の最小割合を緩める
        \setlength{\dblfloatsep}{4pt plus 0pt minus 0pt}     % 全幅フロート同士   plus, minusはそれぞれ最大いくつずれていいかを表す
        \setlength{\dbltextfloatsep}{12pt plus 2pt minus 2pt} % 全幅フロートと本文

        \begin{table*}[t]
          \caption{
            Comparison between single-mode and multimode cases. The elementary link distances are set to (i) $L_\indexrm{EL}=\qty{25}{km}$, (ii) $\qty{50}{km}$, and (iii) $\qty{100}{km}$. For each distance, the reference mean photon numbers are specified as: (i-1) $\mu_0=0.075$, (i-2) $0.054$, (i-3) $0.010$; (ii-1) $\mu_0=0.075$, (ii-2) $0.044$, (ii-3) $0.010$; and (iii-1) $\mu_0=0.075$, (iii-2) $0.038$, (iii-3) $0.010$. (\subref{subtab:meanProbFidelity_Fs61Fi83}) Values of the mean photon number and heralding probability for both the single-mode (SM) case ($\mu_0, \prob_{\indexrm{single, 0}}$) and the multimode (MM) case with 101 modes ($\mu_\indexrm{multi}, \prob_\indexrm{multi}$). The fidelity is represented by $\fidelity_0$, which corresponds to the minimum value across all modes. (\subref{subtab:improveFactor_Fs61Fi83}) Improvement ratios for the mean photon number and heralding probability, expressed as $\mu_\indexrm{multi}/\mu_0$ and $\prob_\indexrm{multi}/\prob_{\indexrm{single}, 0}$, respectively.
          }
          \label{tab:data_Fs61Fi83}
          \vspace{-1.75ex}
          \begin{tabular}{c}
            \begin{subtable}[t]{0.65\hsize}
              \centering
              \setlength{\belowcaptionskip}{2mm}
              \caption{}
              \label{subtab:meanProbFidelity_Fs61Fi83}
              % % \setlength{\doublerulesep}{0pt} %hline hlineで太線にするために間隔を0に
              % \setlength{\doublerulesep}{2pt}   %hline hlineで2重線にするために間隔をデフォの2に
              \begin{ruledtabular}            
                \begin{tabularx}{\textwidth}{cccccc}
                  % \hline \hline
                  & & & Mean photon number  & Heralding prob. (\%) & Fidelity\\
                  \hline
                  \addlinespace[0.4ex]
                  \multirow{6}{*}{
                    \parbox[c]{14ex}{\centering
                      \vspace{0.75ex}
                      (i)\\
                      $L_{\mathrm{EL}}\!=\!\qty{25}{km}$
                    }%
                  } & \multirow{2}{*}{(i-1)} & SM & 0.010 & 1.00 & \multirow{2}{*}{0.9804}\\
                  & & MM & 0.711 & 51.2 &\\
                  \addlinespace[0.5ex]
                  & \multirow{2}{*}{(i-2)} & SM & 0.054 & 5.18 & \multirow{2}{*}{0.9014}\\
                  & & MM & 3.83 & 97.8 &\\
                  \addlinespace[0.5ex]
                  & \multirow{2}{*}{(i-3)} & SM & 0.075 & 7.05 & \multirow{2}{*}{0.8672}\\
                  & & MM & 5.31 & 99.5 &\\
                  \hline
                  \addlinespace[0.5ex]
                  \multirow{6}{*}{
                    \parbox[c]{14ex}{\centering
                      \vspace{0.75ex}
                      (ii)\\
                      $L_{\mathrm{EL}}\!=\!\qty{50}{km}$
                    }%
                  } & \multirow{2}{*}{(ii-1)} & SM & 0.010 & 0.566 & \multirow{2}{*}{0.9761}\\
                  & & MM & 0.711 & 33.2 &\\
                  \addlinespace[0.5ex]
                  & \multirow{2}{*}{(ii-2)} & SM & 0.044 & 2.44 & \multirow{2}{*}{0.9010}\\
                  & & MM & 3.12 & 82.8 &\\
                  \addlinespace[0.5ex]
                  & \multirow{2}{*}{(ii-3)} & SM & 0.075 & 4.09 & \multirow{2}{*}{0.8397}\\
                  & & MM & 5.31 & 94.9 &\\
                  \hline
                  \addlinespace[0.5ex]
                  \multirow{6}{*}{
                    \parbox[c]{14ex}{\centering
                      \vspace{0.75ex}
                      (iii)\\
                      $L_{\mathrm{EL}}\!=\!\qty{100}{km}$
                    }%
                  } & \multirow{2}{*}{(iii-1)} & SM & 0.010 & 0.180 & \multirow{2}{*}{0.9723}\\
                  & & MM & 0.711 & 12.0 &\\
                  \addlinespace[0.5ex]
                  & \multirow{2}{*}{(iii-2)} & SM & 0.038 & 0.679 & \multirow{2}{*}{0.9003}\\
                  & & MM & 2.70 & 38.4 &\\
                  \addlinespace[0.5ex]
                  & \multirow{2}{*}{(iii-3)} & SM & 0.075 & 1.33 & \multirow{2}{*}{0.8159}\\
                  & & MM & 5.31 & 61.4 &
                  % \hline \hline
                \end{tabularx}
              \end{ruledtabular}
            \end{subtable}
            \hspace{0.02\hsize}
            \begin{subtable}[t]{0.275\hsize}
              \centering
              \setlength{\belowcaptionskip}{2mm}
              \caption{}
              \label{subtab:improveFactor_Fs61Fi83}
              % % \setlength{\doublerulesep}{0pt} %hline hlineで太線にするために間隔を0に
              % \setlength{\doublerulesep}{2pt}   %hline hlineで2重線にするために間隔をデフォの2に
              \begin{ruledtabular}            
                \begin{tabularx}{\textwidth}{ccc}
                  % \hline \hline
                  & $\mu_\indexrm{multi}/\mu_0$ &  $\prob_\indexrm{multi}/\prob_{\indexrm{single}, 0}$\\
                  \hline
                  \addlinespace[0.4ex]
                  (i-1) & 71.1 & 51.1 \\
                  \addlinespace[0.8ex]
                  (i-2) & 70.9 & 18.9 \\
                  \addlinespace[0.8ex]
                  (i-3) & 70.8 & 14.1 \\
                  \addlinespace[0.5ex]
                  \hline
                  \addlinespace[0.8ex]
                  (ii-1) & 71.1 & 58.7 \\
                  \addlinespace[0.8ex]
                  (ii-2) & 71.0 & 33.9 \\
                  \addlinespace[0.8ex]
                  (ii-3) & 70.8 & 23.2 \\
                  \addlinespace[0.5ex]
                  \hline
                  \addlinespace[0.8ex]
                  (iii-1) & 71.1 & 66.9 \\
                  \addlinespace[0.8ex]
                  (iii-2) & 71.0 & 56.5 \\
                  \addlinespace[0.8ex]
                  (iii-3) & 70.8 & 46.1 \\
                  % \hline \hline
                \end{tabularx}
              \end{ruledtabular}
            \end{subtable}
          \end{tabular}
        \end{table*}
      \endgroup
      % \clearpage  % ←フロートだけページの後に本文を再開させる
      % figure, tableのオプションをtからpに変えることで対応できた
    % }

    \subsubsection{\label{subsubsec:L25}$L_\indexrm{EL}=\qty{25}{km}$}
      For an elementary link distance of $L_\indexrm{EL} = \qty{25}{km}$ with $\mu_0 = 0.010$, Table~\ref{tab:data_Fs61Fi83}(\subref{subtab:meanProbFidelity_Fs61Fi83}) shows that even the reference mode---which exhibits the lowest fidelity---achieves $\fidelity_0 = 0.980439$, significantly exceeding the benchmark of $0.9$. While this high fidelity comes at the cost of a low single-mode heralding probability ($\prob_{\indexrm{single},0} \simeq \qty{1}{\%}$), the use of approximately 100 multiplexed modes boosts the overall heralding probability to $\prob_\indexrm{multi} \simeq \qty{51}{\%}$ while maintaining a high fidelity of $\fidelity_k \geq 0.9804$ across all modes. As shown in Table~\ref{tab:data_Fs61Fi83}(\subref{subtab:improveFactor_Fs61Fi83}), this represents an improvement of approximately 50-fold compared to the single-mode case.
      
      Next, for $\mu_0 = 0.054$, the reference fidelity becomes $\fidelity_0 = 0.9014$, which is comparable to but still above the $0.9$ threshold. In this regime, multiplexing significantly enhances the heralding probability from $\prob_{\indexrm{single},0} \simeq \qty{5}{\%}$ to a near-unity value of $\prob_\indexrm{multi} \simeq \qty{98}{\%}$, all while ensuring that the fidelity remains above $0.9$ for every mode.
      
      In the case of $\mu_0 = 0.075$, the heralding probability is further improved from $\prob_{\indexrm{single},0} \simeq \qty{7}{\%}$ to $\prob_\indexrm{multi} \simeq \qty{99}{\%}$, rendering entanglement generation nearly deterministic. However, this occurs at the expense of fidelity; for instance, the fidelity of the reference mode drops to $\fidelity_0 = 0.8672$, with several adjacent modes also falling below the $0.9$ threshold.
      
      In summary, for $L_\indexrm{EL} = \qty{25}{km}$, frequency multiplexing enables nearly deterministic entanglement generation even when the fidelity is maintained near $0.9$. Furthermore, even under stricter requirements where a high fidelity of over $0.98$ is necessary, the scheme still allows the success probability to be enhanced to approximately $\qty{50}{\%}$.

    \subsubsection{\label{subsubsec:L50}$L_\indexrm{EL}=\qty{50}{km}$}
      For an elementary link distance of $L_\indexrm{EL} = \qty{50}{km}$ with $\mu_0 = 0.010$, we find that $\fidelity_0 = 0.9761$. As with the $\qty{25}{km}$ case, this represents a high fidelity significantly exceeding $0.9$. In this scenario, multiplexing improves the heralding probability from $\prob_{\indexrm{single},0} \simeq \qty{0.6}{\%}$ to an overall value of $\prob_\indexrm{multi} \simeq \qty{33}{\%}$.
      
      Next, for $\mu_0 = 0.044$, the reference fidelity is $\fidelity_0 = 0.9010$, which is marginally above $0.9$. Here, frequency multiplexing enables all modes to maintain a fidelity above $0.9$ while substantially enhancing the heralding probability from $\prob_{\indexrm{single},0} \simeq \qty{2}{\%}$ to $\prob_\indexrm{multi} \simeq \qty{82}{\%}$.
      
      In the case of $\mu_0 = 0.075$, the heralding probability increases from $\prob_{\indexrm{single},0} \simeq \qty{4}{\%}$ to $\prob_\indexrm{multi} \simeq \qty{94}{\%}$. However, the fidelity suffers a notable decline; the reference fidelity drops to $\fidelity_0 = 0.8397$, and more than half of the available modes fall below the $0.9$ threshold.
      
      In summary, for $L_\indexrm{EL} = \qty{50}{km}$, frequency multiplexing allows for entanglement generation with a high success probability of approximately $\qty{80}{\%}$ while keeping the fidelity above $0.9$. Furthermore, even under the stricter requirement of maintaining a high fidelity above $0.97$, the scheme still achieves an improved heralding probability of approximately $\qty{30}{\%}$.

    \subsubsection{\label{subsubsec:L100}$L_\indexrm{EL}=\qty{100}{km}$}
      For an elementary link distance of $L_\indexrm{EL} = \qty{100}{km}$ with $\mu_0 = 0.010$, we obtain $\fidelity_0 = 0.9723$. Similar to the cases of $25$ and $\qty{50}{km}$, this indicates a high fidelity significantly above the $0.9$ benchmark. In this scenario, multiplexing improves the heralding probability from $\prob_{\indexrm{single},0} \simeq \qty{0.2}{\%}$ in the single-mode case to $\prob_\indexrm{multi} \simeq \qty{12}{\%}$. While this absolute value is lower than those for shorter distances, it represents an approximately 66-fold improvement---the most significant enhancement among the distances investigated. 
      
      Next, for $\mu_0 = 0.038$, the reference fidelity is $\fidelity_0 = 0.9003$, remaining just above the $0.9$ threshold. In this case, multiplexing ensures that all modes maintain a fidelity of at least $0.9$ while increasing the heralding probability from $\prob_{\indexrm{single},0} \simeq \qty{0.7}{\%}$ to $\prob_\indexrm{multi} \simeq \qty{38}{\%}$.
      
      For $\mu_0 = 0.075$, the heralding probability increases from $\prob_{\indexrm{single},0} \simeq \qty{1}{\%}$ to $\prob_\indexrm{multi} \simeq \qty{61}{\%}$. However, the fidelity suffers considerably; the reference fidelity drops to $\fidelity_0 = 0.8159$, and nearly all other modes also fall below the $0.9$ threshold.
      
      In summary, for $L_\indexrm{EL} = \qty{100}{km}$, frequency multiplexing enables entanglement generation with a practical success probability of approximately $\qty{30}{\%}$ while keeping the fidelity marginally above $0.9$. Furthermore, even when a high fidelity of over $0.97$ is required, the scheme enhances the heralding probability to approximately $\qty{10}{\%}$, achieving a 66-fold improvement compared to the single-mode case.

    \subsubsection{\label{subsubsec:ComparativeAnalysis}Comparative analysis}
      Comparing the cases of $L_\indexrm{EL}=25, 50$, and \qty{100}{km}, Figs.~\ref{fig:plot_compare}(\subref{subfig:plot_compare_prob_25})--(\subref{subfig:plot_compare_prob_100}) reveal that for a fixed reference mean photon number, the heralding probability of each mode decreases as the distance increases. Consequently, the overall heralding probability of the multiplexed system also decreases. This is attributed to the reduction in transmission efficiency over longer optical path lengths.
      
      The fidelity also exhibits a degradation as the distance increases. This can be explained by the fact that transmission loss makes heralding events from single-photon pairs less frequent; meanwhile, the relative contribution of heralding events originating from multi-photon pairs increases. As a result, the fraction of multi-photon states in the conditioned signal-photon state rises, leading to a lower fidelity.
      
      Furthermore, as a consequence of this distance-dependent fidelity degradation, the reference mean photon number required to sustain sufficient fidelity must be reduced for longer distances. Consequently, the multiplexed heralding probability under such fidelity-constrained conditions decreases significantly as the distance increases.
      
      Additionally, as shown in Table~\ref{tab:data_Fs61Fi83}(\subref{subtab:improveFactor_Fs61Fi83}), the ratio of the multiplexed heralding probability to the single-mode probability increases with distance, indicating that the advantages of multiplexing become more pronounced at longer ranges.
      
      Regarding the mean photon number, the multiplexed value is consistently approximately 70 times larger than that of the single-mode case across all investigated conditions. Despite this increase in the total photon number, our results demonstrate that frequency multiplexing combined with spectral demultiplexing allows for entanglement generation at practical rates while maintaining sufficient fidelity.

  \subsection{\label{subsec:calcProbFidelityLowFinesse}Calculation (low finesse)}    
    We consider an example in which the finesse is lowered while maintaining the free spectral range (FSR). Specifically, we examine the case where $\finesse_\indSignal = \finesse_\indIdler = 30.0$. This value is selected as a representative case of lower finesse under conditions where the approximation in Eq.~\eqref{eq:approx_Airy} remains valid, assuming an FSR of approximately \qty{121}{MHz} and a full width at half maximum (FWHM) of approximately \qty{4}{MHz}, which is consistent with the bandwidth of the AFC in Pr:YSO used as the quantum memory.
    
    The signal spectrum for this finesse is shown in Fig.~\ref{fig:spectrum_signal_finesse30}. Comparing Fig.~\ref{fig:spectrum_signal_finesse30}(\subref{subfig:spectrum_signal_mode0_finesse30_all}) with Fig.~\ref{fig:spectrum_signal}(\subref{subfig:spectrum_signal_mode0_all}), it is evident that the cluster width increases and the reduction in spectral height near the center modes becomes more gradual.

    % \afterpage{%
      % \clearpage % ←「次ページ頭」でフロートを吐き出す（ここがポイント）
      \begingroup
        % このブロック内だけ効かせる
        \setcounter{dbltopnumber}{7}            % 上部に置ける全幅フロート数
        \setcounter{totalnumber}{9}             % 1ページの総フロート数
        \renewcommand{\dbltopfraction}{1.0}    % 上部を占められる最大割合
        \renewcommand{\dblfloatpagefraction}{0.00} % 全幅フロートだけで構成される「フロート専用ページ」を作ってよいと判定するための最低充足率。
        \renewcommand{\textfraction}{0.00}      % 本文の最小割合を緩める
        \setlength{\dblfloatsep}{4pt plus 0pt minus 0pt}     % 全幅フロート同士   plus, minusはそれぞれ最大いくつずれていいかを表す
        \setlength{\dbltextfloatsep}{12pt plus 2pt minus 2pt} % 全幅フロートと本文

        \begin{figure*}[t]
          \begin{tabular}{c}
            \begin{subfigure}[t]{0.5075\hsize}
              \centering
              \caption{}
              \vspace{-2ex}
              \includegraphics[width=\columnwidth]{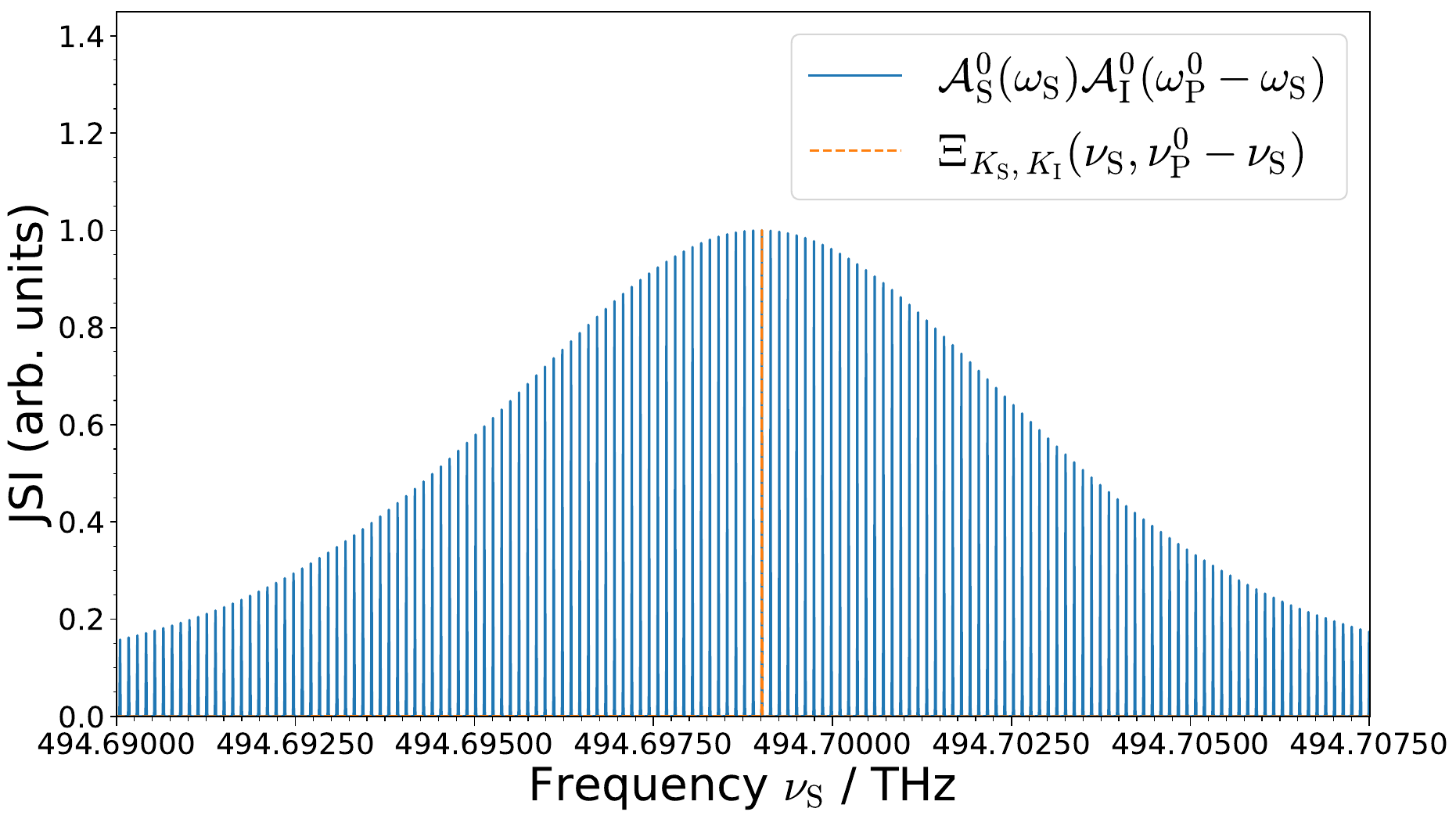}
              \label{subfig:spectrum_signal_mode0_finesse30_all}
            \end{subfigure}
            \hspace{-0.015\hsize}
            \begin{subfigure}[t]{0.5075\hsize}
              \centering
              \caption{}
              \vspace{-2ex}
              \includegraphics[width=\columnwidth]{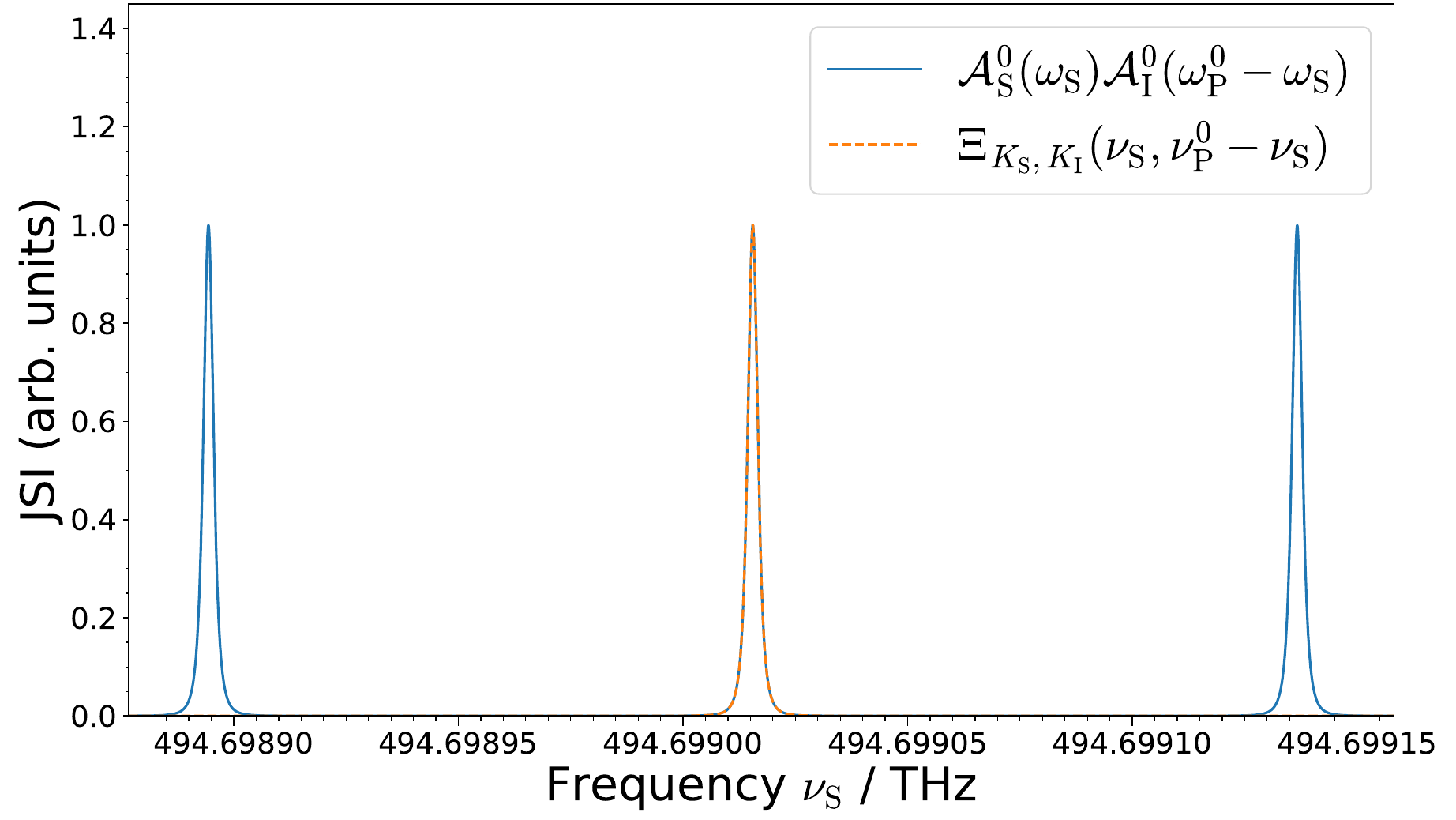}
              \label{subfig:spectrum_signal_mode0_finesse30_center}
            \end{subfigure}
          \end{tabular}
          \vspace{-5.5ex}
          \caption{
            Signal spectrum. The plots are shown for $\finesse_\indSignal = \num{61.0}$ and $\finesse_\indIdler = \num{83.0}$. The solid blue line and the dotted orange line represent $\Airy[^0]{\indSignal}{\omega_\indSignal}\Airy[^0]{\indIdler}{\omega_\indPump^0-\omega_\indSignal}$ and $\Xi_{K_\indSignal, K_\indIdler}(\nu_\indSignal,\nu_\indPump^0-\nu_\indSignal)$, respectively. Panel (\subref{subfig:spectrum_signal_mode0_finesse30_all}) shows the main cluster, while (\subref{subfig:spectrum_signal_mode0_finesse30_center}) provides a magnified view of the three modes around the central peak.
          }
          \label{fig:spectrum_signal_finesse30}
          \vspace{-2ex}
        \end{figure*}
        \begin{figure*}[t]
          \centering
          \includegraphics[width=0.8\linewidth]{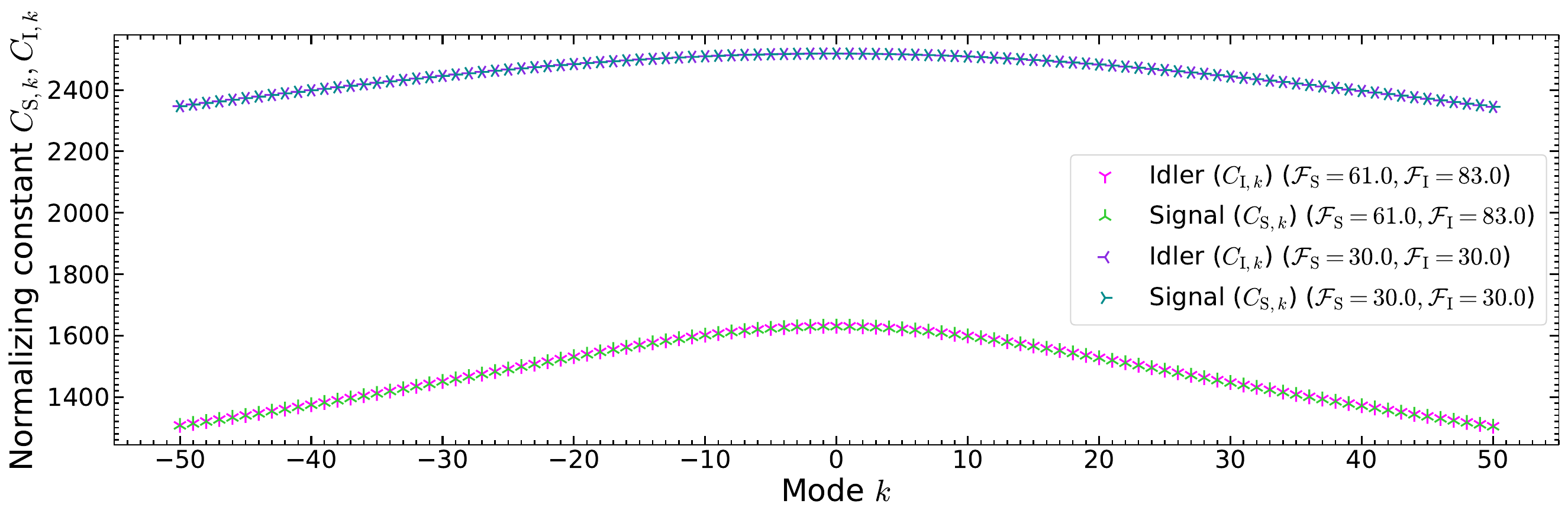}
          \vspace{-3.5ex}
          \caption{
            Calculated normalization constants obtained using Eq.~\eqref{eq:normalizationConstant}. The values are plotted for two cases: $(\finesse_\indSignal, \finesse_\indIdler) = (61.0, 83.0)$ and $(30.0, 30.0)$.
          }
          \label{fig:plot_normConst_finesse30and6183}
          \vspace{-3ex}
        \end{figure*}

        \begin{figure*}[t]
          \begin{tabular}{c}
            \begin{subfigure}[t]{0.333\hsize}
              \centering
              \caption{}
              \vspace{-3.5ex}
              \includegraphics[height=0.975\columnwidth]{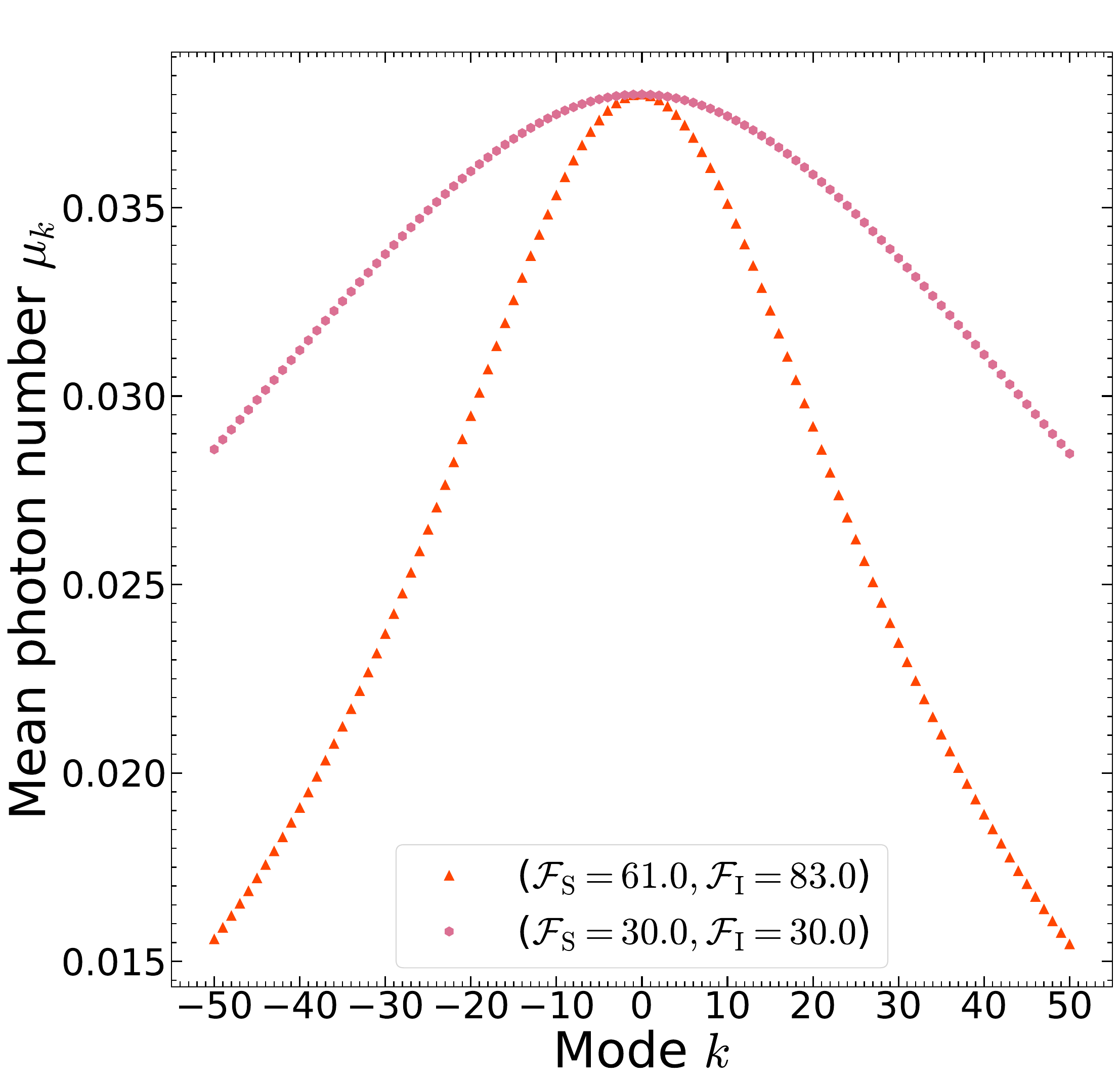}
              \label{subfig:plot_finesse30and6183_0_038_mean}
            \end{subfigure}
            \hspace{-0.01\hsize}
            \begin{subfigure}[t]{0.333\hsize}
              \centering
              \caption{}
              \vspace{-3.5ex}
              \includegraphics[height=0.975\columnwidth]{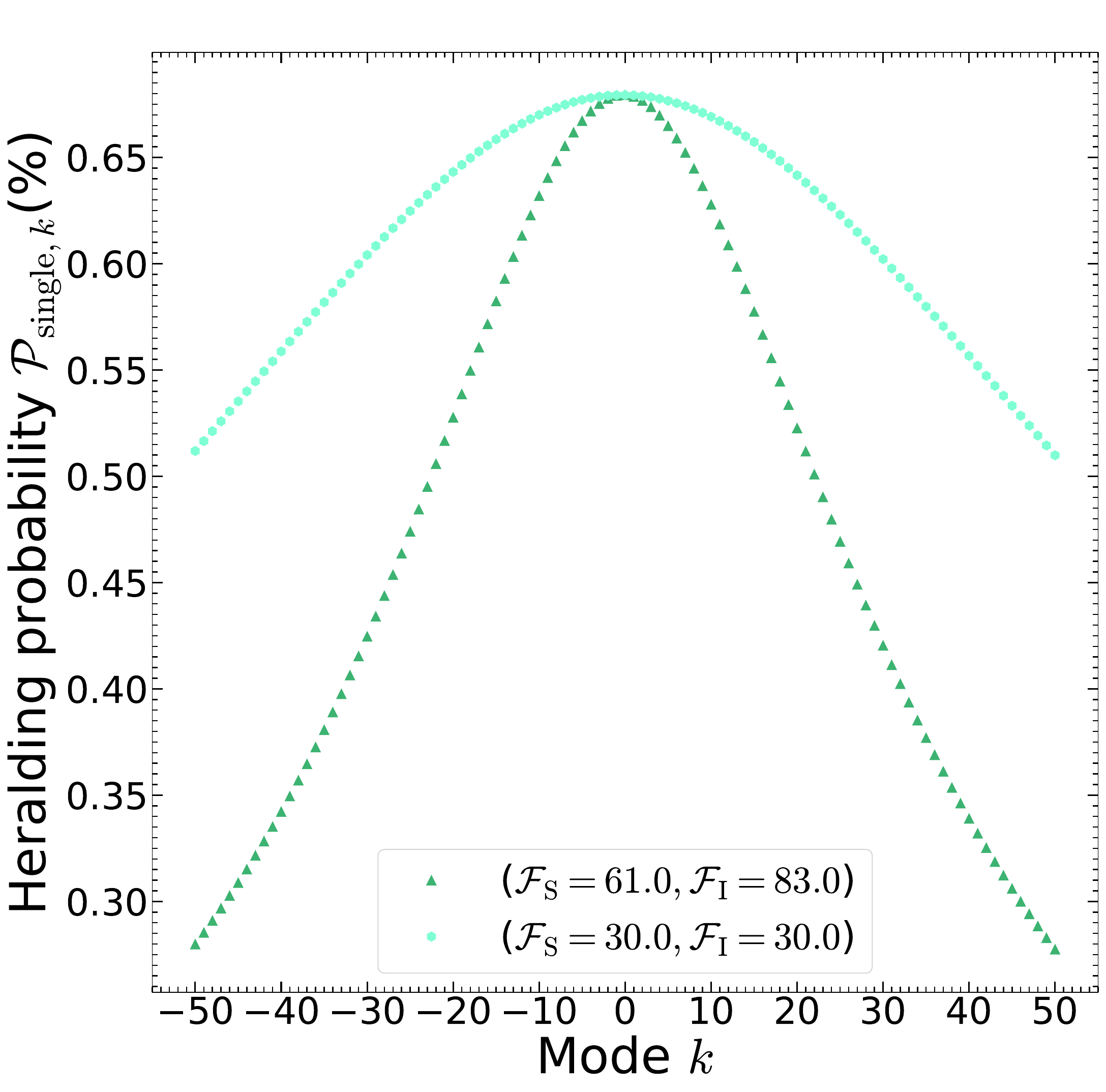} % heightをcolomnwidthの9/16倍
              \label{subfig:plot_finesse30and6183_0_038_prob}
            \end{subfigure}
            \hspace{-0.01\hsize}
            \begin{subfigure}[t]{0.333\hsize}
              \centering
              \caption{}
              \vspace{-3.5ex}
              \includegraphics[height=0.975\columnwidth]{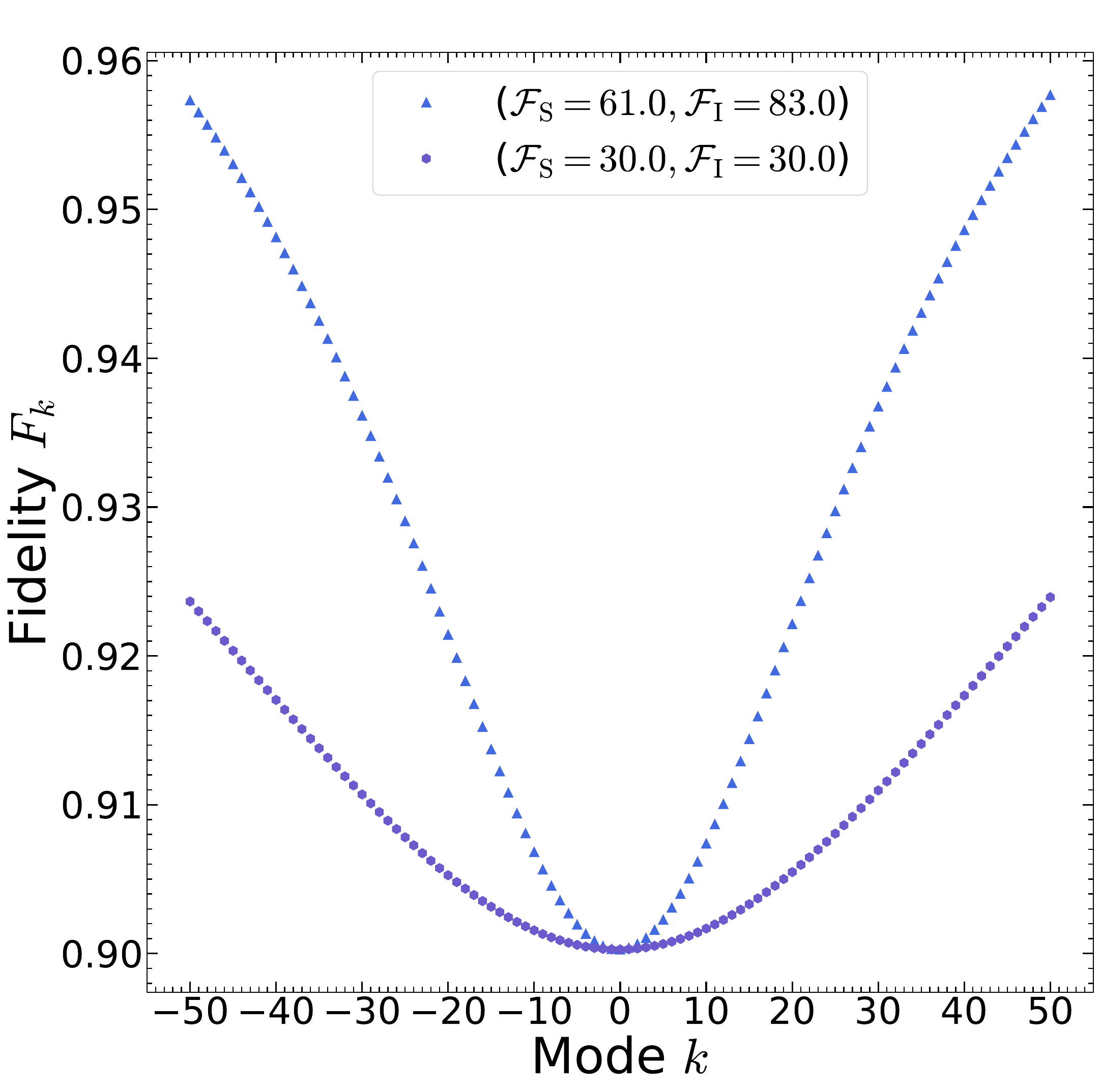}
              \label{subfig:plot_finesse30and6183_0_038_fidelity}
            \end{subfigure}
          \end{tabular}
          \vspace{-5.25ex}
          \caption{
             Comparison of numerical results for mean photon number $\mu_k$, heralding probability $\prob_{\indexrm{single}, k}$, and fidelity $\fidelity_k$, similar to Fig.~\ref{fig:plot_compare}. The results are shown for $L_\indexrm{EL}=\qty{100}{km}$ and $\mu_0=0.038$ in both the high-finesse (HF) case $(\finesse_\indSignal=61.0, \finesse_\indIdler=83.0)$ and the low-finesse (LF) case $(\finesse_\indSignal=30.0, \finesse_\indIdler=30.0)$.
          }
          \label{fig:plot_compare_Fs61Fi83andF30}
          \vspace{-2ex}
        \end{figure*}
      \endgroup
      % \clearpage  % ←フロートだけページの後に本文を再開させる
      % figure, tableのオプションをtからpに変えることで対応できた
    % }    
    
    The calculated normalization constants $C_{\indSignal,k}$ and $C_{\indIdler,k}$ for this finesse are presented in Fig.~\ref{fig:plot_normConst_finesse30and6183}, showing a more moderate variation compared to the case with $\finesse_\indSignal=61.0$ and $\finesse_\indIdler=83.0$ (Fig.~\ref{fig:plot_normConst}).

    To evaluate the impact of this lower finesse on the heralding probability and fidelity in a multiplexed system, we performed specific calculations for $L_\indexrm{EL}=\qty{100}{km}$ and $\mu_0=0.038$. These results are compared with those of the previously discussed $\finesse_\indSignal=61.0, \finesse_\indIdler=83.0$ case. All other conditions are identical to those described in Sec.~\ref{subsec:calcProbFidelity}.

    % \afterpage{%
      % \clearpage % ←「次ページ頭」でフロートを吐き出す（ここがポイント）
      \begingroup
        % このブロック内だけ効かせる
        \setcounter{dbltopnumber}{5}            % 上部に置ける全幅フロート数
        \setcounter{totalnumber}{9}             % 1ページの総フロート数
        \renewcommand{\dbltopfraction}{1.0}    % 上部を占められる最大割合
        \renewcommand{\dblfloatpagefraction}{0.00} % 全幅フロートだけで構成される「フロート専用ページ」を作ってよいと判定するための最低充足率。
        \renewcommand{\textfraction}{0.00}      % 本文の最小割合を緩める
        \setlength{\dblfloatsep}{4pt plus 0pt minus 0pt}     % 全幅フロート同士   plus, minusはそれぞれ最大いくつずれていいかを表す
        \setlength{\dbltextfloatsep}{12pt plus 2pt minus 2pt} % 全幅フロートと本文

        \begin{table*}[t]
          \caption{
            Comparison between single-mode (SM) and multimode (MM) cases for the high-finesse (HF) $(\finesse_\indSignal=61.0, \finesse_\indIdler=83.0)$ and low-finesse (LF) $(\finesse_\indSignal=30.0, \finesse_\indIdler=30.0)$ settings at $L_\indexrm{EL}=\qty{100}{km}$ and $\mu_0=0.038$. (\subref{subtab:meanProbFidelity_Fs61Fi83andF30}) Values of the mean photon number and heralding probability for both SM ($\mu_0, \prob_{\indexrm{single, 0}}$) and MM ($\mu_\indexrm{multi}, \prob_\indexrm{multi}$) cases. The fidelity is represented by $\fidelity_0$, which corresponds to the minimum value among all modes. (\subref{subtab:improveFactor_Fs61Fi83andF30}) Improvement ratios for the mean photon number and heralding probability, defined as $\mu_\indexrm{multi}/\mu_0$ and $\prob_\indexrm{multi}/\prob_{\indexrm{single}, 0}$, respectively.
          }
          \label{tab:data_Fs61Fi83andF30}
          \vspace{-1.95ex}
          \begin{tabular}{c}
            \begin{subtable}[t]{0.515\hsize}
              \centering
              \setlength{\belowcaptionskip}{0mm}
              \caption{}
              \label{subtab:meanProbFidelity_Fs61Fi83andF30}
              % % \setlength{\doublerulesep}{0pt} %hline hlineで太線にするために間隔を0に
              % \setlength{\doublerulesep}{2pt}   %hline hlineで2重線にするために間隔をデフォの2に
              % \vspace{-1.5ex}
              \begin{ruledtabular}            
                \begin{tabularx}{\textwidth}{cccc}
                  % \hline \hline
                  & Mean photon number  & Heralding prob. (\%) & Fidelity\\
                  \hline
                  \addlinespace[0.4ex]
                  SM & 0.038 & 0.679 & \multirow{3}{*}{0.9003}\\
                  MM (HF) & 2.70 & 38.4 &\\
                  MM (LF) & 3.46 & 46.3 &\\
                  % \hline \hline
                \end{tabularx}
              \end{ruledtabular}
            \end{subtable}
            \hspace{0.02\hsize}
            \begin{subtable}[t]{0.275\hsize}
              \centering
              \setlength{\belowcaptionskip}{0mm}
              \caption{}
              \label{subtab:improveFactor_Fs61Fi83andF30}
              % % \setlength{\doublerulesep}{0pt} %hline hlineで太線にするために間隔を0に
              % \setlength{\doublerulesep}{2pt}   %hline hlineで2重線にするために間隔をデフォの2に
              % \vspace{-1.5ex}
              \begin{ruledtabular}            
                \begin{tabularx}{\textwidth}{ccc}
                  % \hline \hline
                  & $\mu_\indexrm{multi}/\mu_0$ &  $\prob_\indexrm{multi}/\prob_{\indexrm{single}, 0}$\\
                  \hline
                  \addlinespace[0.4ex]
                  (HF) & 71.0 & 56.5 \\
                  (LF) & 91.1 & 68.1 \\
                  % \hline \hline
                \end{tabularx}
              \end{ruledtabular}
            \end{subtable}
          \end{tabular}
        \end{table*}
      \endgroup
      % \clearpage  % ←フロートだけページの後に本文を再開させる
      % figure, tableのオプションをtからpに変えることで対応できた
    % }
    
    The resulting mean photon number, heralding probability, and fidelity for each mode are shown in Fig.~\ref{fig:plot_compare_Fs61Fi83andF30}. Both the mean photon number and the heralding probability for each mode are generally higher than in the previous higher-finesse case. Regarding the fidelity, since we only need to consider the minimum value, the fidelity of the center mode $\fidelity_0$ remains unchanged for both cases. Under these conditions, the overall mean photon number and heralding probability are given in Table~\ref{tab:data_Fs61Fi83andF30}(\subref{subtab:meanProbFidelity_Fs61Fi83andF30}). Notably, the heralding probability $\prob_\indexrm{multi}$ shows an improvement of approximately \qty{10}{\%} compared to the case with $\finesse_\indSignal=61.0$ and $\finesse_\indIdler=83.0$.
    
    These results suggest that if there is sufficient room to enhance brightness by means other than increasing the finesse (for instance, by increasing the pump power), it is effective to avoid excessively high finesse. This holds as long as the approximation in Eq.~\eqref{eq:approx_Airy} remains valid---that is, within the range where the frequency modes are sufficiently resolved.

\section{\label{sec:conclusion}Conclusion and discussion}
  In this study, we performed an analysis of doubly resonant cavity-enhanced SPDC (cSPDC) under continuous-wave (CW) pumping, with the goal of representing the quantum states of signal and idler photons generated near the main cluster while incorporating the effects of multi-photon pair generation. 
  
  By focusing on the joint spectral intensity (JSI) and joint spectral amplitude (JSA) and performing approximate transformations, we derived a discrete state representation as shown in Eq.~\eqref{eq:cSPDC_state}, where each frequency mode is independently represented as a two-mode squeezed vacuum (TMSV) state.
  
  Based on this representation and using the frequency and cavity parameters employed in the practical operation of cSPDC as a photon-pair source, we evaluated the performance of a frequency-multiplexed system. Specifically, we calculated the mean photon number, the heralding probability---which is proportional to the entanglement generation rate between nodes---and the entanglement fidelity for each mode. 
  
  Our results demonstrate that even when the overall mean photon number is approximately 70 times larger than that of the single-mode case, these photons are distributed across multiple frequency modes, which effectively suppresses the degradation of fidelity in each individual mode. Furthermore, we confirmed that by utilizing approximately 100 multiplexed modes, the heralding probability can be significantly improved while maintaining a fidelity above $0.9$: for an elementary link distance of $L_\indexrm{EL} = \qty{25}{km}$, the success probability reaches a near-unity value, and even at $L_\indexrm{EL} = \qty{100}{km}$, where high transmission losses typically make entanglement generation difficult, the heralding probability is improved to approximately $\qty{40}{\%}$. In particular, while practical entanglement generation is difficult in the single-mode case for $L_\indexrm{EL} = \qty{100}{km}$, it is significant that we have specifically shown that this challenge can be overcome through multiplexing.

  In summary, this study has quantitatively demonstrated that frequency multiplexing of cSPDC-based photon-pair sources can significantly improve the entanglement generation rate. Moreover, we have established a framework to quantitatively assess the generation rate and fidelity with respect to major practical parameters, such as cavity design, optical path loss, and detector efficiency.

  On the other hand, the current calculations do not account for imperfections in the storage and retrieval processes of quantum memories. Future evaluations under more realistic conditions will need to incorporate these factors.

  Additionally, to include the effects of multi-photon pair generation, the current analysis does not employ a rigorous representation of the fine frequency structure within each mode, as illustrated in Fig.~\ref{fig:plot_JSI}. While the present approach is sufficient for a general assessment of the improvement in heralding probability through multiplexing, a more rigorous treatment of the spectral structure may be necessary when highly precise numerical values are required.

\begin{acknowledgments}
  We are grateful to Rikizo Ikuta at the University of Osaka for providing insightful comments that motivated this work.
  We would also like to thank Masahide Sasaki, Mikio Fujiwara, and Yoshiaki Tsujimoto at NICT, 
  Kazufumi Tanji at Keio University,
  and Akira Ozawa, Tomoki Tsuno, and Daisuke Yoshida at Yokohama National University for helpful discussions.

  This work was supported by 
  JSPS KAKENHI Grant Number JP20H02652, 
  Ministry of Internal Affairs and Communications R\&D of ICT Priority Technology Project (JPMI00316), 
  and JST Moonshot R\&D (JPMJMS226C).
\end{acknowledgments}

\section*{\label{sec:dataAvailability}Data Availability}
  The data that support the findings of this article are openly available \cite{KomatsudairaHorikiri2026_Data}.

\appendix

\section{Cavity configuration and Airy function}
  This section follows the treatment in Ref.~\cite[Chap. 4]{YarivYeh2007}.
  \subsection{Airy function}
    We consider a bow-tie cavity, as illustrated in Fig.~\ref{fig:Bow-tie_cavity}, and the generation of spontaneous parametric down-conversion (SPDC) light within it.
    
    For each mirror $j\ (j\in\{1,2,3,4\})$, let $r_j$ and $t_j$ denote the reflection and transmission coefficients for light incident from outside the cavity, and $r_j^\prime$ and $t_j^\prime$ denote those for light incident from inside the cavity. Furthermore, let $c$ be the speed of light in vacuum, $L_\indexrm{opt}$ be the optical path length for a single round trip in the cavity, and $L_\indexrm{init}$ be the optical path length from the crystal exit to mirror 4. We define $G$ as the intracavity loss coefficient, where $G=0$ corresponds to complete extinction and $G=1$ to a lossless condition. Note that $G$ is sometimes expressed as $G=e^{-a L_\indexrm{c}}$ in terms of the absorption coefficient $a$ and the length of the medium $L_\indexrm{c}$ \cite{LuoSilberhorn2015, Prakash2021phd}.

    We consider light with an angular frequency $\omega$. Here, we assume that the reflection and transmission coefficients, as well as the optical path lengths, exhibit negligible variation near the center frequency of the light.
    The light within the cavity is divided into an infinite number of partial waves through successive reflections and round trips. The phase difference between two partial waves with an optical path difference of one round trip (i.e., the phase shift accumulated during a single round trip), denoted as $\delta_\indexrm{loop}(\omega)$, is expressed as
    \begin{align}
      \delta_\indexrm{loop}(\omega) = \omega \frac{L_\indexrm{opt}}{c}.
    \end{align}

    Let $E_{\text{int}}$ and $E_{\text{out}}$ represent the internal electric field just after exiting the crystal and the output electric field immediately after transmitting through mirror 4, respectively. Then, the following relationship holds:
    \begin{widetext}
      \begin{align}
        E_\mathrm{out}(\omega) &=  r_2^\prime r_3^\prime t_4^\prime \sqrt{G} e^{-i\delta_\mathrm{loop}(\omega) \frac{L_\mathrm{init}}{L_\mathrm{opt}}}\left[
          1+r_1^\prime r_2^\prime r_3^\prime r_4^\prime \sqrt{G} e^{-i\delta_\mathrm{loop}(\omega)} + \left(r_1^\prime r_2^\prime r_3^\prime r_4^\prime \sqrt{G} e^{-i\delta_\mathrm{loop}(\omega)}\right)^2 + \cdots
        \right] E_\mathrm{int}(\omega)\notag\\
        &= r_2^\prime r_3^\prime t_4^\prime \sqrt{G} e^{-i\delta_\mathrm{loop}(\omega) \frac{L_\mathrm{init}}{L_\mathrm{opt}}} \frac{1}{1-r_1^\prime r_2^\prime r_3^\prime r_4^\prime \sqrt{G} e^{-i\delta_\mathrm{loop}(\omega)}} E_\mathrm{int}(\omega).
        \label{eq:electricFieldTrans_cSPDC}
      \end{align}
    \end{widetext}

    Also, we let
    \begin{align}
      A(\omega) \coloneq \frac{E_\mathrm{out}(\omega)}{E_\mathrm{int}(\omega)}
      =
      \frac{r_2^\prime r_3^\prime t_4^\prime \sqrt{G} e^{-i\delta_\mathrm{loop}(\omega) \frac{L_\mathrm{init}}{L_\mathrm{opt}}}}{1-r_1^\prime r_2^\prime r_3^\prime r_4^\prime \sqrt{G} e^{-i\delta_\mathrm{loop}(\omega)}} .
    \end{align}

    Here, when the phase shifts introduced by transmission and reflection are negligible compared to the round-trip phase shift $\delta_\indexrm{loop}(\omega)$, the reflection and transmission coefficients can be expressed as $r_j \simeq \sqrt{R_j}$ and $t_j\simeq \sqrt{T_j}$ using the reflectivity $R_j\coloneq\absolute{r_j}^2$ and transmissivity $T_j\coloneq \absolute{t_j}^2$. By applying these expressions, the relations for the dielectric interface of lossless mirrors ($r_j=-r_j^\prime$ and $t_j = t_j^\prime$), and the conservation of energy $r_j r_j^\ast+t_j t_j^{\prime\ast}=1$, $A(\omega)$ can be expressed as
    \begin{align}
      A(\omega) \coloneq
      \frac{ \sqrt{R_2 R_3 (1-R_4) G} e^{-i\delta_\mathrm{loop}(\omega) \frac{L_\mathrm{init}}{L_\mathrm{opt}}}}{1-\sqrt{R_1 R_2 R_3 R_4 G} e^{-i\delta_\mathrm{loop}(\omega)}}. 
    \end{align}

    Also, the ratio of the electric field intensities is given by
    \begin{align}
      \Airy{{}}{\omega} &\coloneq \absolute{A(\omega)}^2\notag\\*
      &= \frac{R_2 R_3 (1-R_4) G}{(1\!-\!\sqrt{R_1 R_2 R_3 R_4 G})^2 \!+\! 4 \sqrt{R_1 R_2 R_3 R_4 G} \sin^2 \frac{\delta_\indexrm{loop}(\omega)}{2}}\notag\\*
      &= \frac{R_2 R_3 (1-R_4) G}{(1-g)^2 + 4 g \sin^2 \frac{\delta_\indexrm{loop}(\omega)}{2}}.
      \label{eq:Airy_bow-tie}
    \end{align}
    This is called the Airy function, where we have defined $g\coloneq \sqrt{R_1 R_2 R_3 R_4 G}$.

  \subsection{FSR, FWHM, and finesse}
    From Eq.~\eqref{eq:Airy_bow-tie}, the transmittance reaches its maximum when the phase shift satisfies $\delta_\indexrm{loop} = 2\pi m$ for $m \in \mathbb{Z}$. Denoting the corresponding frequency (resonance frequency) as $\nu_m$, we have
    \begin{align}
      \nu_m = \frac{c}{L_\indexrm{opt}} m.
    \end{align}
    The interval between these resonance frequencies is called the free spectral range (FSR), given by
    \begin{align}
      \fsr\coloneq \nu_{m+1}-\nu_m = \frac{c}{L_\indexrm{opt}}.
    \end{align}
    
    Moreover, the interval between the two frequencies $\nu_m^{(+)}$ and $\nu_m^{(-)}$ at each peak $\nu_m$ where the transmittance drops to half of its maximum value is called the full width at half maximum (FWHM).
    
    For these frequencies, Eq.~\eqref{eq:Airy_bow-tie} leads to
    \begin{align}
      \frac{R_2 R_3 (1\!-\!R_4) G}{\left(1\!-\!g\right)^2 \!+\! 4g\sin^2\frac{\delta_\mathrm{loop}(\omega_m^{(+)})}{2}}
      &\!=\! \frac{1}{2} \!\times\! \frac{R_2 R_3 (1\!-\!R_4) G}{\left(1\!-\!g\right)^2 \!+\! 4g\sin^2\frac{\delta_\mathrm{loop}(\omega_m^{\vphantom{(+)}})}{2}}.
    \end{align}
    Rearranging this equation, we obtain
    \begin{align}
      \sin \left(\frac{\delta_\indexrm{loop}(\omega_m^{(+)}) - \delta_\indexrm{loop}\left(\omega_m\right)}{2}\right)
      &= \frac{\left(1-g\right)}{2\sqrt{g}}.
    \end{align}
    Since $\delta_\indexrm{loop}(\omega_m^{(+)}) - \delta_\indexrm{loop} \ll 1$, it follows that
    \begin{align}
      \nu_m^{(+)} - \nu_m
      &\simeq \frac{c}{\pi L_\mathrm{opt}}\frac{\left(1-g\right)}{2\sqrt{g}}.
    \end{align}
    Thus, the FWHM $\fwhm$ is
    \begin{align}
      \fwhm &\coloneq \nu_m^{(+)} - \nu_m^{(-)} = 2 (\nu_m^{(+)} - \nu_m)\notag\\
      &\simeq \frac{1-g}{\pi\sqrt{g}} \fsr.
    \end{align}

    Given this proportional relationship between FSR and FWHM, we define the finesse $\finesse$ as
    \begin{align}
      \finesse \coloneq \frac{\fsr}{\fwhm} \simeq \frac{\pi\sqrt{g}}{1-g}.
    \end{align}

  \subsection{Enhancement factor}
    Rearranging the Airy function in Eq.~\eqref{eq:Airy_bow-tie} using the FSR and finesse, we obtain
    \begin{align}
      \Airy{{}}{\omega} &= \frac{\frac{R_2 R_3 (1-R_4) G}{(1-g)^2}}{1+\frac{4g}{(1-g)^2}\sin^2\frac{\delta_\indexrm{loop}(\omega)}{2}}\notag\\
      &= T_\indexrm{enh} \frac{1}{1+\left(\frac{2\finesse}{\pi} \right)^2\sin^2\left(\frac{\pi}{\fsr}\nu\right)},
    \end{align}
    where
    \begin{align}
      T_\indexrm{enh} \coloneq \frac{R_2 R_3 (1-R_4) G}{(1-g)^2}
    \end{align}
    is the maximum value of the original Airy function and is referred to as the enhancement factor, which represents the enhancement relative to the generated light in the absence of a cavity. When $R_2 R_3 G \simeq 1$, we have $g = \sqrt{R_2 R_3 R_4 G} \simeq \sqrt{R_4} \eqcolon g_0$. Using this value, $T_\indexrm{enh}$ is expressed as
    \begin{align}
      T_\mathrm{enh} &= \frac{R_2R_3(1-R_4)G}{(1-g)^2}\notag\\
      &=\frac{1-g_0^2}{(1-g)^2}.
    \end{align}

    Furthermore, for a large finesse $(\finesse \gtrsim 10)$~\cite[p.79]{Andreas2019phd}, we have
    \begin{align}
      g &= \frac{(2\finesse^2+\pi^2)-2\finesse\pi\left(1+\frac{\pi^2}{8\finesse^2}\right)}{2\finesse^2} \notag\\
      &\simeq \frac{2\finesse^2-2\finesse\pi}{2\finesse^2} =1-\frac{\pi}{\finesse}.
    \end{align}
    Thus, when the finesse $\finesse$ and the finesse $\finesse_0$ corresponding to $g_0$ are comparable $(\finesse \simeq \finesse_0)$ and both are large $(\finesse, \finesse_0 \gtrsim 10)$, $T_\indexrm{enh}$ can be approximated as
    \begin{align}
      T_\indexrm{enh} \simeq \frac{2}{\pi}\frac{\finesse^2}{\finesse_0} - \frac{\finesse^2}{\finesse_0^2} \simeq \frac{2}{\pi}\frac{\finesse^2}{\finesse_0}. 
    \end{align}

    While the spectrum of standard SPDC is given by $\absolute{\alpha}^2 S(\omega_\indSignal, \omega_\indIdler)$, that of doubly resonant cSPDC is expressed as $T_{\mathrm{enh},\indSignal} T_{\mathrm{enh},\indIdler} \absolute{\alpha}^2 \mathcal{A}_\indSignal^0(\omega_\indSignal) \mathcal{A}_\indIdler^0(\omega_\indIdler) S(\omega_\indSignal,\omega_\indIdler)$. Consequently, the intensity of cSPDC is enhanced by a factor of
    \begin{align}
      T_{\mathrm{enh},\indSignal} T_{\mathrm{enh},\indIdler} = \frac{4}{\pi^2}\frac{\finesse_\indSignal^2\finesse_\indIdler^2}{\finesse_{0,\indSignal}\finesse_{0,\indIdler}},
    \end{align}
    indicating that the enhancement in cSPDC is proportional to the square of the finesse relative to standard SPDC.

\section{Proof that the approximated JSA yields the approximated JSI \label{sec:proof_approx}}
  We show that the approximated representation of the JSA in Eq.~\eqref{eq:JSA_cav_approx}, given by
  \begin{widetext}
    \begin{align}
      \begin{split}
          f_\indexrm{cav}^\indexrm{(approx)} (\omega_\indSignal,\omega_\indIdler)
          &\coloneq  
          \sum_{k=-M}^{M} 
          \left(
            \frac{1}{1+i\frac{2}{\fwhm_\indSignal} \left(\rule{0pt}{2ex}\nu_\indSignal -(K_\indSignal + k) \fsr_\indSignal\right)}
            \frac{1}{1+i\frac{2}{\fwhm_\indIdler} \left(\rule{0pt}{2ex}\nu_\indPump^0 - \nu_\indSignal -(K_\indIdler - k) \fsr_\indIdler\right)}
          \right)^{\! 1/2}\\
          &\hspace{10ex} \times
          \left(
            \frac{1}{1+i\frac{2}{\fwhm_\indSignal} \left(\rule{0pt}{2ex}\nu_\indPump^0 - \nu_\indIdler -(K_\indSignal + k) \fsr_\indSignal\right)}
            \frac{1}{1+i\frac{2}{\fwhm_\indIdler} \left(\rule{0pt}{2ex}\nu_\indIdler -(K_\indIdler - k) \fsr_\indIdler\right)}
          \right)^{\! 1/2}
        \end{split}
      \label{eq:JSA_cav_approx_rep}
    \end{align}
  % \end{widetext}
  results in the form of the JSI in Eq.~\eqref{eq:JSI_cav_approx}.

  First, to simplify the calculation, we define the functions $u$ and $v$ as
  \begin{subequations}
    \begin{align}
      u_k(\nu_\indSignal)&\coloneq 
        \frac{1}{1+i\frac{2}{\fwhm_\indSignal} \left(\rule{0pt}{2ex}\nu_\indSignal -(K_\indSignal + k) \fsr_\indSignal\right)},
    \end{align}
    \begin{align}
      v_k(\nu_\indIdler)&\coloneq 
        \frac{1}{1+i\frac{2}{\fwhm_\indIdler} \left(\rule{0pt}{2ex}\nu_\indIdler -(K_\indIdler - k) \fsr_\indIdler\right)}.
    \end{align}
  \end{subequations}
  Then, we have
  % \begin{widetext}
    \begin{align}
      f_\indexrm{cav}^\indexrm{(approx)} (\omega_\indSignal,\omega_\indIdler)
      &= \sum_{k=-M}^{M}
      \left(\rule{0pt}{2.25ex}
        u_k(\nu_\indSignal)v_k(\nu_\indPump^0-\nu_\indSignal)
      \right)^{1/2}
      \left(\rule{0pt}{2.25ex}
        u_k(\nu_\indPump^0-\nu_\indIdler)v_k(\nu_\indIdler)
      \right)^{1/2}.
    \end{align}
  % \end{widetext}
  
  Here, we define the real-valued functions $w_{\indSignal,k}(\nu_\indSignal), w_{\indIdler,k}(\nu_\indIdler), \vartheta_{\indSignal,k}(\nu_\indSignal)$, and $\vartheta_{\indIdler,k}(\nu_\indIdler)$ as
  \begin{subequations}
    \begin{align}
      w_{\indSignal,k}(\nu_\indSignal)&\coloneq \absolute{u_k(\nu_\indSignal)v_k(\nu_\indPump^0-\nu_\indSignal)},
    \end{align}
    \begin{align}
      w_{\indIdler,k}(\nu_\indIdler) &\coloneq \absolute{u_k(\nu_\indPump^0-\nu_\indIdler)v_k(\nu_\indIdler)},
    \end{align}
  \end{subequations}
  and
  \begin{subequations}
    \begin{align}
      \vartheta_{\indSignal,k}(\nu_\indSignal)&\coloneq \Arg\left[u_k(\nu_\indSignal)v_k(\nu_\indPump^0-\nu_\indSignal)\right],
    \end{align}
    \begin{align}
      \vartheta_{\indIdler,k}(\nu_\indIdler)&\coloneq\Arg\left[u_k(\nu_\indPump^0-\nu_\indIdler)v_k(\nu_\indIdler)\right],
    \end{align}
  \end{subequations}
  respectively, where the principal value of the argument is taken in the interval $[0,2\pi)$.

  In this case,
  % \begin{widetext}
    \begin{align}
      \begin{split}
        \absolute{f_\indexrm{cav}^\indexrm{(approx)} (\omega_\indSignal,\omega_\indIdler)}^2
        &= \left\{
          \sum_{k} \left(\rule{0pt}{3ex}w_{\indSignal,k}(\nu_\indSignal)\exp[i\vartheta_{\indSignal,k}(\nu_\indSignal)]\right)^{1/2}
          \left(\rule{0pt}{3ex}w_{\indIdler,k}(\nu_\indIdler)\exp[i\vartheta_{\indIdler,k}(\nu_\indIdler)]\right)^{1/2}
        \right\}\\
        &\qquad \times
        \left\{
          \sum_{j} \left(\rule{0pt}{3ex}w_{\indSignal,j}(\nu_\indSignal)\exp[i\vartheta_{\indSignal,j}(\nu_\indSignal)]\right)^{1/2}
          \left(\rule{0pt}{3ex}w_{\indIdler,j}(\nu_\indIdler)\exp[i\vartheta_{\indIdler,j}(\nu_\indIdler)]\right)^{1/2}
        \right\}^\ast
      \end{split}\notag\\
      \begin{split}
        &= \sum_{k,j} 
        \left\{
          \left(\rule{0pt}{3ex}w_{\indSignal,k}(\nu_\indSignal)\exp[i\vartheta_{\indSignal,k}(\nu_\indSignal)]\right)^{1/2}
          \left(
            \left(\rule{0pt}{3ex}w_{\indSignal,j}(\nu_\indSignal)\exp[i\vartheta_{\indSignal,j}(\nu_\indSignal)]\right)^{1/2}
          \right)^\ast
        \right\}\\
        &\qquad \times
        \left\{
          \left(
            \left(\rule{0pt}{3ex}w_{\indIdler,k}(\nu_\indIdler)\exp[i\vartheta_{\indIdler,k}(\nu_\indIdler)]\right)^{1/2}
            \left(\rule{0pt}{3ex}w_{\indIdler,j}(\nu_\indIdler)\exp[i\vartheta_{\indIdler,j}(\nu_\indIdler)]\right)^{1/2}
          \right)^\ast
        \right\}.
      \end{split}
    \end{align}
  % \end{widetext}
  Here, the domain of the square root of a complex function varies depending on the choice of the principal value of the argument, and the resulting value may differ by a factor of $e^{i\pi} = -1$. However, regardless of the choice, it can be expressed as
  % \begin{widetext}
    \begin{align}
      \left(\rule{0pt}{2.5ex}w_{\indSignal,k}(\nu_\indSignal)\exp[i\vartheta_{\indSignal,k}(\nu_\indSignal)]\right)^{1/2}
      =\sqrt{w_{\indSignal,k}(\nu_\indSignal)} \exp[i\frac{\vartheta_{\indSignal,k}(\nu_\indSignal)}{2}] \exp[i\kappa_{\indSignal,k}(\nu_\indSignal)\pi],
    \end{align}
    by using a function $\kappa_{\indSignal,k}(\nu_\indSignal)$ that appropriately takes a value of either $0$ or $1$ according to the choice of the domain, and $\vartheta$ defined on $[0,2\pi)$. Consequently, we have
    \begin{align}
      \begin{split}
        \left(\rule{0pt}{2.5ex}w_{\indSignal,k}(\nu_\indSignal)\exp[i\vartheta_{\indSignal,k}(\nu_\indSignal)]\right)^{1/2}
        \!\left\{\!
          \left(\rule{0pt}{2.5ex}w_{\indSignal,j}(\nu_\indSignal)\exp[i\vartheta_{\indSignal,j}(\nu_\indSignal)]\right)^{1/2}
        \right\}^\ast\!
        &= \sqrt{w_{\indSignal,k}(\nu_\indSignal)} \exp[i\frac{\vartheta_{\indSignal,k}(\nu_\indSignal)}{2}] \exp[i\kappa_{\indSignal,k}(\nu_\indSignal)\pi]\\
        &\quad \times 
       \sqrt{w_{\indSignal,j}(\nu_\indSignal)} \exp[-i\frac{\vartheta_{\indSignal,j}(\nu_\indSignal)}{2}] \exp[-i\kappa_{\indSignal,j}(\nu_\indSignal)\pi].
      \end{split}
    \end{align}
    Since
    \begin{align}
      w_{\indSignal,k}(\nu_\indSignal) &= \left(\absolute{u_k(\nu_\indSignal)v_k(\nu_\indPump^0-\nu_\indSignal)}^2\right)^{1/2}\notag\\
      &=\left(
        \frac{1}{1+\frac{4}{\fwhm_\indSignal^2} \left(\rule{0pt}{2ex}\nu_\indSignal -(K_\indSignal + k) \fsr_\indSignal\right)^2}
        \frac{1}{1+\frac{4}{\fwhm_\indIdler^2} \left(\rule{0pt}{2ex}\nu_\indPump^0 - \nu_\indSignal -(K_\indIdler - k) \fsr_\indIdler\right)^2}
      \right)^{1/2}
    \end{align}
    is a function that takes non-zero values only in the vicinity of $\nu_\indSignal=(K_\indSignal+k)\fsr_\indSignal$, we obtain
    \begin{align}
      \begin{split}
        \left(\rule{0pt}{3ex}w_{\indSignal,k}(\nu_\indSignal)\exp[i\vartheta_{\indSignal,k}(\nu_\indSignal)]\right)^{1/2}
        \!\left\{\!
          \left(\rule{0pt}{3ex}w_{\indSignal,j}(\nu_\indSignal)\exp[i\vartheta_{\indSignal,j}(\nu_\indSignal)]\right)^{1/2}
        \right\}^\ast\!
        &\simeq \delta_{kj} \sqrt{w_{\indSignal,k}(\nu_\indSignal)} \exp[i\frac{\vartheta_{\indSignal,k}(\nu_\indSignal)}{2}] \exp[i\kappa_{\indSignal,k}(\nu_\indSignal)\pi]\\
        &\quad \times 
        \sqrt{w_{\indSignal,j}(\nu_\indSignal)} \exp[-i\frac{\vartheta_{\indSignal,j}(\nu_\indSignal)}{2}] \exp[-i\kappa_{\indSignal,j}(\nu_\indSignal)\pi].
      \end{split}
      \label{eq:JSAfuncInner}
    \end{align}
    Applying the same argument to the idler part, we have
    \begin{align}
      \begin{split}
        \absolute{f_\indexrm{cav}^\indexrm{(approx)} (\omega_\indSignal,\omega_\indIdler)}^2
        &\simeq \sum_{k,j} \left\{
          \delta_{kj} \sqrt{w_{\indSignal,k}(\nu_\indSignal)} \exp[i\frac{\vartheta_{\indSignal,k}(\nu_\indSignal)}{2}] \exp[i\kappa_{\indSignal,k}(\nu_\indSignal)\pi] 
          \sqrt{w_{\indSignal,j}(\nu_\indSignal)} \exp[-i\frac{\vartheta_{\indSignal,j}(\nu_\indSignal)}{2}] \exp[-i\kappa_{\indSignal,j}(\nu_\indSignal)\pi]
        \right\}\\
        &\qquad \times \left\{
          \delta_{kj} \sqrt{w_{\indIdler,k}(\nu_\indIdler)} \exp[i\frac{\vartheta_{\indIdler,k}(\nu_\indIdler)}{2}] \exp[i\kappa_{\indIdler,k}(\nu_\indIdler)\pi] 
          \sqrt{w_{\indIdler,j}(\nu_\indIdler)} \exp[-i\frac{\vartheta_{\indIdler,j}(\nu_\indIdler)}{2}] \exp[-i\kappa_{\indIdler,j}(\nu_\indIdler)\pi]
        \right\}
      \end{split}\notag\\
      &= \sum_{k=-M}^{M} w_{\indSignal,k}(\nu_\indSignal) w_{\indIdler,k}(\nu_\indSignal)\notag\\
      \begin{split}
        &= \sum_{k=-M}^{M} \left(
            \frac{1}{1+\frac{4}{\fwhm_\indSignal^2} \left(\rule{0pt}{2ex}\nu_\indSignal -(K_\indSignal + k) \fsr_\indSignal\right)^2}
            \frac{1}{1+\frac{4}{\fwhm_\indIdler^2} \left(\rule{0pt}{2ex}\nu_\indPump^0 - \nu_\indSignal -(K_\indIdler - k) \fsr_\indIdler\right)^2}
          \right)^{\! 1/2}\\
          &\hspace{10ex} \times
          \left(
            \frac{1}{1+\frac{4}{\fwhm_\indSignal^2} \left(\rule{0pt}{2ex}\nu_\indPump^0 - \nu_\indIdler -(K_\indSignal + k) \fsr_\indSignal\right)^2}
            \frac{1}{1+\frac{4}{\fwhm_\indIdler^2} \left(\rule{0pt}{2ex}\nu_\indIdler -(K_\indIdler - k) \fsr_\indIdler\right)^2}
          \right)^{\! 1/2}.
      \end{split}
    \end{align}
  \end{widetext}

  This expression is identical to $S_\indexrm{cav}^\indexrm{(approx)} (\omega_\indSignal,\omega_\indIdler)$ in Eq.~\eqref{eq:JSI_cav_approx}. Therefore, we can conclude that adopting $f_\indexrm{cav}^\indexrm{(approx)} (\omega_\indSignal,\omega_\indIdler)$ in Eq.~\eqref{eq:JSA_cav_approx} as the approximate representation of the JSA is appropriate.

\bibliography{komatsudaira_ImpOfEnt_reference}% Produces the bibliography via BibTeX.

\end{document}